\documentclass[12pt]{article}
\usepackage{graphics, color}
\usepackage{graphicx}
\usepackage{amssymb}

\begin{document}
\baselineskip = 18pt

\begin{center}
\bf  \Large Memories of a Theoretical Physicist
\end{center}
 \bigskip
 
 {\bf Joseph Polchinski}

 {\bf Kavli Institute for Theoretical Physics }
 
 {\bf University of California
Santa Barbara, CA 93106-4030 USA}

 \bigskip

{\bf Foreword:}

\bigskip
 \baselineskip = 16pt

While I was dealing with a brain injury and finding it difficult to work, two friends (Derek Westen, a friend of the KITP, and Steve Shenker, with whom I was recently collaborating), suggested that a new direction might be good.  Steve in particular regarded me as a good writer and suggested that I try that.   I quickly took to Steve's suggestion.
Having only two bodies of knowledge, myself and physics, I decided to write an autobiography about my development as a theoretical physicist.

This is not written for any particular audience, but just to give myself a goal.   It will probably have too much physics for a nontechnical reader, and too little for a physicist, but perhaps there with be different things for each.  Parts may be tedious.  But it is somewhat unique, I think, a blow-by-blow history of where I started and where I got to.  

Probably the target audience is theoretical physicists, especially young ones, who may enjoy comparing my struggles with their own.\footnote{I should mention that in 2009 I did an American Institute of Physics oral history, three long interviews with historian of physics Dean Rickles.  I deliberately avoided rereading it until I was done, to make it independent.  Of course, there is much overlap.  I would say that the interview was more focused on the highlights, while this is more about my whole development as a physicist.  Also, the AIP interview badly needs editing by me, I had not realized that before.}
Some disclaimers:  This is based on my own memories, jogged by the arXiv and INSPIRE.  There will surely be errors and omissions.  And note the title: this is about my memories, which will be different for other people.  Also, it would not be possible for me to mention all the authors whose work might intersect mine, so this should not be treated as a reference work.

One of the most difficult decisions was whether to refer to people by first name or last.  But first names are more subject to ambiguity, so I will generally go with the latter, even though it feels unnatural with good friends.

I thank Steven Polchinski and Bill Zajc for careful readings of the draft.  

I think Stanley Deser and Richard Kass for corrections.

Finally, I owe so much to Dorothy for her support over more than 40 years of our lives together, and especially over the last 21 months.  
 
 \tableofcontents
\baselineskip = 18pt

\section{Early years}

\subsection{Family history}

Whenever I am asked where I am from, I always want to answer `Caltech.'  In fact, I did not set foot on the Caltech campus, and barely in the state of California at all, before graduating from high school in Tucson, Arizona.  But Caltech was so formative in my life that it pales for me next to anything that came before.  However, I will start in the usual way, with a bit of family history.  This gives some context for later life, and may provide unexpected insights.

In the town of Hawthorne, in Westchester County, New York, you can find the Joseph Polchinski company, which has sold cemetery monuments since 1883.  It was founded by my great-grandfather, whose name I share.  My father shared the same name, but my grandfather was an `Arthur.'
So I am a `Junior.'   Among family I was distinguished as `Joey,'  and a few of them continue to use this even now. 

My father's grandparents came to the United States around 1870, part of the vast European migration driven by the combination of starvation and ambition.  One of them, Joseph, was from the region between Poland and Germany, while the other three were from Ireland.  Joseph brought his expertise in stonework with him, founding the monument company and the florist next door.  These supported his family for two generations, before they began to spread in the usual American way.  The monument company is now owned by another family, but I am always honored to see that they have kept the name for its historic value.

I know much less about the family of my mother, Joan Thornton.  From a very young age she was raised in a series of foster homes.  She ended up with a warm-hearted German-American family, but she seemed to retain a melancholy from her difficult earlier years.  I got only some basic history about her, and she never felt a desire to learn more.  She was born in Pennsylvania, but her final foster family was in the same New York town as Joseph Polchinski's family and his monument company.  Her ancestry was a mixture of Irish and other parts of the British Isles.

Growing up in the same small town, my parents Joan Thornton and Joe Polchinski married in 1951, when Joan was 19 and Joe was 22.  
I was born three years later, in 1954, and my sister Cindy three years after.  Our family was a rather typical one for the rising American middle class in the 1950's.  My father left the family business to earn a degree in accounting.  He went to work for Schenley, a distiller, commuting by train to his job at the Empire State building.
My mother worked for a few years in an office, then became a full-time homemaker.

Neither of my parents expressed an interest in science.  My father did say that he had wanted to study chemistry but could not because he had not taken German.  But our conversations rarely turned to science.  More common subjects were sports and games, though we did like games like bridge which had some aspect of mathematics.  He was highly competitive, a trait that I picked up.  In other directions, my father's reading tended toward history, and my mother's toward fiction.

\subsection{Early science and math}

My own interest in science appeared early.  When I was six, my passion was the How and Why Wonder Books of Science.  This was a series of several dozen books, each centering on a subject such as Dinosaurs, Atomic Energy, Chemistry, and Rocks and Minerals.  Each was 48 pages long, but in a large format that was packed with information.  The figures were hand-drawn but appealing.  I waited eagerly for each new issue.  Once I misbehaved rather badly, playing with an ember from a campfire, and the new issue was taken away from me for a few days; it was an effective punishment.

A few years later, Isaac Asimov's books in math and science drove me.  So also did science fiction, by Asimov, Clarke, and many others, giving an inspiring if unrealistic picture of what science might do.  Unfortunately, the science books and teachers through high school made little impression.  At that level the subject was too purely descriptive. 

I remember asking my physics teacher, what is the speed of gravity?  He did not understand the question, even though I drew a diagram illustrating how you would measure it.  Another misunderstanding, at an earlier age, was a test question, `Which is strongest: a) Pressure, b) Electricity, c) Gravity, d) Magnetism.'  I knew that the question made no sense, but having good test-taking instincts I knew they wanted the answer `Gravity.'  But this could not be correct: I could lift up my hand even against the gravitational attraction of the entire earth.  So I chose another answer almost randomly, refusing to make the choice that I knew was wrong.  I probably made a token argument with the teacher, but I was used to losing those.  But the smallness of gravity is indeed one of the  principles of physics.\footnote{In retrospect, gravity could have been correct, depending on the context.  Since gravity is the only force that is always additive, a large enough body of matter will attract with great strength.  So in the extreme case, gravity does win. }

One very exciting moment, on the other hand, was reading (no equations at this level) how an electric field can make a magnetic field, and a magnetic field can make an electric field, and these two together made a wave that was the origin of light.  So my future in science was clear, even if it took a few more years to get the details.
Thus, from an early age I was drawn to the basic principles of physics.  I am very fortunate that I have been able to spend my life studying this, and contributing new understandings.

With math, one gets closer to the real subject at a younger age, so the classes were more interesting. I raced through my courses, meeting the New Math in fifth grade.  This program was a response to Sputnik, and the perception that the US was falling behind the Soviets in science (the How and Why books likely had the same origin).  I can remember the school assembly, where all the students and their parents learned about this new thing.  The plan was actually rather bizarre.  Students would first learn such abstract notions as sets and operations, only moving on to arithmetic after the theory was understood.  It is hard to believe that anyone thought this was a good idea, and indeed it faltered in a few years, but it was perfect for me.\footnote{I have just learned, from Wikipedia, that Richard Feynman was on the California State Curriculum Commission at just this time, and was one of those to criticize the New Math.}

Unfortunately I missed the full benefit of the New Math because we moved to Tucson, Arizona a year later.  My father was looking for a better job, as an account manager at Merrill Lynch, a stock brokerage, and Tucson was one of the available openings.
Perhaps too my parents were ready to leave the small town they had grown up in.  So the chance to race ahead in math was delayed a little.  I missed another chance around the same time: my father was second in line among the applicants for a position as business officer at the Institute for Advanced Study, where my connection with science may have been accelerated.

Canyon del Oro (CDO), my combined Junior High/High School, was a new school, and a small one, which would limit me in some ways.  But I had the good fortune that my first math class was with Ed Baceski.  Mr. Baceski loved math, and he made it a game.  For example, 
completing a problem set would lead to a code to unravel (and you could short circuit the problem set by working backwards).
In retrospect, Baceski was a bit like the new math, not ideal for the typical student, but great for me and a few others.  Early on he set the Gauss problem, summing 1 to 100, and after I solved a few of these I was allowed to race on in the textbook on my own. I completed four years in one, through geometry.  My most vivid memory was starting trigonometry, reading on my own, and not getting the point of this `sine' and `cosine.'  But after a couple of days it suddenly fell into place, and it was wonderful.

The next year, I took Advanced Algebra, the highest level offered in this small school.  It was taught by the football coach, leading to more of the sorts of disagreements that a student doesn't win.  In retrospect, there might have been a right way and a wrong way to make such points.

Having run out of math classes, I spent my first high school year commuting evenings to the University of Arizona for calculus, driving with my father or some older students.  Unfortunately, this did not go well.  Part of this was the instructor, who contributed little insight or inspiration.  One day we had a substitute, who regaled us with stories about math, and in particular challenged my precocity with examples of great mathematicians who had accomplished much more much earlier than I (he could see that I was full of myself, and needed this).  But then it was back to the regular teacher.  

The second problem was that I couldn't really grasp calculus, just as earlier I  couldn't get trigonometry for a while.  But in this case it took three years, when I took college physics and found out what calculus was really for.  (Mathematicians might tell you that it has other uses, but they would be wrong.)

Disappointed by the class, I decided I could learn math on my own.  I chose a book on group theory.  Unfortunately I again seemed to lack the knack of the subject, and my effort faded.  I ended up spending most of my last two years of high school studying no math.  Science was similar.  My small school had no advanced courses, so after racing through the sciences that were available I found myself with a year of no math or science classes at all, spending it taking the other required courses to graduate a year early.

\subsection{Family}

My sister Cindy and I seem as different as two people can be, in personality, interest, and career.  Where my passion was physics, hers was animals, horses in particular.  She took only one year of college, and that was to mollify our dad.  She was then a groom on a large stable near Santa Barbara.  Over the years she has owned horses, bred them, competed on them, and most recently served as steward at horse shows all over the country.  

To support her interests, she also served as a police officer for almost 20 years.  This is something I could not imagine doing; for one thing, I can't make quick decisions.  But she did this with aplomb.
Cindy is not academic in her interests, but she is extremely capable.
Yet another difference is that I have always been shy, working up from extreme shyness when young to mere introversion today.  My sister is the opposite, taking great pleasure in meeting and talking with people from many walks of life.

In spite of our differences, we have always gotten along well, and she is a great supporter.  She has often told me that she looks forward to traveling with me to Sweden when I win the Nobel.  That is not going to happen, but I did take her to the LHC a few years ago.

My parents were as helpful as they could be, given that they did not understand what this alien in their family was doing.  My father was the type who always had to be in charge.  When I told him what I was learning in school, especially later on when we got to relativity, he told me that this could not be true.  So my father, I am sorry to say, was a bit of a  crackpot.  The number of people who have never studied science but still feel qualified to present their ideas is remarkably large: notably, 99\%  of them are male.  Indeed, my mother did not have such theories.  She did make it a point, many years later, to tell me that she had been very smart in school.  Unfortunately, the limitations experienced by so many women prevented her from pursuing this.

\subsection{Interests}

I did have some stimulation outside of school, notably science fiction, telescopes, and chess.  I mentioned science fiction before.  It is curious to recall that this was almost entirely through books.  Star Wars was still seven years away, and with a few exceptions like `War of the Worlds' and `2001, A Space Odyssey,' there was not a big market for science fiction movies.  It is remarkable how it now dominates.

My interest in telescopes began with the surprise gift of a 4-inch reflector from my parents when I was 12.  This was an excellent idea.  Tucson was then a rather small city, and we lived on the edge, where well-separated houses trailed off into desert. The seeing (air clarity) and darkness were incredible.  My interest was drawn to picking out galaxies, finding as many of the Mersier catalog as I could.  My interest was mostly visual, I was too young to follow the science.

After exhausting the potential of the 4-inch, I set out to build an 8-inch reflector.  I did not have a large budget or a lot of mechanical aptitude, so the results were mixed.  I made a creditable mirror, working it against another glass using progressively finer grit and measuring my progress with the help of the University of Arizona's astronomy club.
But the mechanical support was built with whatever wood I could get hold of, patterned on a scaled picture of the Hale telescope.  This worked, and was great for showing off, but it was well short of the real capacity of an 8-inch.  Still, finding the Crab Nebula was one of my favorite challenges.  Seeing Andromeda was easy even by eye, and I still can pick it out in Santa Barbara on a good day.

Chess dominated much of my school years.  I learned the moves from my father when I was young (aside from some confusion about the pawns).  After occasional games with my father and a few friends, my interest exploded when I got to CDO and discovered a group to play with.  For the next five years, at almost every lunch period or other break, we would pull out our boards and play.  As I got better, I played in local tournaments, and in larger ones in Phoenix. This was a lot of fun, and virtually my only social life.  
In my last two years, when I had run out of math and science to study in class, I spent many hours studying chess books, about chess openings, and attacks, in particular.

There is an anomaly here, which has always puzzled me.  Based on my progress in physics, first in progressively more advanced courses, then in original research, and finally in significant discovery, you could say that in physics I am the analog of a fairly strong Grandmaster.  In chess, I started out as a beginner, and in a few years had worked my way up to the level of a good recreational player.  In my last two years, working nearly full-time on chess, I expected to continue to improve.  Instead, I came to a virtual standstill.

Chess has a nice numerical system, called ELO.  Based on their wins and losses, each player has a numerical rating.  Grouping them, they are designated
\\
 ... $<  D  <  C  < B < A < $ {\sl Expert $<$ Master $<$ Senior Master $<$ Grandmaster}.
\\
Roughly speaking (the full theory is  more elaborate), if two players are separated by M levels, the  relative probability that the higher ranked player win is $3^M$.
When I started out I was a $D$, a beginner, and after three years I rose to $A$.  But I never quite reached Expert, much less the promised land of Master and beyond.

I have always wondered why.  Are chess and physics so different, that one can be a grandmaster in one, and not even an expert in the other, in spite of similar efforts?  Seeing younger and younger teens achieve grandmaster has always amazed me.

I got one clue when I ran into a high school chess buddy many years later.  When I had first met Keith Nelson in school, he challenged me to a game.  Having faced such challenges often, I expected a quick victory, but he beat me.  I was sure that with a bit more concentration, I would set things right.  But he beat me again!  Over time I won a share of the games, but he was clearly the better player.  So, perhaps twenty years later, I ran into Keith again.  I had not known of his interest in science, but he had in fact become a  professor of experimental chemistry at MIT.  And as we began to reminisce, he astonished me by recounting in detail our first two games, which I could remember only dimly . 
Evidently he had phenomenal memory, at least compared to mine.

Indeed, I have always felt that I did not have an especially good memory.  In one of my first classes in college, the instructor told us that you do not need a good memory to do physics, because you can derive everything from first principles.  If I had had any doubts that this was the right field for me, that sealed it!

Beyond the issue of memory,  I did not have a real knack for chess.  I was conservative, using a few basic attacks and waiting for the opponent to make a mistake.  I did not like to advance pawns, because the effect is irreversible.  This is not the way that Grandmasters think!  Likely with training I could have done much better, but not been a prodigy.  I am curious, what distinguishes these different mental strengths?

\subsection{Traits}

One thing I want to do is to recall some of my development as a physicist.  There are a number of traits that have played a role here.  Many of these have already come up in the discussion of my early life.

To start with, my parents and relatives could see from a very early age that I was not a normal kid.  I could solve puzzles and games at a level far above my age, and my general knowledge was advanced.  So from a young age, this was my identity: being very smart.  It has stayed with me as I have moved from level to level, all the way to string theory.

On the other hand, I have noted that I was painfully shy all through school.  I tried to keep conversations as short as possible, so as not to bore people.  Only gradually, in college and beyond, did this fade.

I also think I have some lack of common sense.  My poor telescope design was one example.  Another was my two year gap in high school math: with common sense I should have looked for advice.  And my approach to chess also seems to have a lack of common sense. 

In a sense, shyness and lack of common sense were two sides of the same coin.  If you talk to other people you learn things.  If you don't, you have to figure everything out yourself.  Even after maturing from shyness to introversion, I tend not to ask questions or seek help.  This may be one of the reasons that my science didn't really reach its peak until rather late.

\section{Caltech, 1971-1975}

\subsection{Arrival}

The information available to college applicants today of course dwarfs what we had then.  For me, the principal sources were Reader's Digest and the World Almanac and Book of Facts, but these were enough.  A few years before college, I read a Reader's Digest article about the famous Caltech Rose Bowl prank.  

The Washington team had planned a show in which several thousand fans in the stands would manipulate a set of cards, producing stadium-sized images.  This went well until the last two cards, when `Huskie' was replaced by `Beaver,' and then `Washington' was replaced by `Caltech.'  This had required an elaborate scheme by the Techers, culminating on the night before the game, when they replaced the thousands of individualized instructions.  The article also mentioned the unique scientific environment of Caltech.

Having never heard of this place before, I went to my other source, the Almanac.  It had a nice table of colleges, from which I learned that the student/faculty ratio at Caltech was around 2-1, compared to 20-1 at any other college.  Pranks and top faculty ---  I was hooked.  I never even read up about  other schools --- probably a lack of common sense, but it worked out this time.  I applied to Caltech
on early decision, and then waited with  bated breath.  Coming from a small school in a small state, I had no idea of where I stood.  But when the letter came it was positive, and I sat back and laughed for a long time.  This is what I had been waiting for.

So, in September of 1971, my family drove up from Tucson.  This was a good day's drive, but my parents both enjoyed driving. 
And so we rolled up to my dorm, Blacker House, where I would spend the next four school years.  We said goodby, my mother crying on one of the few occasions I can remember, and my new life began.

\subsection{Zajc}

During my first week on campus, I met three remarkable people: Richard Feynman, Kip Thorne, and William (Bill) Zajc.  Feynman had won the Nobel Prize when I was still in grade school, more than 50 years ago now.  Thorne may win the Nobel Prize next year.  Zajc, who is probably least known to most of you, is a distinguished scientist as well.  He led the development of the PHENIX heavy ion detector at Brookhaven.  This may not  lead to a Nobel Prize (though who knows?), but it did reveal a connection between the entropy in nuclear physics and that in string theory.  At the time we met, Feynman was a star, Thorne was a rising young star, and Zajc was, like me, a young whippersnapper setting his first steps on the Caltech campus.

Zajc, like me, was in Blacker House, so we met on Day 1.   Almost as quickly, he had an enormous effect on my life.
As I've noted, I left high school with no knowledge of what physics really is.  But  at Caltech, I was immersed in it.  Zajc was a big part of this.
In high school in Wisconsin, he had already read some of the Feynman lectures, Feynman's three volume introduction to physics.  In fact, this was to be used as the course for the `advanced track' for introductory physics.  So Zajc got to Caltech with a much clearer picture of what physics was.

Zajc was outgoing (for a Techer), and he loved to talk about physics.  I quickly learned what I had been missing.  As Zajc told me amazing facts, and we worked through some key equations, I could see for the first time how calculus and physics fit together.  I saw that this was what I was designed for.  I had brought my chessboard with me to school, planning to practice an hour a day, but it was never opened.  

For four years Bill and I took most of the same classes, working together as we learned physics.  But as much as the physics, I remember the ways that Bill, our other friends, and I blew off steam in between.  These included drives for Tommy's burgers and various sports.  A group of us became avid cyclists, riding to the beach at Santa Monica and exploring LA on rides as long as a century (100 miles).  Our special challenge was the ride to the top of Mount Wilson.

\subsection{Pranks}

Exploring the Caltech campus to discover its secrets was another pastime.  At the most benign, this meant wandering the hallways of physics and other departments to see what went on there.  More interesting were the night-time excursions, clambering into construction sites, negotiating Caltech's elaborate steam-tunnel network, or simply applying the knowledge of lock-picking that had been passed down from generation to generation of Tech students (the physics building, Bridge, was known to be particularly easy).  Finding some interesting, perhaps historic, piece of scientific equipment was a Prize.

In this way we carried out (though not on the scale of the Rose Bowl!) the kind of pranks that first lured me there.  Our most notable one involved the roof to the library.  The Caltech library, at nine stories, loomed over the campus.  It was a natural location for undergraduate activities, which usually involved throwing, or floating, objects that may or may not have been flammable, from the roof.  The door to the roof was locked, but this could be picked by the more talented Techers.

Caltech security did not think this was a good thing, and so the lock was replaced by an `unpickable' one.  Indeed, our best locksmiths could not open it.  This was an outrage, and a Plan B was necessary.  There was a ventilation shaft nearby, opening on each floor and then out on the roof.  The frame could be unscrewed, so all that was needed was someone with no common sense to climb the shaft from the top floor to the roof.
Having a particular talent in this area, late one night I found myself wriggling up the last twelve feet of the shaft, with 9 stories beneath me and tethered by a climbing robe that I had never tried.  Fortunately it was an easy climb, aside from a tense moment when I was stuck and unable to move in either direction.  I got to the top and opened the door from the back 
 
Having defeated the unpickable lock, we had to celebrate.  We removed the door from its hinges, and a dozen of us carried it down to the library basement and then through the underground steam tunnels to the Caltech security office.  The lock-pick experts  opened the office and we left the door, but only after the group of us, including Zajc and me, signed it.  This might seem to be a foolish thing, but we knew that Security understood Techers and was easy on us; in fact, there ended up being no penalty at all (Warning: things are different now. Do not attempt.).  And there was a nice bonus.  Security replaced the door without
removing or covering our signatures.  So as of a few years ago, 40 years after the fact, my signature, Zajc's, and the others (with maybe a fake or two) could still be seen on the door to the library roof. 

\subsection{Feynman}

I first met Feynman as an idol, not a person.  In the courtyard of Dabney house, next to Blacker, a large bas-relief depicting the great scientists of history had been built many years before.  When Feynman was awarded the Nobel Prize  with Julian Schwinger and Sinichiro Tomonaga in 1965, the Dabney's students replaced the dominating figure of Galileo with that of Feynman.  Thus we were surrounded by Feynman all the time, from his image, his books, and many other reminders.
For many of us, Feynman was who we wanted to be. 

I got a chance to meet the man himself before too long.  Once a week, Feynman led Physics X, where freshman and sophomores could  
ask their questions about physics, or if we ran out of questions he would talk about some of his ideas.  One example of this was, how do you take the square root of the Fourier transformation, so that acting on a function twice with the operation would be the same as the Fourier transform.  For those interested, the answer is in the footnote,\footnote{In phase space, the Fourier transformation $x \to p \to -x$ is a $90^{\circ}$ rotation.  So rotate by $45^{\circ}$ (or $225^{\circ}$, it's nonunique).} but try it first.  
This kind of happy creativity was fascinating to see.  Another question was, what is a negative probability?  Unfortunately, my main contribution to the class was falling asleep one day, in the front row, which has delighted some of my classmates to this day.

One of the notable sights in Bridge Hall was a pair of small objects, a miniature page of text and a miniature motor.  Both were inspired by Feynman's lecture `There's Plenty of Room at the Bottom.'  He foresaw the smaller and smaller scales that physics and technology could reach.  In addition to his powerful calculational ability and his outsized personality, Feynman's ability to think far outside the box was awesome.  Another example was the idea of quantum computation, where the `Plenty of Room' comes not from space but from Hilbert space.

So, many of us wanted to emulate Feynman.  As I began to stand out in my classes, a couple of my classmates proclaimed me the next Feynman.  I was happy to hear this, of course, but I knew better.  In contrast to Feynman's striking originality, I have always felt myself to be weak in this area.  This is not just me being self-effacing; you can judge it as we go along, but my impulse has simply been to follow my nose.

I was too shy to take more advantage of the time with Feynman, though I saw him often on that small campus.  I did hear his stories at one faculty party, some of the same stories that later appeared in his book.  Most exciting, when we were seniors, Zajc and I, along with two other seniors, were asked to grade Feynman's junior quantum mechanics homework.  My strongest memory of the class is the very  beginning, when he started, not with some deep principle of nature, or some experiment, but with a review of Gaussian integrals.   Clearly, there was some calculating to be done.

I did get over my shyness one time, to ask him about the infinities that appear in quantum field theory (QFT): do they have a physical interpretation?  Feynman said `no.'  In retrospect, he must have known more, from the work of Wilson, Weinberg, and others.
But perhaps it did not satisfy him, since he had not derived it himself.  But this question tugged on me for the next eight years, and was my first deep result.\footnote{Zajc reminds me of another interaction we had with him, asking about whether rotating bodies produce gravitational radiation, something we were puzzling over.}

\subsection{Thorne}

Kip Thorne was my designated freshman advisor, so we met every quarter.  His first order when I went to his office was `Call Me Kip.'  This I could not do, so I spent the year addressing him without using his name.  

Thorne's office door was covered with interesting artifacts. Most notable was a bet between Kip and Stephen Hawking, as to whether the radio source Cygnus X-1 was a black hole: Thorne bet yes, and Hawking no.  If yes, it would be the first confirmed black hole in nature.  Actually both wanted, and expected, the answer to be yes, but Hawking was covering his bet: if he was disappointed by the black hole, at least he'd win a magazine subscription from Thorne.  But indeed it was soon confirmed.

Thorne was a leader in general relativity, with a particular interest in the rich astrophysics of black holes.  
This required a team, so one always saw him with a  group of enthusiastic grad students and postdocs.  
As we have seen, when Thorne began black holes were still theoretical, though the theory was compelling.  Soon there was Cygnus X-1, and in time an enormous number more, from quasars down to collapsed stars.  Most recently, Thorne capped off his remarkable career as one of the leaders of the LIGO project.  This billion dollar experiment gave the first observation of gravitational radiation, predicted nearly 100 years ago, and the first observation of coalescing black holes.

I did not have much interaction with Thorne as a student, aside from auditing his general relativity class.  The research was too advanced for an undergrad.  I did have an interesting science/sci-fi interaction with him several years later, which I will get to.

\subsection{Classes}

Thus surrounded by these and many other outstanding scientists, my education went forward.  The Feynman lectures were one of the highlights of the first two years.  They were not perfect: as with everything Feynman did, he redid the subject from his own approach.  This was inspiring, but challenging.  There was a shortage of examples and calculations, but these were supplemented by a  set of problems authored by two other high level professors, Robert Leighton and Rochus Vogt. 

There were also a variety of other subjects - Astronomy, Chemistry, Advanced Calculus, Electronics.  I took too many courses, this was common at Caltech.  There was so much to learn, one wanted to cram it all in.  Happily, only one non-science course per quarter was required.  In my four years I did not take much math: advanced calculus, complex variables, and a course on group theory that I again failed to grasp.  I think I was influence by an offhand remark from Feynman that one did not need to know much math, but it worked for me.

\subsection{Tombrello}

Having so much fun in school, I did not want to leave during the summers.  Today, undergraduate research is expected, but back then it was more hit-or-miss.  Happily, Tom Tombrello was there.  Tombrello was a nuclear and atomic physicist, working in particular on measuring the nuclear decay rates needed to understand the formation of the chemical elements.  He was also a remarkable people person.  When he saw that four of the top physics students (Bill, me, Roland Lee, and Ken Jancaitis) were looking for research projects, he took all of us on!

This was heaven: four of us sharing a basement office in Bridge, with a modest stipend, talking physics all day and unwinding at night.  And Tombrello did not just put the four of us on some large project.  We each had our own problem (which might be part of some larger collaboration), coming from Tom's many interests.  Over time I worked on estimation of waveguide shapes, calculating nuclear decay rates, and finding methods to study intermediate energy atomic collisions.  He even showed one of my plots to Hawking when he was visiting Caltech.

Tombrello was one of the rare physicists who did both theory and experiment.  He used the Van de Graff generator to study nuclear interactions, and so I got some time learning to run that.  He also guided my senior thesis, attempting to zone-refine gallium in order to detect solar neutrinos. 

Tombrello told me I should follow his path, and that of Fermi, doing both experiment and theory.  But I was set on theory by nature.  I remember spending a few hours moving some lead blocks with Tom, and thinking I did not want to spend my career moving lead blocks.  But of course theory has its own drudgery, such as searching for factors of two.  But I may have disabused Tom when I managed to break both the multi-channel analyzer and some expensive glassware on my senior project.

Even after graduation, Tombrello kept in touch.  He sent me a copy of Hawking's information paradox paper, written while a guest at Caltech, together with a note `Joe, you should work on this!'  He was right, but it took me a few years to get there.

Tombrello took a break after the four of us graduated, but several years later he instituted Physics 11 as a regular undergraduate research course.  Tom passed away unexpectedly a few years ago.  At his memorial, it was remarkable to hear about all the aspects of his life.  The number of people he had affected, and especially his talent for bringing people together, were wonderful to hear about.

\subsection{QFT, GR, QCD}

Senior year, physics got even more interesting.  I took QFT from Frautschi, and General Relativity from Thorne.  I did not end up with a good grasp of either subject.  These days it is rather routine for seniors in theoretical physics to take these courses, but the subjects then were more difficult.

QFT was undergoing rapid development, and the textbooks had not yet caught up.  The two volumes of Bjorken and Drell were the text of general choice.  This was a beautiful book when it was written, but ten eventful years had gone by, and a new text was needed.  

General Relativity did have a new text, and that was the problem.  Charles Misner, Thorne, and John Wheeler had just rewritten the subject in an epic text known widely as the Big Black Book.  Unfortunately, it was almost impossible to learn from, especially by me.  It was intended to present the subject in a geometric way, which most people would take as a good thing, but it went too far, so that it seemed like reading pictures.  Robert Wald, several years later, presented the geometric story in a more conventional way.  For me, Weinberg's book, following particle physics as much as possible, was ideal, and I learned this way as a grad student.  Weinberg explicitly downgraded the role of geometry in gravity, never even mentioning black holes.\footnote{It occurs to me that even today, our most precise description of black holes is gauge theory, not geometry.}  

In field theory we had a notable guest lecture from David Politzer, a new member of the department.  Three years earlier, David Gross and Frank Wilczek from Princeton, and Politzer from Harvard, had discovered the principle of asymptotic freedom.  This showed that due to quantum loops, the interaction strength of the strong nuclear force could grow larger at larger distances.  This then explained how the weak force seen between quarks at high energy was consistent with quark confinement at long distances.  Thus they had determined the nature of the strong nuclear force, so-called QCD, and established QFT as the correct framework for particle physics.

Unfortunately, no one at Caltech had been working on this.  Feynman and Gell-Mann each liked to follow their own directions, though ironically asymptotic freedom explained the relation between Feynman's partons and Gell-Mann's quarks.  So Politzer was the first direct connection with the new physics.

Another source of particle physics excitement was the discovery of a new long-lived heavy particle, something that had not been seen before, the $J/\psi$.  This was big enough news that even the undergrads knew they should attend the colloquium.  After the observation was described, various faculty members put forth their theories.  Feynman thought it might be free quarks, while a young professor, Glennys Farrar, proposed that it was a bound state of the charmed quark with its antiquark.  Fairly quickly, the latter was confirmed.  In fact, the existence of the charmed quark, as well as its mass and other properties, had been correctly predicted by Sheldon Glashow, John Iliopoulos, and Luciano Maiani several years earlier, a great success of theory.

For nonspecialists, here is a handy list of acronyms:\\
{\bf AdS/CFT: equivalence between quantum gravity in a certain curved space and a supersymmetry gauge theory in one less dimension.}\\
{\bf AdS/CM: use of AdS/CFT to model strongly coupled condense matter systems.}\\
{\bf AdS/QCD: use of AdS/CFT to model strongly interaction nuclear system.\\}
{\bf BFSS: matrix theory, exact description of M theory.}\\
{\bf BHC: black hole complementarity.}\\
{\bf BPS: partially supersymmetric state.}\\
{\bf CGHS: a model of gravity in two dimensions.}\\
{\bf GR: general relativity.}  Gravity arising from the curvature of space and time. \\
{\bf GUTs: grand unification.} \\
{\bf KKLT: first full model of stabilized string vacua.}\\
{\bf QCD: quantum chromodynamics.} The theory of the strong nuclear force.  The `chromo' comes from Gell-Mann's whimsical labeling of the three kinds of quark as red, green, and blue.\\
{\bf QED: quantum electrodynamics.} The quantum theory of electromagnetism.\\
{\bf QFT: quantum field theory.}  Quantum theory in which the basic variables are fields.  Confirmed in 1971 as the basic framework of quantum mechanics and matter.  The particles appear from the solutions for the quantum mechanics of the fields.\\
{\bf SUSY: supersymmetry.}\\

\section{Berkeley, 1975-1980}

\subsection{Moving on}

After four years, it was time to choose a graduate school.  I knew that I wanted to do theoretical physics, and my choices came down to Berkeley, Stanford, Harvard, and Princeton.  I tried to be scientific about my choice, but had a strong leaning for Berkeley: several friends, including Zajc, were going there and I liked California.  Also, I had a nice Hertz fellowship, which at the time was restricted to a small number of schools, including Berkeley.  

At the time, Hertz had a strong defense orientation.  One might think that liberal Berkeley would be ruled out, but it also had a close connection with nearby Lawrence Livermore Lab, designer of nuclear weapons.  I took the fellowship and went to Berkeley.  Participation in defense was not at all mandatory, aside from a pledge to aid the US in time of need.  My own connection was limited to a one-day tour of Lawrence Livermore (but only outside the security fence, because I did not have a security clearance).  

There was one off-note in my grad school search.  Paul Martin, the Harvard physics chair, was passing through and stopped to do some recruiting.  His main argument that I should go to Harvard was that East Coast physics was better than  West Coast physics.  I had never heard of such a thing, Berkeley after all had a great history.  So I did not take this seriously, and off to Berkeley it was.  If he had been more direct, things might have gone differently.

\subsection{Finding a major, and the East-West divide}

I was one of those students who thought he had to learn everything before starting research.  Retaking QFT and relativity were essential, of course.  Field theory was taught again from Bjorken and Drell.  So I learned the equations this time, but I still thought that the basic principles were not clear.  At least relativity was taught from Weinberg, so I got that subject down, though from the point of view of a particle theorist, not a relativist.

I had a vague notion that I should learn more about the other possible majors, but I really knew that I was going to end up in particle physics.  And so somewhat belatedly I began to look for an advisor.  And then I learned what the Harvard recruiter was saying.
With Caltech, the dominance of two charismatic professors, Feynman and Gell-Mann, had slowed the reaction to the discovery of asymptotic freedom and all that it implies.  At Berkeley, another charismatic professor had the same effect.

In 1948, Feynman, Schwinger, Tomonaga, and Dyson had found the correct theory that incorporated quantum mechanics, special relativity, and electromagnetism.  The theory, known as quantum electrodynamics (QED), was based on the principle of QFT.  It made predictions of enormous accuracy.  Thus it was natural to look for the same kind of theory for the nuclear force.  It would have to be much stronger: the coupling constant in QED is a convenient 1/137, for the strong force it would need to be around 1.

But this did not seem to  work, to describe the strong force in this way.  Thus it was that throughout the 1960's there was a search for new approaches.\footnote{It is interesting to look back at the 1967 Solvay meeting.  The speculations were fairly radical, like breakdown of spacetime inside the nucleus.  Few were near the mark.}  One idea that attracted wide attention was the bootstrap: the idea that one did not need fields, but just a few principles like Lorentz invariance and crossing.  This idea was suggested by some of the data, in particular the presence of a large spectrum of massive excited states.
The leader of the bootstrap program was Geoffrey Chew, of Berkeley.  He was said to be religious in his fervor for the idea.  Even notoriously strong willed field theorists such as David Gross and Steven Weinberg, who went through Berkeley as grad student or postdoc, were afraid to mention QFT.  

The bootstrap did lead to some interesting work.  It was elaborated into the Dual Resonance model, which in turn led to string theory.  But this was the solution to the wrong problem, and too soon.  Thus, when asymptotic freedom was discovered and the strong interaction was understood, the work was done not at Berkeley, but on the East Coast.  Even four years later, when I was looking for an advisor, there was almost no one working on QFT.  One prospective advisor told me `You should have gone to Harvard.'  Another told me he had stopped supervising students.  But fortunately there was one choice.

\subsection{Mandelstam}

If Geoff Chew was the spiritual leader of the bootstrap, Stanley Mandelstam was its engine, solving the difficult problems that were required.  When the world tilted from the bootstrap to QFT, it took Mandelstam a little while to catch up, but when he did he did it in his own powerful fashion.  It was never Mandelstam's style to start with some easy problem to learn from.  He chose the most important, and most difficult, problems.\footnote{Whether by inclination or by example, I tend to be the same way.  Frank Wilczek once told me, `Joe, an important problem doesn't have to be hard!'} 

The problem that he focused on in QCD was the nature of quark confinement.  The fundamental fields of QCD, the quarks and gluons, are not seen directly in experiment.  Rather, we see mesons and hadrons, which are bound states of the quarks.  Somehow the quarks are prevented from escaping to be seen as individual particles.  Asymptotic freedom points to the idea: the growing coupling at long distance binds the quarks.  But a more explicit demonstration was needed.

The idea that Mandelstam hit upon was suggested by superconductivity.  In superconductors, the Meissner effect repels magnetic fields from the superconductor.  If one inserted a magnetic source into a superconductor, it would be confined into a tube and grow linearly.  So what was needed to confine quarks, which are electric sources, was a `dual superconductor,' with electric and magnetic fields interchanged.  Indeed, this suggested a duality symmetry, the equivalence of electric and magnetic theories under change of variables.

So Mandelstam seemed like the only choice, and I asked him if I could do research with him.  He hesitated, probably because he had just taken on two other students, but then agreed.

Students are generally started off with a warmup problem.  This is for the student to get oriented to the advisor's research, and for the advisor to gauge the student.  But as Mandelstam only worked on the hardest problems, he naturally gave the same to his students.
My warmup was to find a QFT that had both electrically and magnetically charged particles.  This is an contrast to the known theory of QED, which has only electric charges.

I was unable to solve this problem, and I gave Mandelstam an argument why it was impossible.  In a theory with both electric and magnetic charges, Dirac showed that quantum mechanics requires the electric charge $e$ and the magnetic charge $g$ had to be quantized, $eg = 2\pi$.  For QED, $e = 1/\sqrt{137}$ is very small, and so we can use perturbation theory, expanding around $e=0$.  But when $e$ goes to zero, $g$ goes to infinity.  In this limit, we had no way to calculate, or even to know if the theory made sense.  

In fact there is a solution, though I don't know if this is what Mandelstam had in mind.  It requires the electric and magnetic objects to be different: the electric charges would be ordinary field quanta, but the magnetic charges would be solitons --- sort of like bound states of many particles.  I am not sure if this was within the intended bounds of the problem, but it involved too many new ideas for me.

In fact, it now seems that {\it every} quantum theory with electric charges also has magnetic charges.  I may not have been the first to enunciate this principle, but I have made frequent use of it, especially with D-branes, and I will of course return to that.  What about ordinary QED?  I would bet, at high odds, that it does in fact have magnetic charges.  But it is unlikely that either side of the bet could ever collect, because one needs to get to the Planck scale to be sure.

So Mandelstam gave me another project, which was intended to be my thesis.  This was to construct the 't Hooft vortex operator. 
Like the earlier problem, it was a part of Mandelstam's broader program to understand confinement in terms of electric-magnetic duality.  Ken Wilson had shown that one could distinguish different states in a gauge theory, in particular the confining state, by measuring a certain operator.  This operator, defined as the integral of the vector potential along a one-dimensional path, had come to be known as the Wilson loop.
Gerard 't Hooft, who was pursuing the idea of confinement from duality independently of Mandelstam, noted that there should be a dual to the Wilson operator, with the electric potential replaced by a magnetic potential.  It was my goal to fill in the details of this.

So I met with Mandelstam to discuss this about once a week for a year.  Mandelstam was always generous with his time.  But he was a difficult advisor, because his thinking was deep, but his explanations were often oracular.  So I was never sure if I was making progress.  Sometimes, in response to a question, he would turn to the blackboard and just think for several minutes before responding.  I never knew whether this meant that this was a good question or a dumb one.

I have always thought that my project was unsuccessful.  But on reviewing it for the first time in a very long while, I realized that I had basically solved the problem.  One needed an operator  whose physics effect was only in one dimension, and which also had a singular gauge piece acting in three dimensions.
But I also had a lot of irrelevant stuff mixed in.  I had not really mastered path integrals, which had not made it into the standard texts.  So I was using canonical methods, which were very clumsy for this kind of problem.  

Rather strikingly, my central problem was not clearly solved until 25 years later, by Anton Kapustin.
This required several new ideas, such as conformal invariance, that had not yet been applied.  He was also able to treat the nonabelian case.   This was typical of Mandelstam, how far ahead he was in much of his thinking.  Another example, the first paper that one studies in the Langlands program today is the first paper that Mandelstam gave me to read 40 years ago.   

Another striking example was eleven-dimensional supergravity.  I have this distinct memory that on several occasions we would be discussing QFT and quark confinement, and Mandelstam would make some remark about eleven-dimensional supergravity or string theory.  This was a bit mind-blowing for someone struggling with four-dimensional QFT, and for whom string theory was assumed to be an artifact from the past.  I did not know what to make of this, so I just waited for Mandelstam to come back to earth.  But a few years later string theory was back in the center of things, so Mandelstam was as usual well ahead of time.

\subsection{Colleagues and visitors}

Mandelstam's two other grad students at that time were Susan Elma Moore and Omer Kaymakcalan.  Each of Mandelstam's students had a different project, but all were connected in various ways through QFT and QCD.  Susan's project was to find variational states of heavy quarks, based on a Wilson loop model of the states.  In his own work, Mandelstam was never following the herd, and he guided his students the same way.  This could be challenging, as I  have noted.  In Moore's case, after writing her dissertation she changed fields --- first spending some time trying acting, and then ending up as a doctor.

Kaymakcalan's graduate project was to understand the non-abelian properties of the Higgs phase.  After his thesis, Omer went to Syracuse and worked on various projects with the members of the Syracuse group.  Several of his papers, dealing with chiral Lagrangians, proton decay, and strongly coupled Higgs dynamics, were well-cited.  Sadly, Kaymakcalan was diagnosed with cancer at around this time, and passed away shortly after.

One notable classmate was Dan Friedan.  Friedan stunned me, and I think everyone else, at his Ph.D. seminar, when he showed that Einstein's equation, the basic equation of general relativity, could be reinterpreted in terms of one of the basic objects in QFT, the $\beta$ function that governs the energy scale.  I did not see what this could possibly mean, but a few years later it showed up as one of the key ideas in string theory.  (I don't think that this connection was known to Friedan at the time --- at least it is not mentioned in his thesis). 
He also taught me a key idea in QFT, one that did not appear in textbooks at the time.  This was the idea that one could separate operators into `relevant,'  `irrelevant,' and `marginal' operators, which is central in organizing QFT.  

Friedan had also had some difficulty finding an advisor, but he had solved the problem by having a nominal physics advisor but working in fact with Isidor Singer, from the math department.\footnote{Another Berkeley student from that era, Andy Strominger (who we will hear much more about later), solved the problem by transferring to MIT.}  Singer was a famous geometer, best known for the Atiyah-Singer index theorem.   He was beginning to take a strong interest in QFT, recognizing that it was going to lead to rich connections between math and physics (a development that would soon accelerate with string theory).  So he was always happy to talk about QFT, and was a good sounding board for me.

Another helpful professor was Korkut Bardacki.  Like Mandelstam, he was in the middle of making the transition from dual models to QFT.  Not as deep as Mandelstam (few are), but some times clarity is more useful than depth.  Two other professors I recall mostly for their advice:  my strongest memory of David Jackson, whose famous text I will come back to much later, was that ``It is not enough to be smart, you have to work hard.''  It was good advice, and much-needed given my lack of common sense.  The other was Robert Cahn, a new professor at LBL, the lab affiliated with Berkeley.  He also helped to fill the gaps in my common sense, especially when it can to finding my next job.  Also, a growing group of postdocs  at LBL focused on field theory and particle phenomenology, including Howie Haber, Eliezer Rabinovici, and Ian Hinchliffe.  They added some liveliness. 

Visiting speakers brought in new ideas that were missing at Berkeley, and these are still some of my strongest memories.  Steven Shenker, a student at Cornell and already a deep and broad thinker, taught me key things about quantum field theory and lattice gauge theories.  Sidney Coleman, the famed quantum field theorist and pedagog from Harvard, spoke about magnetic monopoles, and I was please to see that I had independently found some of his results.  Lenny Susskind from Stanford, about whom we will hear much more, spoke about `Hot Quark Soup.'  The connection between QFT and high temperature was largely new to me, but the Feynman-like presentation was the most memorable part.  Another Stanford visitor, the postdoc Stuart Raby, brought the idea of technicolor, strong coupling instead of a Higgs field.

One final visitor was Edward Witten, a postdoc at Harvard.  Witten asked me probing questions about Mandelstam's program.  This was startling to me, first because he was the first person I'd met who understood Mandelstam's unconventional and technical approach, and second because he understood it better than I did after years of study.  I would learn that this was a common reaction to Witten.

\subsection{Dorothy}

This is intended as a scientific autobiography, not a personal one.  I have included personal bits only to give general background.  But of course I have to tell you about Dorothy, who I met the year I arrived at Berkeley, and married five years later when I graduated.  Dorothy had several Caltech connections, including a brother and a former boyfriend.  She started as a student at Occidental College, near Caltech, but we did not meet until we were both grad students at Berkeley, she in the German department.  We actually met through another Caltech connection, who was now also at Berkeley and active in organizing volleyball games and parties.  I decided that I liked her, and spent several difficult months convincing her of the same.  So a few days after turning in my dissertation, I was in Hawaii, where most of Dorothy's family lived, and we were married.

Tom Tombrello, whenever we met, would remind me what a good choice I had made in Dorothy.  Although I had learned a few social skills at Caltech, I still had many rough edges.  Having Dorothy straightened many of these out.

I was always afraid of her asking me why I loved her, because the first answer that came to mind was always that she was the sanest person I knew.  It seemed not so romantic, though she had many other wonderful 
features.  But having been around for a while now, I think that in making a list of qualities in a spouse, being the sanest person you know should be near the top.\footnote{Her answer for me was, `because you make me laugh.'  I guess laughter is a good complement to sanity.}
So, while we each worked our way through grad school, we unwound with food, volleyball, skiing, and friends and family.

\subsection{Other physics}

In between working on Mandelstam's project, I spent most of my time trying to understand what this quantum field theory is.  I read whatever references I could find.  There was a short text that had been written by F. Mandl.  This nicely complemented Bjorken and Drell, leaving out many of the technical questions to focus on the physics.  A book by Nishijima was also good for some points.  Notable was ``PCT, Spin and Statistics, and All That,'' by Streater and Wightman, to which I will return.

For renormalization theory,  the cancellation of infinities that is needed to get the physics out, the classic source was Bogoliubov and Shirkov, which was massive and very technical.  Renormalization was presented much the same way in Bogoliubov and Shirkov and in Bjorken and Drell. It was basically a combinatoric argument, summing up the Feynman graphs at each order and showing that the infinities in each cancels.  But as an illustration of the difficulty, a previous version of the proof worked for six or fewer loops, but failed for seven.  It bothered me that  such a key principle depended on such a complicated thread, especially after Friedan and Shenker had shown me that much of renormalization was just dimensional analysis.

But there were plenty of new wonders in QFT, while this one festered.  't Hooft and Polyakov discovered magnetic monopoles in QFT, and Polyakov and his collaborators discovered instantons and their nonperturbative effects.  Coleman (rediscovering earlier work in condensed matter) showed that bosons and fermions, the two kinds of quantum statistics, could be turned into one another in 1+1 dimensions.
He also provided wonderful reviews of these and other QFT topics (spontaneous symmetry breaking, large $N$, topological solitons, ...).  His lectures, first presented over several years at the Erice summer school, and them collected in his text `Aspects of Symmetry,' did much to bridge the pedagogical gap during the 70's.  The papers of Steven Weinberg, beautiful for their clear and systematic presentations, were also a great resource.

\subsection{And moving on}

After five years, it was time to write a dissertation.  In theoretical physics, the custom was simply to combine one's published papers, often written with one's advisor or others, and insert some amount of overview.  But I had a problem: I had written no papers at all (the undergrad papers under Tombrello didn't count).  This is extremely rare.  The only similar example is the great Ken Wilson, who went years without publishing.  But he was busy recasting the nature of QFT.  I was simply suffering from a lack of common sense and of any collaborative instinct, and an advisor who was much the same.  Somehow I cobbled together 130 pages about what I had understood about vortex operators, and related issues of field theory.  But it is not something that anyone else would benefit from.

Dorothy had two more years to finish her own Ph.D., so my first choice was Stanford, though I applied to the several dozen departments that did this kind of physics.
Not having written any papers was not a fatal flaw.  The letters of recommendation from faculty really carry more information.  I am sure too that phone calls from my `godfather,' Bob Cahn, played a role in allaying concern about this applicant with no papers.  So I ended up with my desired position, though not before some agonizing weeks while higher-ranked candidates made their choices. 

In all, this was a chastening time for me.   I was used to being at the top of things.  But 
quite a few of my cohort had already published significant work (some with their advisors, but many on their own), including Dan Friedan, Steve Shenker, Mark Wise, John Preskill (the selections for the prestigious Harvard Junior fellowship), Larry Yaffe, Steve Parke, Subhash Gupta, with Edward Witten a few years older.  Most of these have become good friends, and some collaborators, but at the time it was a difficult experience.

The postdoc period is valuable for giving young people exposure to different approaches to physics.  For me, with my native lack of sense, my next two stops, at Stanford and Harvard, were particularly important.

\section{SLAC/Stanford, 1980-1982}

\subsection{Starting out at SLAC}

My postdoctoral appointment was actually at the Stanford Linear Accelerator Center, SLAC, about two miles from the physics building.  Although this was an experimental lab, at the time most of the theorists were housed there as well.  The theory director was Sid Drell, co-author of the QFT text that I had spent so much time studying.  My first meeting with Drell was a near-repeat of my first meeting with Thorne nine years earlier:
``Hello, Professor Drell.''  ``{\bf WHAT DID YOU CALL ME?}''  And so he was Sid from then on.
Sid was very involved in arms control at that time, but did some physics in collaboration with his colleagues Helen Quinn and Marvin Weinstein on a variant form of lattice gauge theory.  Their interests were not close to mine, but they were a friendly group and good to talk to.  Stan Brodsky was another faculty member who was fun and energetic, but whose interests then were different from mine.  But 20 years later I would remember it, and it would lead to a nice body of papers.  There was also Fred Gilman, who had mentored Mark Wise to the Junior fellowship, but whose interest in weak interaction phenomenology was not near mine.  

I should also mention Stephen Parke, another postdoc who went on to important work.  We talked physics and socialized a lot, but did not work together --- it would have been natural, but my collaborative sense was not yet developed.  Some other members of the postdoc group, though more phenomenological, were Geoff Bodwin, Eve Kovacs, and Tom Weiler.

The postdoc years are a chance to learn new things.  One should generally not just continue working on their dissertation problem.   So of course, this is just what I did, for a while.  I believed that I could prove what Mandelstam wanted to show, that quarks were confined with infinite strength.  I had studied several important results in QFT where things could be proven exactly.  Coleman's theorem, forbidding  spontaneous symmetry breaking in 1+1 dimensions, was a prime example.  I had the idea that in 3+1 dimensions, a similar mechanism would forbid free quarks.  Streater and Wightman, proving the PCT theorem (parity$\times$ charge conjugation$\times$time symmetry) and the spin-statistics theorem, gave further examples of the power of rigor.

I did get one little paper out of this.  I was very interested in 't Hooft's classification of possible phases of the electric and magnetic fluxes.  A lattice gauge theory model had appeared that seemed to contradict this classification.  This should not be, and indeed, a closer look revealed that the modified theory had twice as many conserved fluxes as normal, two electric and two magnetic.  When these were taken into account, 't Hooft's conditions were properly satisfied.  The moral was that one had to be sure to include all conserved quantities to understand the phases.

This was a nice little paper, though not especially significant.  Still, it should have gotten more than three citations in 35 years.  But it would take me a while to realize that it is important not just to write papers but to give talks about them --- not only to get attention, but to be forced to clarify your work, think it through, and get valuable feedback.  Even eight years later, I gave no talks about the first D-brane paper.  If I had, history might have moved faster.

But after a few months, I had not made real progress.  Moreover, I came to realize that trying to prove things was not generally a profitable approach in QFT.  It had seemed like a good idea.  With the basic nature of QFT still apparently mysterious, making things rigorous would seem to provide a desirable base.  But as I looked at other proofs, I realized that in many cases, what one could understand was often far greater than what one could prove.  One example was Polyakov's argument that in 2+1 dimensions, instantons lead to quark confinement.  The argument takes a few lines, and is convincing.  But the proof, due to Gopfert and Mack in 1981, ran to 100 pages.  For confinement in 3+1 dimensions, it seemed likely that the proof would have to be much longer.  My interest was the physics, the simple physics argument, not the 100 page details.
Also, an argument by 't Hooft showed that proving confinement might be difficult.  He was studying the phase structure of QCD.  He pointed out that if one could have a phase transition between states of free and confined quarks, as seemed to be the case, then there could not be a simple principle that said that quarks are always confined.
So I was not completely without common sense, and started looking around for other directions.

\subsection{Susskind}

About once a week, a whirlwind would settle on SLAC.  Lenny Susskind and his group of visitors and senior postdocs (Willy Fischler, Peter Nilles, and Stuart Raby) would meet in Susskind small office and talk physics for most of the day.  Unlike most of the theorists, who were at SLAC, Susskind was officially at Stanford, but he had a SLAC office as well.

Where I was the extreme introvert, Susskind was the extreme extrovert.  Even when I learned how to collaborate, my style was still to talk, perhaps for an hour, and then go away for a few days to think about things.  Susskind, on the other hand, seemed to be able to work by talking, without a break, and to make progress in this way.
In the many years that I have known him, he has almost  always been surrounded by young people, talking through his current puzzle.

Although our personalities were very different, our interest in physics was much the same: we wanted to understand the basic principles.  Neither of us were drawn to mathematics for its own sake: we used only enough to solve the problem at hand. Of course my own approach was still developing, and was surely influenced by Susskind.\footnote{Dimitri Skliros recently noted that my work on perturbative string theory reminded him of Chern.  Indeed, a short book by Chern was one of the few math books that I enjoyed, and even then I was not able to finish it.}

\subsection{Supersymmetry}

What Susskind and his friends were excited about when I got there was supersymmetry.  So I will start with a short review of grand unification and supersymmetry.

Both the strong and the weak nuclear forces had been understood around the time I got to Caltech.  Together with QED, these three forces (or four, if you count the Higgs field as a force) seemed to account for all of particle physics, a theory known as the `standard model.'  Of course, there was still gravity, but it is extremely weak on particle scales, and could be neglected at first.  All three particle forces were based on the framework of QFT, and more specifically on gauge theories.  These were like electromagnetism, but with the fields extended to matrices:  $3 \times 3$ for the strong force, and $2 \times 2 + 1 \times 1$ for the combined weak and electromagnetic interactions.

The similarity of these forces suggested some more unified origin.  Georgi and Glashow, in 1974, noted that the three forces fit nicely into a $5 \times 5$ matrix, so called Grand Unification (GUTs).  Moreover it made at least two predictions.  One was the ratio of electromagnetic and weak forces, the weak mixing angle, which came out to pretty good accuracy.  The other prediction was proton decay.  The baryon number is a symmetry of the Standard Model, but not of the additional fields needed for $5 \times 5$.  These fields mediate such processes as $qq \to \bar q  l$, turning three quarks into a lepton.  This had not been seen, but seemed within reach.

Supersymmetry (SUSY) was another idea, which nicely complemented GUTs.  With GUTs, there could be symmetries within spin 0 (scalar), spin 1/2 (fermions), and spin 1 (gauge fields), but not between different spins.  A complete theory 
might be expected to relate different spins.  A theorem showed that this was consistent with QFT --- not a trivial result --- and so this was beginning to be explored right around the time I was finishing at Berkeley.

SUSY actually solved at least three problems with GUTs.  First, the weak mixing angle was a little bit off with GUTs.  With SUSY added, it came out better.  Second, searches for proton decay were starting to exclude the GUT prediction, but with SUSY the energy scale was higher and so the decay rate much slower.  The third problem was more theoretical.  In the standard model, all the interactions are dimensionless except the one that gives mass to the Higgs field.  Quantum corrections (loop effects) will naturally generate a large scale for the Higgs, such as the GUT scale, much higher than the scale of the standard model where it is expected.  But supersymmetry can lead to cancellations that do not follow from symmetry alone, and might cancel the mass correction to the Higgs.  Of course, SUSY implied twice as many particles as had been seen, but this was consistent with supersymmetry breaking. 

{\it A narration from the future:}  For the most part I am writing this chronologically, but this requires a comment.  As we all know, the last 40 years have so far not discovered supersymmetry.  In fact, the only `fundamental' discovery, in my narrow sense of the word, is the cosmological constant, which I will get to.  But for now we will relive the glorious time in the past, when all things seemed possible.

\subsection{$D$-terms}

It was this cancellation of quantum corrections that drew the attention if many theorists, including Susskind and his friends.  The latter's particular interest was the $D$-term, a sort of special mass term in supersymmetric theories (the $F$ term is the more common one).\footnote{The nonspecialist should not confuse this with the D of D-brane!  Note that the font is different.  D is short for Dirichlet, while as far as I know the $D$ was an arbitrarily chosen label.}
Witten had shown that the corrections to the $D$-term canceled for any supersymmetric theory that was embedded in a non-Abelian theory at high energy.  If this was the whole story, it would imply that physics at low energy depended on the spectrum at much higher energies.  This was not normally seem in QFT, and might have important consequences.

So Susskind and friends hung out in his office thinking about how to calculate the quantum corrections to the $D$-term.  They were happy to have a newcomer listening in.  I had taught myself how to do some of the main calculations in SUSY, it was clearly an exciting direction.  And after a bit I was able to go from skulking to making suggestions.  Before long we had solved the problem, and I had made  substantial contributions.  

The result was that the condition for the correction to $D$ to vanish was simply that the sum of the charges of all the particles in the spectrum vanish, Tr$(Q) = 0$.  This was a simple result, just five pages.  It meant that some new high-energy/low-energy connection was not needed.  It was my first real contribution to theoretical physics, and my first exposure to doing science collaboratively.

\subsection{SUSY breaking}

A few months later, Raby, Nilles, and Fischler left for faculty positions.  Susskind and I continued to talk regularly, but neither of us had a particular project in mind.  This changed when Mark Wise gave a talk at SLAC.  Mark, along with Mark Claudson and Luis Alvarez-Gaume, and in parallel with several other groups, was constructing realistic supersymmetric models of physics.  

Susskind saw that their model was a good place to study the relations between three important scales.  The first was the observed weak interaction scale.  The second was the boson-fermion energy difference.  This could not be too much larger than the weak scale, if the cancellations were to hold.  The third scale was that set by the spontaneous breaking of supersymmetry.  What Susskind noted was that in Wise's model, the boson-fermion difference and the supersymmetry breaking scale are not the same but differed by a parameter $\alpha =$ SUSY/boson-fermion.  Susskind asked Wise if the parameter could be large, and Wise said it could.

Susskind wanted to know whether this situation was really stable, or whether quantum corrections would destroy the separation of scales.  Happily, I was able to make substantial contributions.  This kind of problem, with physics at several scales, is today  routine, but at the time was rather new.  But all my time thinking about QFT had prepared me for it.\footnote{A few years later, Steven Weinberg commented that he could tell which parts of the paper were by me and which by Susskind.  I was pleased because he had greatly influenced my understanding of the subject.}

So we carried out the analysis, and indeed found that the multi-scale structure was stable.  This meant that whatever was responsible for the boson-fermion splitting at visible energies could actually arise at a much higher scale, such as 
(weak $\times$ GUT)$^{1/2}$, or even (weak $\times$ Planck)$^{1/2}$.  Indeed, the idea of connecting SUSY with gravity led to much work in the following years.

One of the pleasures of the project was writing up the paper.  Susskind is my only collaborator who has done this in real time, the two of us sitting in his living room, each with a small glass of wine,  writing the paper line by line.

After this, I wanted to work out some more explicit models.  But Susskind, having understood the key point, was not interested in details.  So I wrote two papers on my own.  They dealt with the spectra of gauge fermions, and of the scalar potential, in the models we had studied.  Looking back on these early papers, I remember some nice bits, that I am still pleased by today.

We did spend some time trying to make a lattice theory of SUSY.  We tried a number of approaches, and I had thirty or forty pages of a paper written, but the two structures, SUSY and the lattice, did not want to come together and we did not finish it.  Some of these ideas have been rediscovered and taken further, but they do not seem to be the best way to capture the rich strong-coupling dynamics of SUSY.

So the next papers Susskind wrote, the year after I left, were about gravity.  This included his famous paper with Tom Banks and Michael Peskin, showing that black hole information loss, as suggested by Stephen Hawking, would imply large energy nonconservation.  

\subsection{Times up!}

Two years, the standard term at the time for a postdoc, was really only 1.25 in postdoc years, because one had to apply for one's next job in the Fall.  At that point I had three papers: a condensation of my Ph.D. dissertation (written just as I was leaving Berkeley), my little-known lattice phase paper, and my $D$-term paper with five authors.  Unimpressive even for a first-time applicant, much less a second-timer.  But again, I seem to have impressed my letter writers enough to get a position at Harvard.  I wanted to spend a few years on the East Coast, to be exposed to new ideas and new people.

Dorothy had just finished her Ph.D\@.  In her field it was typical to next do a few years of short-term teaching.  She applied, and got a short term position teaching German at MIT.  

So shortly after our wedding, we came back from Berkeley and hopped into our Datsun B210 for the drive east.  Along the way we spent several weeks at the Aspen Center for Physics.  As anyone who has been there knows, it is a remarkable combination of science and recreation.  I met many excellent scientists there for the first time.  I actually wrote a paper there, in collaboration 
with Mary K. Gaillard, Larry Hall, Bruno Zumino, Francisco del Aguila, and Graham Ross, a distinguished and varied group. 
We spent the time at Aspen discussing our common interest, the mass scales of supersymmetry breaking models, and after two weeks we felt that we had enough that we could write it up. 

Sidney Coleman, my soon-to-be supervisor, was also at Aspen then, so we got an early start on discussions of QFT.  Sidney was an avid walker and hiker, a frequent visitor to Aspen, and so gave some of us first-timers advice on reaching the nearest 14,000 foot peak.  This nearly led to Dorothy, me, Tom Weiler, and Lawrence Hall tumbling off Castle Peak with a large boulder, when we misinterpreted his instructions. 

After Aspen, we had a memorable trip east: a hot drive through the Midwest (which Dorothy still jokes nearly led to a divorce), spectacular thunderheads not familiar to Californians, staying with my collaborator Raby and his wife in Ann Arbor, a drive across Niagara Falls, and a stay with Peter Galison (Junior Fellow in particle physics and history) while getting settled in Cambridge.

\section{Harvard, 1982-1984}

\subsection{Wise}

I start with Mark Wise, my effective mentor.  Mark was a great physicist and a wonderful person.  As a student, he had already done important work on weak interaction physics, and as a postdoc he would have important results on supersymmetry and on cosmology.  As a person, he was generous and self-effacing, and had a great sense of fun.  When he turned 60 a few years ago, a time when most prominent scientists are celebrated by a major conference, Mark insisted that his be celebrated by renting a skating rink and going curling.

My first meeting with Wise went something like this:  ``So you're Polchinski.  We've been hearing about this guy who doesn't write any papers.  Let's write a paper.''  And we did.  Unlike Mandelstam, and Susskind, and I supposed myself, Wise did not insist that every paper be
Important.  Of course, many were, but he was happy to think about any physics puzzle.  His puzzle here had to do with the masses of the bosons and fermions in theories of broken supersymmetry.  They need not be equal, but in many models there was a sum rule on the masses-squared, roughly $\sum_i b_i^2 = \sum_j f_j^2$.  This had been shown in tree-level models and a few others, and the question was how general it was.  So we analyzed the problem --- it was a good blend of our skills --- and found that the sum rule applied to first order in the supersymmetry breaking but exactly in the other interactions.  

It was a nice result, finished, refereed, and in print two months after my arrival.  However, it did not have much impact, as it did not apply to the most interesting models.  In fact, it has received only three citations in 35 years, something that Wise chuckles over whenever we meet.  But apparently I passed, and he said we should look for a bigger problem.

\subsection{Coleman}

There were only three tenured professors of high energy physics at Harvard, with Steven Weinberg having left for Austin.\footnote{On Weinberg's blocky 1980 office computer was a sign~{\sl Contrary to appearances, this is not Steven Weinberg}.  The suspect was Paul Ginsparg.}
  Glashow did not interact much with students, so Sidney Coleman and Howard Georgi each supervised half the students and postdocs, meeting weekly as the `Coleman family' and the `Georgi family.'  Both were delightful people, interested in the students and postdocs and generous with their time, though I believe that Georgi was much more burdened by administration.  As a rather formal student, I was in the Coleman family.  But we could, and did, cross over to the other family when it was interesting.

Coleman was as much a delight in person as in his writing.  He came to the office promptly at 1 pm each day, and stayed late.  He was always ready to talk, but the highlight of the week was the family meeting.  This might be a student or postdoc presenting his work for critique, or Coleman himself holding forth on some point, or a general discussion.

Coleman often took students and postdocs walking in the area, or hiking up Mount Monadnock in New Hampshire.  He also had us over frequently for dinner.  He was fond of telling us that in the last year he had just gotten married, bought his first house, learned to ride a bike, and been diagnosed with diabetes.  I went with him on one of his early bike outings, helping him negotiate some of the difficult Cambridge turns.

I thought that it was unfortunate that Coleman did not get involved in supersymmetry, because it meant that we could not look forward to his insights on the subject.  But perhaps he felt that there were enough people working on it already, and there were still interesting questions for him in pure QFT.  Indeed, in the time I was there he expanded his overview of magnetic monopoles, worked with a student on the 't Hooft anomaly cancellation, and discovered a surprising new class of topological solitons.  

This last has been widely influential for possible particle and cosmological models.  I had a tiny role in it.  Coleman was happily describing the construction, which was based on a charge he labeled $q$, and he was constantly referring to $q$ as he described this round object.  And then he said, `I just need a good name for it.'  Having listened to him at length, the only possible name came to mind at once: Q-ball (as in `cue ball'), and so it became.

\subsection{The kids run the circus}
 
Sometimes, one gets a group of young people who are so outstanding that they run the show.  Such it was here.  Sidney was excellent, but not interested in the latest sensation.  But this was more than made up by the students and postdocs.  I still marvel over the excellence of that group, many of whom are still leaders today. 

So in this section I am simply going to list the members of this outstanding group (those whom I remember),   where they are now, and what they do.

Untenured faculty: John Preskill (Caltech), Lawence Hall (Berkeley, phenomenology)

Junior fellows: Mark Wise (Caltech), Paul Ginsparg (Cornell, QFT and creator of arXiv), Laurence Kraus (Arizona State, cosmology and public speaker/writer), Luis Alvarez-Gaume (CERN, fields and strings),
Peter Galison (Harvard, physics and history)

Postdocs:\footnote{The difference between a junior fellow and a postdoc is that the fellows got free lunches at the law school, and once a week they had fancy dinners, capped off with excellent wine, chocolate, and cigars.  But I'm not bitter, really. I don't even like cigars.}
Tadeusz Balaban (Rutgers, mathematical physics), Steven King (Southampton, phenomenology), Steven Sharpe (Washington, lattice), and me

Grad students:\footnote{I have not included some that I did not know at the time, such as Boris Shraiman, Catherine Kallin, and Subir Sachdev.}
Robert Brandenberger (McGill, cosmology), Andrew Cohen (Boston, strings and phenomenology), Ben Grinstein (San Diego, phenomenology), David Kaplan (Washington, QFT and nuclear), Greg Kilcup (Ohio State, lattice), David Kosower (Saclay, amplitudes), Anish Manohar (San Diego, phenomenology), Robert Mawhinney (South Florida, lattice), Ian McArthur (Western Australia, string theory), Greg Moore (Rutgers, mathematical physics and strings), Ann Nelson (Washington, phenomenology), Phil Nelson (Penn, mathematical physics and biology), Stephen Della Pietra (Renaissance, mathematical physics), Lisa Randall (Harvard, phenomenology), Jacques Distler (Austin, mathematical physics and strings), Richard Woodard (Florida, relativity).

It seems to me that this is a group of young people that has rarely been equaled.  With all of them I can recall discussions of physics and other fun times.  We will see many of them later in this and other chapters.

\subsection{Low energy supergravity}

Early attempts at constructing supersymmetric models were based on classical string actions.  A Fayet-Iliopoulos $D$-term allowed for a simple means of SUSY breaking.  Unfortunately, it did not seem to give the right symmetry breaking and spectrum.

The idea that supersymmetry could be broken at a much higher scale, as put forward by Susskind and me and several other groups around that time, allowed for other possibilities.  The focus was on supergravity models, assuming the highest possible SUSY breaking scale.  It was exciting to be thinking about a coupling between particle physics and (super)gravity, but it was very indirect.  The gravitational field was very weak, and so played no direct role other than setting a scale.

So Wise, his previous collaborator Alvarez-Gaume, and I set out to build a realistic model along these lines.  We began with the simplest possible assumptions: (1) Begin with the standard model. (2) Extend to SUSY: this adds a field of opposite spin for each one present, and an extra Higgs multiplet.  (3) Add soft SUSY breaking.  Soft breaking means that one includes all gauge invariant terms with positive mass-squared.  (4) This leaves dozens of parameters, so we adopted a simplifying assumption that had been introduced by Weinberg. The SUSY breaking was taken to  be the same for all scalar fields.  Weinberg came to this from a guess about unification.  From the later context of string theory, one can say that it was not well-motivated, but it did make things simple, and we followed it.

So, take this starting point, set the renormalization group running, and over a wide range of parameter space with no further choices, out came the standard model.  The trickiest part was that one needed exactly one scalar, a Higgs, to get a negative potential and break its symmetry.  But this happened automatically: a loop of a heavy fermions generates a negative mass-squared for a scalar coupled to it.  The heaviest fermion was the top, so the scalar most strongly coupled to it broke its symmetry.
It was very elegant: a large class of the simplest SUSY models flows exactly to the standard model plus broken SUSY.\footnote{I have to make an apology here.  Several groups were making supergravity models, but we asked a slightly different question.  Rather than a specific high energy model, we integrated out the high energy theory, replacing it with soft SUSY breaking subject to Weinberg's assumption.  So we did not pay too much attention to the specific models.  But the nicest part of the result, the symmetry breaking from the top mass, had been discovered and published first by Luis Ibanez and Graham Ross.  We should have cited it, but we did not learn about it until after our paper was published.}

We followed this up with more detailed studies of the electric dipole moment and (with student Ben Grinstein) the decays of W and Z bosons. 

\subsection{Renormalization and effective Lagrangians}

From Susskind, Wise, Coleman, and others, I had learned how to work on a variety of physical questions, and put aside the more formal ones that I had been drawn to as a student.  But the questions were still there, and one day came roaring back.  John Preskill was teaching QFT, and I was sitting in the back, auditing the lectures on renormalization.  At the end of the discussion of the cancelation of the infinities, he said, `I think there should be a better way to do this.'  And instantly I knew there was.

Again, what bothered me was that the proofs that renormalization works seemed extremely combinatoric and technical, but the results 
in the end came down to statements of dimensional analysis.  What I realized was that things would become nearly trivial if, instead of describing the path integral order by order in perturbation theory, as nearly always done, we described it scale-by-scale in energy.  As soon as I thought those words, I knew I could prove them.  Of course, I mentioned earlier having put aside proofs, but this was a special case.

It took just three weeks for me to work out the proof and write it up.  Probably it would have been better to take a little longer and make it a bit clearer, but I was afraid of being scooped --- probably silly, but you never know.  I think I presented it fairly well --- first the idea, then a 2-component model, then a precise statement of what was needed to be proved, and finally the proof.
What made this proof clearly different is that it did not need graph combinatorics or Weinberg's theorem.\footnote{Weinberg's theorem was an intricate statement about the high energy limits of multiloop integrals.  I had nothing against Weinberg, or his theorem which I had spent much time studying.  But I wanted an argument that depended only on dimensional analysis.}

Organizing the path integral scale by scale is messy: one has a differential equation that has to include, not just the small number of normalizable couplings, but the full infinite nonrenormalizable set.  But this point of view was much more flexible than the traditional quantum field theory.  In many situations one only knows what the theory is up to some energy --- that is, it is an effective field theory --- and this makes that notion precise.

This work was very exciting for me.  For the first time, I felt that I had changed the way that people think about the world.  Of course, aside from the details of the proof, most of the ideas were already known, especially from the work of Wilson and of Weinberg.\footnote{The notable field theorist Edouard Brezin was visiting when I presented my talk.  I was happy that he followed and appreciated the subtleties of the argument, and also when he informed me he had heard Wilson say that he regarded his own work as a proof.  I agreed, though not many others would have seen the point: once you say to work scale-by-scale, the rest is just bookkeeping.}  And in the next few years the idea of an effective field theory would become universal.

My paper thanked three names for inspiration.  The first was Dan Friedan, for teaching me about the idea of effective field theory.  The second was John Preskill, for his comment in class.  If not for that, I might never have put together what I knew.  The third was C. Arabica.  I am disappointed that no reader of the paper has ever asked who that is.  In fact, C. Arabica is the caffeine plant.  I was distinctly buzzed that day in John's class, and that too played an essential role.

This work led me also to discussions with Tadeusz Balaban, a postdoc in Arthur Jaffe's group.  Balaban was in the middle of proving asymptotic freedom, which he eventually did over the span of several Communications in Mathematical Physics adding up to 300 pages.  One could say that it is a landmark result, but it is almost unknown. It is another example where a convincing one page physics argument can require orders of magnitude more to prove.

\subsection{Monopole catalysis}

I am not going to discuss every paper I've written, but only those that have some story.  The paper that I wrote right after the the one on renormalization did not have much impact, but there are a couple of stories.

Over the years, a number of physicists have commented to me that my papers are distinctive, in that many of them are written not to discover new things, but to explain what we already know in a new way.  The renormalization paper is a prime example.  It is not that I always set out to do this, but that I have to understand things, and can't proceed if I don't.  My next paper was of the same type.

Grand unified theories like $SU(5)$ will have magnetic monopoles, as shown by 't Hooft and Polyakov.  These are soliton states, not pointlike but with a size set by the GUT scale.  We know that baryon number is not conserved in GUT theories, so there should be baryon-number violating scattering processes involving monopoles.  The surprise, as shown by Valery Rubakov and Curt Callan, was that the rate for this process was determined not by the GUT size, but by the much larger baryonic size.

This conflicted with my understanding of soliton amplitudes.  The symmetry (anomaly) argument of Rubakov and Callan was too indirect for me, but eventually I found a toy model that allowed me to understand the details of the process, and see that they were right.  Normally I have a pretty good intuition for QFT, but Rubakov has twice done things that I thought were impossible.  

The other came a few years later, also with baryon number violation.  't Hooft had shown that instantons could violate baryon number, but it was very slow, taking more than the lifetime of the universe.  This seemed like a simple calculation with the Euclidean path integral.  But Rubakov and collaborators showed that {\it if you heat the system up,} then rather than the very slow Euclidean tunneling process, you could move thermally with little suppression.  Again this seemed to conflict with my intuition, and again I wrote a paper, 
with collaborators Michael Dine, Willy Fischler, Olaf Lechtenfeld, and Bunji Sakita, to explain it to ourselves.

So don't bet with Valary Rubakov on quantum field theory!  (Another surprise from Rubakov and collaborators was the `braneworld,' well before it became popular).

\subsection{Phenomenology}

I had chosen a fortunate time to be a postdoc at Harvard, and do my one stint of phenomenology at a place where the lines between theory and phenomenology where particularly thin.  During the time I was there, I had the excitement of seeing the discovery of the 40 GeV top quark, supersymmetry, and the $\zeta_{8.3}$. The less credulous readers will point out that the top was discovered more then ten years later at $170$ GeV, supersymmetry has not been found, and {\it what} is the $\zeta_{8.3}$ anyway?  Such was life in the Wild West (or should I say Alternate Truth?) days of particle physics.

The culture at Harvard seemed to be that one expected most new experimental observations to be wrong, so you should write your papers about them before they are withdrawn.  There were even a couple of maxims about this from Howard Georgi: ``Not more than one half of an idea per paper,'' and ``Don't hide your light under a bushel basket.''  The latter was apparently a Biblical injunction against slow publication.

In the case of the 40 GeV top quark, our result actually appeared at the same time as the experiment, and did not agree with it.  At the 4th annual Supersymmetric Unification workshop in Philidelphia, Carlo Rubbia presented his evidence for the discovery of a 40 GeV top quark at CERN.  I reported on our minimal SUGRA model.  It favored a heavy top, around 150 GeV, though this was rough because other parameters could be tuned.  I said in my talk that our result argued against the 40 GeV top, but the ideas of a random theorist had no weight compared to an experimental result.  This result was soon withdrawn, I believe due to a better understanding of the statistics.  But our model also did not last long.  The very minimum assumptions that we had made put an upper bound on the superpartner masses, which was soon ruled out.

Shortly after, Rubbia reported evidence for several monojet candidates, just a single jet with unbalanced momentum.  The missing momentum could be carried by various kinds of new particles, particularly a gluino or squark pair.  Assuming that we were seeing superpartners, the first question was whether they were squarks or gluinos.  Lawrence Hall and I realized that for high-scale SUSY breaking, the RG flow would have a large effect on these.  In particular, squarks would gain mass but not gluinos, so the latter would almost certainly be lighter.  A nice prediction, but spoiled by the fact that the monojets were found to be misidentified standard model particles.

The third observation was at the Crystal Ball detector at DESY.  This was designed to study bottom-antibottom states $\Upsilon, \ldots$, around 10 GeV.  It reported several events at 8.3 GeV, produced in the process $\Upsilon \to \gamma + \zeta$.  There was no obvious candidate; it might be a light Higgs, or even a colored state.   A puzzle was that the $\zeta$ was produced at very different rates in different states $\Upsilon,\Upsilon'$.  Stephen Sharpe, Ted Barnes, and I thought that this might be a wavefunction effect, and set out to calculate the relativistic wavefunctions; Pantaleone, Peskin, and Tye did this independently.  It was a fun calculation, one that I had never done before.  It did not solve the problem, but this went away by itself when it was realized that the apparent events were actually due to a flaw in the detector.

So, two exciting years of ambulance chasing.  But it was good: I learned a lot of physics that I would not have learned otherwise.

\subsection{Time to grow up}

It had been a great two years on the East Coast.  I had learned a lot of new physics, and new ways to do physics, and had written a couple of significant papers.  I had met a large number of excellent scientists, both at Harvard and on various visits around the East, many of whom I still interact with to this day.  Now it was time for the next step, a faculty position.

Job openings in physics are often cyclic, driven largely by the economy.  At this time, the job market was not good.  But I was confident, based on my work in the past two years.

In fact, I had turned down a faculty job the year before.  At Harvard, Princeton, and a few other places, it was understood that untenured faculty were glorified postdocs.  There was no expectation of a later promotion to tenure.  So from my perspective, all it meant was teaching and less time for research.  I think this startled them, few if any had turned it down before.  But I was aware that I had been slow to get into research, and now that I was making progress I did not want to cut it short.  Also, I had a two-body problem.  Dorothy's MIT position was over, and the next position she found was at Urbana-Champaign.  So I wanted time free to visit her.

We had our two-body problem again, so we each looked at the jobs that were advertised in our field, and there was very little overlap.  Dorothy had an excellent opportunity to remain at 
UC, moving from her temporary job to a tenure track position.  John Kogut, lattice gauge theorist at UC, believed that he could bring me in on a spousal hire.  If I had been a condensed matter theorist that would have been great, but there was virtually no one in high energy theory, and I would have felt tremendously isolated.  For me, there was a possibility of an untenured position at Princeton, and perhaps something at SLAC, but there was nothing for Dorothy at either one.

Fortunately, Texas came to the rescue, with a position for me in Weinberg's group, and a lecturer's position in German with the promise of a later tenure track job.  This was neither of our first choices, and not one that we had expected, but it was an excellent compromise.  Texas had a long history in German, going back to its early settlers, and had a large department.

\section{Austin part 1, 1984-1988}

I have been dividing each section according to where I worked at the time.  My stay in Austin ran for eight years, which would make for a much longer section.  After some thought, I have come up with an event that singles out the midpoint of my time in Austin: it is when I started my book on string theory.  So this separates Parts 1 and 2.

\subsection{The group}

Austin had a strong history in theoretical physics.  Alfred Schild, an early relativist who had recently passed away, had been a leader of the group.  There was Bryce Dewitt, one of the first to develop quantum gravity, and a proponent of the many worlds interpretation of quantum mechanics.  John Wheeler had gone to Austin when he reached Princeton's mandatory retirement age (and he returned a few years later, when they changed their rule).  George Sudarshan, co-discoverer of the V-A theory of the week interaction was there, as was Yuval Ne'eman, co-discoverer of the $SU(3)$ color symmetry of the strong interaction.  Duane Dicus, phenomenology and cosmology, and Richard Matzner, relativity, were also good colleagues.

A few years earlier, Weinberg and his wife Louise were looking to solve a two-body problem, and Austin was looking for ways to spend more of its oil money.  So Weinberg and Louise were soon in Austin.  Steve's salary was a subject of much speculation, but he never spoke of it.  He also had an agreement that he could hire four faculty colleagues, with ample postdoc and student support. 
His first three hires were Phillip Candelas (a relativist who would very soon be one of the leaders of the first superstring revolution),\footnote{Looking at the record, I see that Candelas was at Austin before Weinberg, so there must have been some negotiation that he would count as one of Weinberg's group}
Willy Fischler (my collaborator at Stanford, and one of the inventors of the invisible axion), and me.  Vadim Kaplunovsky would join a few years later.

Weinberg had always been rather solitary.  For example, most of his papers were single-author.  But he was proud of his group.  He instituted a weekly family meeting, as at Harvard, and he took his whole group to the faculty club every week.  He tended to hold court at lunch, and I used to joke that he had three subjects of conversation: English history, Israeli politics, and DOS versus Windows.  On the last, Steve was a notably text-oriented thinker; for example, he used very few figures in his books and papers, so he was one of the last DOS holdouts.

There was also a remarkably good group of grad students there, for a place rather out of the way.  Perhaps the very low tuition, the same for in-state, out-of-state, and international, played a role.  Indeed, there were many international students.  The students added greatly to the energy in the group.  Finally there was Adele Traverso, the group secretary.  She was delightful but tough-minded, making sure to dress down each new group member for any infraction, so they would know who was in charge.

All in all, it was an exciting place to be.

\subsection{Weinberg and physics}

I had studied Weinberg's relativity book and papers at length, and heard some talks, but did not interact with him until Austin.  I am embarrassed to say that my first impression of him was that he was a little slow: in talking with him, he seemed to get stuck on things that seemed obvious.  But before long I realized that this was part of his genius.  By not assuming things that everyone else took for granted, he would time and again discover possibilities that had been overlooked.  

A minor example, which he was working on when I arrived, was `quasi-Riemannian geometry.'  When gravity is written in terms of a connection, the curvature appears both in the metric and in the vierbein.  Normally these are essentially the same, but he asked what happens if they are independent fields.  As far as I know, this did not lead to much, though it did foresee some aspects of string compactification.  But more interesting examples will come up later. 

Weinberg's focus on his physics was famous.  When he needed to learn something that I might know, he would question me in detail.  But when my knowledge was exhausted, and I changed the subject, his eyes would visibly glaze over, and I knew that the meeting had ended.  Many others had the same experience.  

But I held nothing against him for this: this is what made him great.  Even with his public interactions and other distractions that came with the 1979 Nobel, he continued to be creative.  In the years after getting the Prize, Weinberg published five papers with more than 1000 citations, including one with more than 3400.  And over time I had ample opportunity to interact with him, as did all the group members and most notably the students.

\subsection{String theory}

Just as I was getting settled in Austin, the first superstring revolution struck.  I had known very little about string theory before.  When I went back to Caltech for a conference, John Schwarz told me I should read his latest papers.  I tried, but it was all written in a noncovariant way, which I could not get past.  Lenny told me that there was a new formulation of strings by Polyakov, which was more covariant, but it was a lot to absorb.  And Witten was starting to write papers that hinted at string theory, such as his work with Alveraz-Gaume on anomalies in higher dimensions, and his work on equivalence of different string actions under bosonization, which he presented at the same GUTS meeting where Rubbia and I spoke.  

But it all came to a head in the fall of 1984, when Green and Schwarz found a new anomaly cancelation mechanism, Gross, Harvey, Martinec, and Rohm found the heterotic string, and Candelas, Horowitz, Strominger, and Witten found the Calabi-Yau solutions.  Together, these gave a close connection between string theory and the standard model.  I had spent the last several years on unification.  My work was focused on supersymmetry, but I also informed myself about GUTs and Kaluza-Klein theory.  Together they implied unification between fermions and bosons, between different gauge groups, and between gauge fields and gravity, while constraining the spectrum of particles.  Moreover these three ideas were nicely compatible with one another, and it was plausible that they were all part of some larger structure.

But there was one thing missing, even when all were taken together.   Each had a lot of arbitrariness, in choice of gauge field, matter spectrum, masses, and coupling constants.  A unified theory should have a uniqueness, and it was hard to see how this could come out of these frameworks.  But string theory apparently did this.  For example, there is no free gravitational coupling constant; rather, its value is determined by the value of the dilaton field.  All other constants would be determined in the same way, as the values of fields, which are determined in part by field equations.  So this does not solve everything, but rather transmutes it, from freedom in the theory to freedom in solving the field equations of a fixed theory.\footnote{Tom Banks emphasized this to me.}  This is the kind of progress one normally sees in physics, with equations that are often written in a few lines like Maxwell's or Einstein's, but have many solutions.  But we will return to this later.

Looking over the papers I wrote while in Austin, the earlier characterization of my work still fits.  Most of them seem to be written not to discover new things, but to explain what we already know, perhaps in a clearer way.  This led to a lot of fairly forgettable papers, but also some nice ones, though none that changed the direction of the field.

As I was learning string theory I first zeroed in on the question, why are strings forced to live in the critical dimension, 26 for bosons,
when we knew that there were string solitons in any dimension, such as 4.  Of course I realized that the stringy solitons were an effective description, valid only at long distance, while the critical strings presumably had zero width.  But I guess I thought there should be some unified description of critical and solitonic strings.  But after a few months I got stuck and moved on to greener pastures. 
Several years later, Andy Strominger was visiting and we started thinking about the puzzle again.  This time we were clearer minded, and we found a nice construction in terms of conformal symmetry, which has been somewhat useful.

My next bit of self-pedagogy was the Polyakov path integral.  Previous string amplitudes were based on light-cone calculations, but the Polyakov theory promised a covariant starting point.  So I carried out the path integral: a straightforward exercise, but useful.  My favorite part of it was that it allowed me to apply it to the amplitude with no particles (vertex operators), thus determining the cosmological constant (which was nonzero in this bosonic theory) and the finite-temperature partition function.  So I could  connect some interesting physics to the Polyakov calculation.  Several later papers also will make novel use of vacuum amplitudes.\footnote{In passing I mention two other papers from this period on Polyakov path integral technology, one on the vertex operators and one on the factorization of the amplitudes.} 

I had several followup papers with Andy Cohen, Greg Moore, and Phil Nelson, three of the outstanding students from my Harvard stay. We had talked a lot then, but did not write a paper together until meeting at a conference and finding a common interest in the Polyakov path intergral.  Our first project was to construct off-shell string amplitudes.  We thought we had succeeded, but I don't think that we had gotten the gauge symmetry right, since we now know that only physical amplitudes are gauge invariant.  Our most explicit example, where the ingoing and outgoing strings were contracted down to points, I think were in fact what we now interpret as $D(-1)$-branes, D-instantons.  This began ten years of getting close to D-branes and not getting the point.  Another paper with Moore and Nelson was an extension of the Polyakov path integral to the supersymmetric case, but here I was more of a follower.\footnote{In order to participate in the projects, I had to go to the computer center and get access to something called `bitnet,' which would allow us to communicate via our computers.}

\subsection{Hughes, Liu, and Cai}

I had remained a postdoc for as long as possible, but now I had responsibilities.  Supervising graduate students turned out to be a great thing.  The common pattern with a student was that I would suggest an idea and we would meet weekly.  Usually the idea turned out to be too hard for the student, so we would end up working together.  I am pleased that with almost all of my students I ended up writing one or or more significant papers.  So the students got a great research experience, and many times I got to work out good ideas that I otherwise might have let slide.

My first three students were Jim Hughes, Jun Liu, and Yunhai Cai.  The nine students that I supervised in Austin happened to come in groups of three, so I always think of them that way.  Each had their own projects, but they often ended up collaborating.  Jun and Yunhai both came to Austin through the CUSPEA program run by T. D. Lee.  This brought large numbers of excellent students from China to many US institutions for graduate work.

SUSY phenomenology was based on broken $\mathcal{N}=1$ supersymmetry.  There was an argument that one could not, staying within four dimensions, have a partial breaking such as $\mathcal{N}=2$ to $\mathcal{N}=1$.  But I knew this had to be false, by construction.  It was known that there were vortex solutions in which $\mathcal{N}=2$ broke to $\mathcal{N}=1$, found by Lancaster.  These were not counterexamples yet, because they also involved Lorentz breaking from $D$ to $D-2$.  But by taking the low-energy limit, this became $D-2 \to D-2$ while $\mathcal{N}=2 \to \mathcal{N}=1$, violating the argument.  Of course, I am talking about BPS states, a universal idea now, but at the time it was rather new, and usually applied to monopoles rather than vortices.

So I gave Jim the problem of working out the four-dimensional action that breaks $\mathcal{N}=2$ to $\mathcal{N}=1$, with two dimensions as a warmup.  As would be the pattern with many of my students, this was too hard for him, but turned into a great joint project.\footnote{According to the acknowledgements, the problem was suggested by Luca Mezincescu, a postdoc recently arrived from Romania. I do not remember this, and do not know why he was not part of the collaboration.  I think that early on I collaborated more easily with students than with postdocs, and I conjectured that this was because they were better at doing what I told them to do.}  It was educational for both of us, learning the Volkov-Akulov treatment of nonlinear broken supersymmetry and the Green-Schwarz action. Although this was an explicitly QFT problem, it used many ideas from string theory.  

Jim and I worked out the $D=4 \to D=2$ case, and then with Jun we extended it to $D=6 \to D=4$.\footnote{Jun was formally a student of Weinberg, but did all of his work with Jim and me, so I have always thought of him as my own student.  But many students did get very good ideas from Weinberg.}  As we noted there, there were several applications: 1) we completed the construction of $D = 4, \mathcal{N}=2 \to \mathcal{N}=1$ SUSY;\footnote{A short explanation of how the no-go theorem could be violated is that the Haag-Lopuszansky-Sohnius argument on which it is based constrains possible symmetries of the S-matrix, but the action could have additional charges.}  2) We had found a new and more general form of the Green-Schwarz action, based on a scalar field rather than a vector field;  3) This allowed us to construct supersymmetric membrane actions, 3-branes in D=6 being the case we studied.  

I had a bad trait, sometimes, of not following through on my ideas.  Having solved the original puzzle, we moved to new directions, such as writing the string field equation in terms of the renormalization group.  These were fun, but not so notable.
But Bershoeff, Sezgin and Townsend classified all possible supermembranes, finding that the maximum, 2-branes in 11 dimensions, was the same as the maximum dimension of supergravity.  This led to parallel activity for several years, string theory and membrane theory, with little communication between them. Membranes could not be quantized the same way as strings, and so most string theorists, myself included, assumed that they were an aberrant offshoot of the real theory.\footnote{Michael Duff made the drive from Texas A\&M to Austin to give us a review of membrane theory.  In my wise-guy way I told him that I had only been joking when I invented supermembranes.  To which he aptly replied `Many a true word is spoken in jest,' an adage that apparently goes back to Chaucer.  So be careful trading quips with Brits.  And he was right about the physics, too.}   But those whose expertise was supergravity knew that there was something important there.  Only with the second superstring revolution, eight years later, did the whole dual picture become clear.

With Yunhai, again the goal was to explain more completely something that was already known.  The open superstring theory was shown by Green and Schwarz to be consistent only if the gauge group was $SO(32)$.  They showed this by a calculation in the effective field theory of $\mathcal{N}=1$, $D=10$ supergravity, where there was an anomaly unless the gauge group was $SO(32)$.  But it should be possible to understand the anomaly in terms of the fundamental string theory, not just the low energy approximation.  This is what Yunhai and I set out to do.  The key terms in the string path integral were again of my favorite type, the vacuum graphs.  Here there were three, the cylinder, the disk with a crosscap, and the sphere with two crosscaps.  These summed to
$(N-32)^2 (\infty)$.  The infinity was from the volume of spacetime, times a normalization.

The three factors in the expansion of the square are from the graphs, with $N$ counting the Chan-Paton factors at the boundary.
It was natural to interpret this as the vacuum-to-vacuum amplitude for the dilaton.  This was correct for the NS-NS (boson-squared) sector of the integrals, but there was an equal contribution from the R-R (fermi-fermi) sector for which there was no corresponding particle.  This had to be a nondynamical 10-form field.  We now know this as the form carried by the $D$9 brane.

I have two regrets about this work.  First, the authors of the paper are ``Joseph Polchinski and Yunhai Cai.''  In high energy theory the convention is nearly universally alphabetical.  In this case, this started as a joint project, and Yunhai made some good comments and some calculations, but it quickly went much farther than the original idea.  I was still rather solitary in how I worked, and when things got really interesting I would race through to the end.  This happened here, so I ended up with a long paper written almost entirely by myself.  So I did not see any other way to sign the paper.  But this would do Yunhai no good, either pedagogically or when he went to apply for jobs.  Of course, the right thing was to slow down just a little and give Yunhai a piece of the project that was his own.  But I can say that I did learn, and became a good advisor before long.

The second regret is that I never gave a talk about the result: my shyness speaking about my work still lingered (I think that I rarely felt that my work was important enough).  I was at a string meeting in downtown Santa Barbara around that time, and did not ask to speak.  When I told Michael Green about the result, he said I should have spoken.\footnote{This conference is famous for its banquet, which was so slow that one table phoned out for some pizzas to be delivered.  It is also known for its after-dinner speech, in which Frank Wilczek explained why he should get the Nobel Prize, which he did fifteen years later.}
  Indeed, the paper has received over 400 citations.  But, like quite a few of my papers from that period (including the two with Jim and Jun), it got rather few at first, but then exploded after the second superstring revolution.  Perhaps if I had been less shy about speaking, physics would have moved faster.

Jim, Jun, and Yunhai each did a few postdocs and then moved on to other things.  Jim is at Microsoft, Jun got a second Ph.D. in finance and is now a professor in this field at UCLA, and Yunhai became a magnet designer at SLAC.  Even after the first superstring revolution, there were no jobs for string theorists.  There was widespread doubt about string theory as physics, so that only a handful of places were willing to hire in it.  It would be best at least if one had accomplishments both in string theory and in `normal' physics.  Only after the second superstring revolution, when the web of connections emerged, did most departments feel that it should have a string theorist or two.  Personally, I think young people should work on a wide range of problems, but I suppose that this is harder now as things become more specialized.

\subsection{More fun with physics, colleagues, and students}

Conformal symmetry came to the attention of many of us through the Polyakov path integral, where it is part of the symmetry algebra.  A conformal transformation is like a position-dependent scale transformation, and there was lore that any scale invariant theory would be conformally invariant as well.  The argument for this seemed weak: the conformal algebra had more elements and so should have fewer invariants.  So I set out to find the truth.  There was previous literature on this, for classical field theory.  But scale and conformal dimensions typically received quantum corrections, so I wanted a quantum argument.

In 1+1 dimensions it turned out to be quite easy to give a proof, using an important result by Zamolodchikov.  He had shown that the scale transformation was monotonic, and a small twist of that gave the result that scaling implies conformal invariance.  There were technical conditions, the most important being unitarity; without this there would be exceptions.  
I  tried to find an argument in other dimensions, especially 3+1, but failed.  I could find neither a proof nor a counterexample.  This work attracted little attention at the time: particularly relevant to string theory.  But in later years there was renewed interest in such questions, and I will return to it then.

Most discussions of strings at the time dealt with low-lying states, small loops.  But one could imagine highly excited states that were very long, perhaps spanning the universe.  Cosmology could even lead to such strings being produced.  Witten had recently considered this for the superstring and found several obstacles to their being produced.  I will return to this later, but for now it was still interesting just to see how strings behave under various conditions.  

An interesting question was, what happens if two strings cross?  Do they pass through one another, or do they reconnect?  My colleague Matzner was studying this question for cosmic strings from GUTS.  He found that in this case, where strings were classical, they always reconnected.  But for fundamental strings the answer would be more quantum mechanical, a probability for each outcome, and it was an interesting exercise to work it out.  So the simplest way to address the question was to introduce some large periodic dimensions to wrap the long strings, and turn the problem into an S-matrix that was readily obtained.  For my newest student, Jin Dai, I gave the same problem with open strings.  Here there were two processes, an open string breaking in two (or the reverse), or a closed string breaking off from an open string.  The calculations were done, a good warmup for Jin.  They had modest relevance when cosmic strings came back for a while, but it mainly was `fun with strings.'

Before moving on, I should mention what my colleagues were doing.  Each of us was working on strings in their own way, and I benefited from all of them.  Phillip Candelas was by far the most successful then, as one of the discoverers of Calabi-Yau spaces, a simple connection between string theory and the standard model.  I should have understood his work better, but our approaches could hardly be more opposite, his geometric and mine field theoretic, with a minimum of geometry (wrapping a line in a periodic dimension was about my limit).  While I was there he discovered mirror symmetry and also the first hint that all Calabi-Yau spaces might be connected through nonperturbative effects.  These were fascinating, but I did not have the tools to follow.

Weinberg was trying to learn string theory much as I was, looking for simple calculations to do.  I do not know why we did not work together, I guess neither of us played well with others (though I improved with time).  But I did find his work interesting.  His first paper was to work out in detail the forms of the vertex operators, with attention to the normalizations and unitarity.  His second was the bosonic open string theory, showing that it was finite for $N = 2^{13} = 8192$ Chan-Paton states (though there would be higher corrections), the analog of the $N=32$ that Yunhai and I studied.  But string theory did not hold Weinberg's interest.  I think it was because he wanted to derive the theory using the same principles as served him so well in QFT, but strings seemed to have new aspects that did not resonate so well with him.  A conference talk he gave, `Strings without Strings', reminded me of his earlier philosophy that geometry is not central to general relativity.

Most early string work focused on supersymmetric states, where the diliton would vanish order by order in string perturbation theory.  But soon we would have to deal with states of broken supersymmetry, and the dilaton energy would have to involve cancellation between different orders.  This would be easy to work out in field theory, but was clumsy without string field theory.  So Fischler, with Susskind, showed how it worked, cancelling string amplitudes against loop divergences.  

So I learned a lot from my colleagues even when not working with them.  I should also mention Clifford Burgess, Anamaria Font, and Fernando Quevedo, three of the international students, who independently wrote a very nice treatment of the low-energy effective action of the superstring.  All went on to successful careers in theory, with Fernando recently finishing a term as Director of the International Centre for Theoretical Physics in Trieste.\footnote{I will not try to make a comprehensive list as I did for Harvard, it would be too hard.  I will mentions those I worked with in the text, but here are just a few others -  Carlos Ordonez, Don Marolf and Scott Thomas (two who got away), and Brian Warr (who died too soon).}

\section{Austin part 2, 1988-1992}

\subsection{The book}

So in the summer of 1988, having realized that I would never be a great scientist, I decided to write a book.

This may come as a surprise.  Weren't things going so well?  Certainly the problems that I was working on were fun, and occasionally I got positive feedback from others I respected.  But I did not have a feeling that I was moving science forward.  The great excitement of the day was connecting the heterotic string to the observed standard model, and I did not seem to have the particular tools for this.  In fact, when I look back, I seemed to have worked almost entirely on what looked like oddities as compared to the real problem.  The only string paper that was fairly well-cited at the time was my first, on the Polyakov path integral, and that was almost all pedagogical.  Meanwhile, many others were making what looked like major progress.

Certainly, the most notable of these was Edward Witten.  For nearly ten years he had driven high energy theory forward with new ideas, the way that Feynman, Gell-Mann, Weinberg, Polyakov, and 't Hooft had done earlier.  I recall the pleasure, even before string theory came along, of reading each new paper by Witten and learning unexpected new aspects of quantum field theory.  But at the same time, it was overwhelming.

In Feynman's Nobel speech, he tells the story of poor Slotnick, whose just-finished Ph.D. dissertation Feynman had reproduced, and more, in a single night.  Not surprisingly, Slotnick never wrote another paper.  And stories had it that Feynman affected others  the same way.  I have earlier mentioned that first meeting with Witten, which was a little bit like Slotnick's meeting with Feynman.  But I don't think that Edward has ever shown the highly competitive streak of Feynman; instead, he is competing with history.  But each new paper from him gave me the joy of reading, and the question, ``why am I needed?''

On a smaller scale, I must have had some of this effect on my classmates at Caltech.  But science is large, and they found their own directions.  Happily, theoretical physics also turned out to be large, but I didn't know it then.  Before going on with the book, a few more bits of schadenfreude for you.  I had recently seen the movie Amadeus, which (a bit inaccurately) described Salieri's torment at being unable to match Mozart's genius.  So I  empathized with Salieri.  I also put a picture of Witten on the back of my office door, to desensitize myself for when we met.  (Yes, I was such a goof).

The reason for my book was that I had just taught a one-year string course based on the Polyakov path integral.  Green, Schwarz, and Witten (GSW) had just written a two-volume book on string theory but it did not include the Polyakov path integral, using mainly the older light-cone methods.  I thought that in a year I could transcribe my course notes, avoiding too much repetition with GSW.  People seemed to enjoy my writing, and I enjoyed it, though I did not account for how the effort would scale between a paper and a book.  And I kept wanting to improve things, and string theory kept moving, and it ended up taking nine years.  During this time I spent about 30\% of each year on it, mostly in the summer.   There was a break of a year when D-branes hit, but the next year I knew that I had to finish, and spent almost the whole year on it.

Having admitted to channeling Salieri, I can also tell you about channelling Michelangelo.  In the Michelangelo story, I cringed at the years that he spent on his commitment to making Pope Julius's tomb.  How could he have wasted so much of his creative life? It was only long after finishing my book that I realized that I had done exactly the same thing. 

I will not make much mention of the book as we go along.  It just did not intersect much with the rest of my life, even my research.  That seems surprising, but the book lagged the research. Just picture me slaving away, 30\% of my time.  But there is one bit of missed physics, right at the beginning, which I have always regretted.  I was thinking again about the monopole catalysis problem, trying to improve the theory.  I had an unusual effective field theory, with the effective fields lighter than the monopole but heavier than the others.  I realized that this would arise in many situations, like heavy-light quarks, and even proton-electron.  So I asked Mark Wise if he had seen this before.  He said it was very interesting, and we should work it out.  But I had just started the book, and was not ready to pause.  So I left this to Wise and Nathan Isgur.  The resulting Heavy Quark Theory was very useful.  So Wise and I joke that he gave me some projects when I got to Harvard, and later I paid him back.

Finally, three influences.  Steve Weinberg, for setting the bar with his beautiful gravitation book, which I hoped to match.  Initially we talked about collaborating on a book, but it would have been very difficult melding my non-historic approach with his.  
The second influence was Edward Witten, for the reasons given above.  The third was Jan Haag, who was our live-in nanny when my first son was one.  She was an interesting woman who had traveled the world, and planned to write her own autobiography.
So I figured that if my nanny could write an autobiography, then so could I. 

\subsection{Fun with duality}

Reflecting now on my work in that period, I was doing in string theory much of what I had done in QFT as a student, trying to understand what the theory really was.  Point particles in quantum field theory had been studied for a century.  Infinitely thin relativistic strings were new.  What special properties might they have?

A striking phenomenon special to strings was $T$-duality.  If you put a particle in a box and make the box smaller and smaller, all that happens is that the excited states get heavier and heavier due to the momentum quantization.   But for strings, after the box gets small enough, lighter and lighter states appear in the spectrum, and there is a perfect symmetry between a very large and very small box.  Apparently there was a minimum length, something one might expect in the ultimate short distance theory.  It was also an example of duality, the equivalence between the quantum theories of the large box and the small, but one that was visible even at weak coupling.  And in more current parlance, it was an example of emergent space.

Almost all of the attention to this subject had been for the closed bosonic string (as a warmup) and the heterotic string, as the putative theory of the real world.  But there were other string theories, the open and unoriented bosonic theories and the SUSY Type I, Type IIA, and Type IIB theories.  And I had two new students, Rob Leigh and Jin Dai, who needed problems.  Initially I divided the strings between them, but before long it became one big project.

 The IIA/IIB case was quickly solved: they are $T$-dual to each other, meaning that they are the same theory in different limits.  The duality transformation flips the sign of one fermion, which flipped the IIA and IIB strings.
It was a nice result, though later we learned that Dine, Huet, and Seiberg (DHS) had found this a few months earlier.  But the rest of our papers were orthogonal.

The other cases were harder.  We found that the $T$-duality flipped the normal Neumann boundary condition with the fixed Dirichlet condition.  This change in the boundary condition meant that on the open string, the endpoints were no longer free to move in some directions, while the interiors, and the closed strings, were still free.  After some thought, we realized that the string endpoints had to be stuck to some lower dimensional object, with any dimension obtainable depending on the number of $T$-duals.\footnote{Petr Horava and Mike Green were both interested in these new $T$-dualities at around the same time, with Horava's work in particular overlapping some of ours.}
  
So this was rather remarkable: starting with a theory of open and closed strings, one finds in the limit of a very small space a new large space, with closed and open strings, but also these new objects, one for each Chan-Paton factor.  Moreover, we reasoned that due to gravity, these objects could not be rigid, and we identified the excitations in the string spectrum.  So they were not stuck in the form given by $T$-duality, but could take any shape and number.  The term {\it $p$-branes} had just been coined by Achucarro, Evans, Townsend, and Wiltshire to describe the membranes of supergravity,\footnote{The word brane, which first appeared in the title of that 1987 paper, has now appeared in 8,500 titles.} so we called our new branes Dirichlet branes, or D-branes for short, to distinguish them.  The fact that a $T$-duality produced something (the D-brane) from nothing (empty space) was nicely resolved by realizing that empty space was actually full of space-filling D$9$-branes.

The T-duality of the unoriented strings led to new puzzles, and a new object.  We can think of $T$-duality as acting on the left- and right-movers as $(x_L, x_R) \to (-x_L, x_R)$, while the unoriented theory is defined by projecting on  the orientation $(x_L, x_R)  \to (x_R, x_L)$.  Conjugating orientation reversal by $T$, we get $(x_L, x_R)  \to (-x_R, -x_L)$.  This new operation is equivalent to a world-sheet reflection times a spacetime reflection.  The unoriented space is thus $T$-dual to an oriented string theory on a half-space.  This new kind of boundary we christened an orientifold, because it was constructed from the product of an orbifold and an orientation reversal.

I seem to like simple compound constructions (Q-ball, D-brane, orientifold, and later enhancon and discretuum).  I have a certain pride that other people would have discovered all these things, but by naming them I have put my stamp on them.  But orientifold was almost a joke, such a clumsy word for something I thought that no one would ever find interesting.  So today I give a private chuckle whenever I hear it.

The last $T$-duality we did not quite get right.  The Type I superstring had both D-branes and orientifolds, but we had mistakenly concluded that these were stuck together into a single object.  This would be true for the minimal D-brane number ($1 \over 2$ from the orientifolding), but with a higher number they had degrees of freedom that could move.  But in all this was pretty nice for a ten-page paper. 

You might think that with this great set of insights I had made it, and I did not need to write that darn book (which was on hold anyway, because I was suffering from research-withdrawal after 4 months of writing).  But I did not appreciate what I had done.  
I thought that the next step had to be finding the D-branes of the heterotic string, since this was assumed to be the real theory.  A nice argument by Dixon, Kaplunovsky, and Vafa showed that the standard model could not be obtained from the Type I or Type IIA,B theories.  But if I had any imagination, I would have realized that with the new possibilities from D-branes, the argument no longer held.  But I persisted, fruitlessly, in trying to find heterotic D-branes.

I gave zero talks about the paper, my lack of confidence and common sense stopping me.  I had forgotten Georgi's maxim, ``Don't hide your light under a bushel basket."  Had I given a few talks, someone in the audience, or even just the effort of writing the talk, might have led to the missing connections.  I have mused over the fact that my paper with Yunhai had shown that the D9-brane sourced a 10-form $RR$ potential, and the paper with Jin and Rob had shown that all the different D$p$-branes were connected by T-duality.  But it took me six years to put these two together.  Or more precisely, perhaps I knew the connection implicitly, but did not know what it meant; I needed someone to ask the right question.

As it was, the paper got two citations in its first five years.  But there was one very nice and important paper, written by Leigh entirely on his own, working out the effective field theory for the D-branes.  Leigh went on to some outstanding work, in strings, particle physics, and QFT, and is now a professor at Champaign-Urbana.  Dai works in the IT industry in Shanghai, currently with a startup, and has over 100 patents filed.

By the way, the title of this section was my intended title for the paper.  But Rob is a serious guy and vetoed it, so we ended up with ``New connections between string theories.''  Keeping score, of the five string theories, K. S. Narain had shown that the two heterotic theories were $T$-dual, we and DHS had found that the type II theories were $T$-dual, and we found that the type I theory was dual to type II, but with the ground state of type I dual to an excited state of type II, with D-branes.  The fact that type I theory was dual to type II with D-branes meant that the D-branes were BPS states, something whose significance may not have been clear, but soon would be.  By considering all the excitations of the D-branes, one could conclude that these were intrinsic excitations of the type II theory.  The last connection, between heterotic and type I-II, would come not from a perturbative $T$-duality, but from a nonperturbative one.

\subsection{Cosmological constant}

I think I first heard about the cosmological constant (CC) problem during a lecture by Sidney Coleman on spontaneous symmetry breaking.  Of course the classic Einstein story was well-known, but the full quantum problem, though known long ago to Pauli, had not penetrated into most discussions of QFT.  But as Coleman explained, spontaneous symmetry breaking pointed up the reality of the vacuum, and that the ground state was not the most symmetric state.  So it would naturally have an energy of order  the spontaneous symmetry breaking scale.  Moreover, even if canceled classically, it would get large quantum corrections.

I gave this a lot of thought as a postdoc, understanding why it was so hard to solve.  Each particle generated a large quantum contribution to the vacuum energy, and somehow all these would have to cancel perfectly.  Normally one would need a symmetry to enforce this.  Supersymmetry could do it, but it is a broken symmetry, so the cancellation should be inexact.  One might also look for a dynamical mechanism, whereby the CC backreacts on the matter fields so that they cancel.  But gravity is an irrelevant interaction (in the renormalization group sense).  This means that a quantum effect at length $l$ acts on spacetime at the much longer and weaker scale $l^2/l_{\rm P}$ (with $l_{\rm P}$ the Planck scale), much too small to cancel the CC.  In effect, what was needed was for some way for long-distance physics to feed back into the short-distance action.

Most string theorists, having seen such remarkable properties as T-duality, expected that string theory had some trick that we had not yet figured out.  So, like the renormalization problem before, the CC was one of the big questions that I always kept in mind.  When I wrote my first string paper, on the Polyakov path integral, my first calculation was the cosmological constant.\footnote{More precisely, what I was calculating was the dilaton potential, rather than a constant.  There was a tendency to conflate these in the early papers, with the expectation that higher order effects would fix the dilaton and produce a constant.}
It showed no particular suppression, although this was just for the toy bosonic theory.  But even with (broken) supersymmetry, studied by others, no suppression emerged.\footnote{A very nice paper by Rob Myers showed that the special values of 10 dimensions and zero vacuum energy were not actually required in string theory.  With Shanta de Alwis, a postdoc who had come to work on Weinberg's generalized gravity, and Rolf Schimmrigk, a student of Candelas, we tried to generalize this to give small values of the cosmological constant and the SUSY breaking, with limited success. \label{Myers}}

At this time some new ideas emerged.  These had nothing to do with string theory, but pure quantum gravity.  They led to great excitement for two or three years, and then it dissipated.  Today it is not even mentioned to students, it is one of those subjects they can save some energy not learning.  

I had thought this would be a good story to tell, but I have found it difficult.  It is not so much fun to remember ideas that rather thoroughly did not work, even after great promise.  So I will try to be brief.
Actually, there are two stories.  One, due to Coleman, was based on axion wormholes.  The second, due to Hawking (and also
Duff and P. van Nieuwenhuizen, and Aurilia, Nicolai, and Townsend) was based on four-form potentials.

The essentials of the Coleman story:  Quantum gravity plus axions give wormhole solutions connecting different points of spacetime.  At first sight, passing through these wormholes would destroy information, but Coleman, Giddings, and Strominger argued that summing over all configurations in the path integral made the wormhole (baby universe) coherent but random.  All the constants of nature would get random contributions, but if one measured the constant repeatedly the same value would be found everywhere.  But by considering the full path integral, with arbitrarily many de Sitter regions coupled through arbitrarily many wormholes, Coleman found that the path integral was infinitely peaked at zero cosmological constant, $\Lambda$, $e^{e^{1/G\Lambda}} \to \infty$ as $g \to 0$.  

This was remarkable.  And it fit the idea that long distance physics needed to feed back to the short distance action, the large Euclidean de Sitter space acting back on the baby universe action.  But there was also doubt.  The double positive double exponential was not like anything in field theory.  One would like to derive the path integral from a Hamiltonian, and this would normally lead to minus signs, or even phases from the determinant.  Several groups looked at this, including Willy Fischler, Igor Klebanov, Lenny Susskind, and me.  Besides finding no evidence for a peak at zero CC, we fulfilled Lenny's ambition of getting the word Googolplexus into the title of a paper (the number of states needed to get a small CC).  And those who studied the predictions for other constants of nature found that they were unphysical, or ambiguous.\footnote{A fun, but ultimately uninstructive, toy model of quantum gravity was to treat the string worldsheet as a 1+1 dimensional spacetime.}

After a couple of years, the subject was dropped as uninteresting, and even now it is painful for me to try to reconstruct the arguments.  I imagined that some features of this idea might return in the future, but they do not seemed to have.  I will return to this in around 16 years, book-time.

Hawking's idea was simpler, in that it used the quantum mechanics of de Sitter space but without the wormholes.  He did add one more degree of freedom, a 4-form field strength.  In four dimensions, such a field is nondynamical: it would be constant over all of spacetime, but with an arbitrary value.  If its value is integrated over the path integral, this will pick out a zero for the CC by a calculation similar to that for Coleman, but now with a single exponential $e^{1/G\Lambda}$.  

But there was another problem.  Even if all else worked, the mechanism would lead to an empty universe.  A universe with excitations, especially a highly excited universe like ours, would be exponentially suppressed.  So Fischler, my student Daniel Morgan, and I wanted to see if there was a process by which energy could appear from virtually nothing by tunneling.\footnote{Daniel also had some nice single-author papers, on forms of the renormalization group (showing Weinberg's and mine to be equivalent), and looking at the behavior of black holes with cutoffs.  He went into public science policy after graduation.}  This had already been argued by Farhi and Guth, based on a path integral saddle point, but there were subtleties.  We confirmed it in a Hamiltonian treatment, but the effect was probably too small for our purpose.\footnote{There was another problem, which did not appear until much later.  Note the resemblance between the $D$=4 4-form and the $D$=10 10-form of my work with Cai.  Hawking's form should be interpreted as a D-brane field, and therefore quantized, rather than the continuous value needed to cancel the CC.}

There was one other new CC idea out there, the anthropic principle.  As soon as you go beyond the standard model to some form of unification, it often happens that there are mechanisms that allow the constants of nature to vary.  We've seen two above, the 4-form field and the Euclidean wormhole.  Other possibilities would be a slowly rolling scalar (Banks), a downward rippling scalar (Abbott) or membrane (Brown and Teitelboim), or any kind of complicated potential with many minima.  In these conditions, it may be that the constants of nature differ over time or space, or in branches of the wavefunction.  

Weinberg, building on ideas of Linde and Banks, essentially said that this is all you need. Under these conditions, essentially all observers will see a very small cosmological constant.  The argument is that for observers, or any kind of organized structure, to form, there had to be a lot of space (bits), and a lot of time (events), and this requires the CC to be very small.  So if this `constant' is actually some sort of variable, the observers will only exist in the few regions of small CC.  

One of Weinberg's striking abilities was to take some new idea, even a very radical one, and turn it into a calculation that could be tested.  By replacing `observer' with `galaxy', he could show that the CC could be no larger than around 100 times the matter density, a large improvement over the prediction that it was set by the Planck scale $10^{120}$ or the weak scale $10^{60}$. 

This was remarkable to me, and upsetting.  This problem that I was spending much of my time on, which was supposed to be the clue as to the nature of quantum gravity, did not need a solution, it was nearly automatic.  But it required giving up the idea that the constants of nature, the lifetime goal for me and for my colleagues, was possible: it depended on details of astrophysics and partly even biology.  

But Weinberg had a prediction: the CC had no reason to be exactly zero.  Rather, it should be given by some random number less than 100 times the mass density.  Of course, the observed upper bound was already pushing down by 1 or 2 orders of magnitude, but 1 or 2 was much better than 60 or 120, and one could imagine a refined calculation. 

 And there were already some signs of a nonzero CC, such as the age problem (stars apparently older than the universe), which would be solved if there were a nonzero CC.  So I spent the next ten years hoping that the evidence for this would go away.  I do not know how many others were in the same state.  To me, Weinberg's argument was so clear, and should have been known to everyone.  But I had the benefit of talking to Weinberg in person, as well as my long history of unsuccessful attempts.  Most others would find it easier to continue their denial. 
 
 My fretting would have been much better spent asking, does string theory produce the dynamics needed for Weinberg's argument?  Fortunately, the question was still there for Raphael Bousso and me ten years later.  It was a measure of the general `anthropic denial' that no one else asked this question first.  And it is a tribute to Weinberg for his unique way of doing science, even asking questions that others might fear.

\subsection{Science fiction and science}

Around this time, I saw a very interesting preprint by Kip Thorne and collaborators.  General relativity has solutions with closed timelike curves, where an observer could return to a past event.  They had a scheme for constructing these. Start with a wormhole with both ends near a point.\footnote{We are now talking about wormholes that exist along the time direction, as opposed to the earlier wormholes that exist only for an instant.}  Then, boost one for a long time before bringing it back to rest.  Due to time dilation, the times are shifted: an observer who enters the unboosted end leaves the boosted end in the past.  Thorne, Morris, and Yurtsever wanted to see whether physics could make sense in such a space, perhaps even having observable consequences. 

As a veteran reader of science fiction, I was well aware of the grandfather paradox, and I was sure that Thorne was as well.  The observer could go far into the past and then kill his grandfather before he himself had been conceived.  So he would not have existed to start the process and kill his grandfather.  But there is a lot of free will involve here, so it did not make for a sharp argument.  But I realized that one could easily do away with that.  You can just replace the observer with a billiard ball, aimed so that it travels from the past through the future wormhole, and then leaves the past wormhole in just such a way as to intersect its previous path and knock it off course.   Then it will never be there to pass through the wormhole.  So there seems to be no consistent answer to whether the ball passed through the wormhole.

So I sent Thorne a message with this, and he got excited.  His next paper was about motion in wormhole spaces.  He had a possible solution to the conflict.  If the ball was deflected just a little bit, then after passing through the wormhole it could meet its former past at just the right place to make it all consistent.  He acknowledged me prominently for my note, which felt very good.  But I wondered whether Thorne would be taken seriously after writing such papers, especially while he was trying to get the billion dollar LIGO project approved.  

I have named a number of things, but I do not have many things named for me.  The Polchinski equation is just a refinement of Wilson's.  So the Polchinski paradox, a science fiction motivated idea for which I wrote no papers, seems to be my claim to fame.

My other bit of science fiction was due to Weinberg.  In his characteristic way, he had asked the question, how do we know that quantum mechanics is linear?  So he designed a generalized theory with nonlinearity,  satisfying certain consistency conditions.  This could be compared with experiment, and he found that any nonlinearity had to be extremely small.  I was studying the structure of his model, and realized that the EPR problem was no longer avoided: information could be sent faster than light (Gisin discovered this as well).  One could find a variant theory that avoided this, but the consequence was that different branches of the wavefunction could communicate.  I summarized this by saying that Weinberg's original theory allowed us to build EPR-phones, which send messages faster than light, while the modified theory allowed us to construct Everett-phones, which can communicate between different branches of the wavefunction.  A few months later, this made it into a column in Fantasy and Science Fiction.  

\subsection{Short stories}

\subsubsection{Nonperturbative strings}

String theory, impressive as it was, was still just a perturbation theory.  Finding the complete description was one of the big questions.  Even for the standard model, we knew that nonperturbative effects led to rich physics and were essential to confinement, chiral symmetry breaking, and more.  With quantum gravity added in, we could expect much more excitement in a nonperturbative theory.

The simplest guess would be to copy the structure of particle physics, replacing quantum field theory with some sort of string field theory.  Effectively, this amounted to breaking the string worldsheet up into pieces corresponding to string propagators and interactions, which would follow from a string field action.  But what worked for particles may not have been the right thing for strings.  The classical action for closed strings required a sum over an infinite number of terms.  Even worse, the quantum theory required an infinite number of additional terms at each order in $\hbar$.  Effectively, the amplitudes were being put into the action by hand --- it seemed to me more like an effective theory, or an action for a composite object like a hadron.  There were some nice constructions in this approach, like Witten's open string action, Sen's soliton solutions, and Zwiebach's BV symmetry, but it did not seem like enough.

A different direction that came up at this time was a solvable matrix quantum mechanics, 
found by Gross and Migdal, Douglas and Shenker, and Brezin and Kazakov,
which was equivalent to string theory in 1+1 dimensions.  In this low dimension, the only degree of freedom was a scalar, but the model was still rich enough to be interesting.   In effect this was an early version of holography, a connection that Polyakov in particular emphasized later, with AdS/CFT.  This was fascinating for me, and I wrote six or seven papers on it.  These were mostly about the connection between the 1+1 dimensional space and the matrix model, including the emergence of space and gravity, the finding of all classical solutions and some of their interesting properties, the attempt to go beyond 1+1 dimensions, and some issues about the nonperturbative definition.

The most interesting lesson from this system came from Steve Shenker, who showed that nonperturbative effects scaled as  
$e^{-C/g}$, where $g$ is the closed string coupling and $C$ a constant.  In quantum field theory, nonperturbative effects scale as $e^{-C/g^2}$.  So string theory had stronger nonperturbative effects than quantum field theory.  What could they be?  Perhaps I should have read my own papers.

\subsubsection{Working with Bryce}

Bryce Dewitt had strong opinions.  He was fun to talk with, about topology change (he was against), about the many world interpretation (he was for), and more.  I had the fun of joining a project with him.

Bryce, with the help of a large team of postdocs and students (Jorge de Lyra, See Kit Foong, Timothy Gallivan, Rob Harrington, Arie Kapulkin, and Eric Myers), was trying to directly answer the question of whether quantum gravity might be nonperturbatively renormalizable by directly integrating the path integral on the lattice.  It was not clear that what he was doing made renormalization group sense, but it was Bryce's characteristic way to choose his direction and plow through it.  Anyway, it was a hard question, and perhaps one would learn something in this way.

At least Bryce had made things easier by replacing the metric with an $O(2,1)/O(2)$ sigma model, with a lattice action he had determined through some reasoning of his own.  So the path integral involved two fields in four dimensions.  I noticed that for his specific action, the theory could be factorized into a free field and a self-interacting one.  So half the integral could be done by hand, with half still having to be done numerically.  This allowed for a lot of checks, and I was able to debug some of the team's long computer calculations using very simple ones.  Most notable was one case where there was a large discrepancy.  I realized that  there was a Schwinger term that needed to be included, and then the numbers fell right in line.  Bryce commented that he had never believed in Schwinger terms until he saw them in the numerical data.

And there was a conclusion: there was no high energy fixed point.  I wonder whether there might be a useful confrontation between this and asymptotic safety.

\subsubsection{Fermi liquids}

When I first learned about the Fermi liquid theory, I was puzzled by how one could neglect the electromagnetic interaction.  This was driven home even more when I taught graduate quantum mechanics using Davydov's book,
which went through BCS superconductivity in detail.  It was claimed that one could calculate things like the gap with great accuracy, while ignoring seemingly much larger effects.  It was said that this worked because we were working with quasiparticles, not electrons.

I had never encountered this word quasiparticle in QFT, and I did not know of any such method that would allow one to just ignore an interaction.  All I knew was effective field theory, so I tried that, and it worked.  The finite fermion density was unfamiliar for a relativistic theorist, but putting in the proper scaling made it just right.  All interactions were irrelevant except the one producing the superconducting condensate, which was marginally relevant.  So superconductivity was due to asymptotic freedom, just like confinement.  What I had done was well-known in terms of the Fermi liquid theory, but expressing it in the language of effective field theory made it more transparent to field theorists.

In my typical way, I was not planning to give any talks about this, or write a paper.  But I was co-organizing the 1992 TASI with Jeff Harvey, and it was suggested that I give a couple of impromptu lectures.  So I gave one lecture on how effective field theory works, including a very efficient summary of my renormalization proof.  The second explained how Fermi liquid theory fit in this framework, including the treatment of the BCS theory.  This has been a fairly valuable review, and I almost did not write it.  I should not hide my light under a bushel basket!

At the same time, the problem of high temperature superconductivity was a great puzzle.  Its low energy behavior, such as the conductivity, did not fit the Fermi theory.  The low energy interactions were larger, but there was no other stable low energy field theory known.  So I thought, maybe my new understanding of Fermi liquids would solve the problem.  I worked at it for several months, trying several things, but eventually decided that I had little to contribute.  It seems that it is still a puzzle.

\subsection{Students}

Finally, let me remember my third triad of grad students, Eric Smith, Djordje Minic, and Makoto Natsuume.  All three of them began working with me in Austin but finished after I moved to UCSB.  Eric and Makoto came with me, while  Djordje stayed with his wife in Austin.  All of them worked on varied subjects.  For most of you, this will be just a laundry list that you can skip.  But I remember many of these projects with pleasure, and am happy to see that all these students are still doing science.

Smith's first project was to show that $T$ duality held for a class of time-dependent solutions, something that had not been obvious in the literature.  He then worked out the light-cone action and spectrum for the 1+1 dimensional string theory.  We talked about my work on condensed matter, and he followed some of his own ideas and moved more in that direction.  He is now at the Santa Fe Institute.

Minic wrote a couple of papers with me and a postdoc Zhu Yang on solutions to the 1+1 dimensional string theory.    He then worked with Duane Dicus on quark dynamics, on his own on the Luttinger liquid, and with Shyamoli Chadhuri on 1+1 dimensional string black holes.  It was a tough time to get a job, and Minic went through many postdocs before getting a faculty position at Virginia Tech, where he has been very successful.  He and his wife Joy made many sacrifices, but his enthusiasm for physics was great, and it is wonderful that it worked out for them.

My first project for Natsuume was a follow up on my work with Strominger on noncritical strings, showing that the effective field theory could be derived from a renormalizable one, and so confirming our construction.  The second was a bit of speculation, seeing if he could make a generalization of the string S-matrix to higher dimensional objects (not much success).  On his own, he did nice work on the S-matrix of 1+1 dimensional string theory.  He also worked out some challenging $\alpha'$ corrections to the string theory black hole.  And together, he and I understood gravity in the 1+1 dimensional string theory.
Makoto ended up at KEK.  Besides his research, Natsuume has written a number of popular and pedagogical books on string theory and AdS/CFT in Japanese.

\subsection{Farewell to Austin}

Dorothy and I were happy in Austin.  Our two sons were born there, Steven in 1986 and Daniel in 1989, and were growing up with Texas accents (though they could drop these when they were just with us).  We enjoyed life in Austin, apart from the weather.  And both of us were in departments that we could thrive in.  So we were in no hurry to look elsewhere, and a few times I got  feelers but was not interested.

But California was still home to us.  Though we were each born about 2500 miles away, in opposite directions, we met in California, and each of us had many formative experiences there.  So if opportunities for both of us were available, we would be very tempted.  But Dorothy's field in particular, German linguistics, was very small, and there were no prospects for openings in sight. 

And then UC Santa Barbara came through.  Universities back then were not as responsive to two-body problems as they are now. But UCSB had a drive to grow to the top.  Dorothy's position was not as ideal as at Austin, going to a smaller department with interests less in tune with her own.  But she could pursue her research, and over time she was able to build an impressive program.
Of course it put us closer to our families.   And for me, it was a great opportunity, with excellent colleagues and a position at the Institute for Theoretical Physics, where I could focus on science.  

So we did not hesitate for long.  For me the most painful part was telling Weinberg.  He is a great man, and he was proud of the group he had built.

\section{D-branes and orientifolds, 1992-1995}

\subsection{UCSB and ITP}

The University of California at Santa Barbara had one of the leading string theory groups in the world, with Andy Strominger and Gary Horowitz, two of the discoverers of Calabi-Yau compactification, Mark Srednicki, one of the inventors of the invisible axion and a creative thinker in particle physics and quantum theory, Steve Giddings, one of the young leaders in string theory, and now me.
It was probably the largest and strongest string group in the world, outside of New Jersey.  Of course, New Jersey had
Princeton/IAS (Witten, Polyakov, Gross, Klebanov, Migdal, Wilczek, Nappi, Callan, Verlinde) and also Rutgers (Banks, Shenker, Seiberg, Friedan, Zamolodchikov).

I have always been astonished to think about the growth of UCSB physics, from becoming a university in 1944 to being a department that in some measures is as high as fifth in the country.  All the other top departments have been around since at least the previous century.  The coup by the gang of four, Jim Hartle (relativity), Ray Sawyer (particle physics), Doug Scalapino (numerical condensed matter), and Bob Sugar (lattice gauge theory), did not start this, but it greatly accelerated it.

In 1978, the High Energy program director at the National Science Foundation, Boris Kayser, saw a need to enhance collaboration between physicists at different institutions and in different fields, and also to support postdocs who were leaving physics for lack of support.\footnote{I am repeating this second hand, and may get corrected.  I hope that the best bits are true.}  He persuaded his superiors to fund this, to the tune of around a million dollars per year.  There was a call for proposals, and the story, as told by the winners, is that all the established departments said, `we know what to do with the money, give it to us.'  But UCSB's gang of four had a unique idea, to use the funds to bring scientists from around the world to interact for as long as six months, rather than the typical week-long conference.  There would be time to conceive new projects and carry out the collaboration there.  And the outstanding Walter Kohn agreed to come to UCSB to be the first director.

The NSF liked the UCSB proposal the best, but it wanted to see a greater contribution from UCSB.  So the gang proposed that UCSB would contribute four faculty positions, a huge bargaining chip that no other group could match.  These `Permanent Members' would mentor the postdocs and help design and run the scientific programs.  But they had to convince their new chancellor, Robert Huttenback, to back them.  Huttenback, just arrived from Caltech, knew about the competition because Murray Gell-Mann had boasted to him that Caltech's proposal would dominate UCSB's.  So Huttenback gave the gang what they asked for, and UCSB got the Institute for Theoretical Physics (ITP), and the gang of four became the Founders.  And so my position exists because of Murray's boast.

The first four PM's were Frank Wilczek and Tony Zee, both broad particle theorists, Jim Langer, condensed matter, and Doug Eardley, gravity.  By the time I arrived, Wilczek had moved to the IAS and Langer had become ITP Director.
Kohn had become a regular member of the physics department after finishing his 5-year term as Director, Robert Schrieffer had moved to Florida State after finishing the next 5-year term as Director, and Jim Langer had become the third Director.  I was the replacement for Wilczek,\footnote{There is a history of the IAS with title ``Who Got Einstein's Office?''  Wilczek did not get the office, but he did negotiate to get Einstein's house when he moved there.  And I got Wilczek's office at the ITP.  So at one point, the  tentative title of this memoir was ``Who got the office of the guy who got Einstein's house?"} and Matthew Fisher replaced Langer as PM the next year.

For its first fifteen years, the ITP operated on the 6th floor of Ellison, the rest of which housed History, Geography, and Political Science.  The whole institute, every office, could be seen from the point in the middle, with one corridor on the left, one on the right, and one perpendicular.  I liked the coziness of that, but it had long outgrown its space.  A year after I arrived, the University completed a dedicated ITP building, soon named for Walter Kohn.  With a beautiful design by Michael Graves, right across from the cliffs above the Pacific, it nearly doubled the ITP's capacity.  

\subsection{Information loss}

The format at ITP was the same for its first 20 years: two programs in the first half of the year and two in the second, each running for 5 months.  Typically there would be one program from each of the three main subfields --- high energy, condensed matter, and astrophysics --- while the fourth might be a new direction, an interdisciplinary area, or a second subject from one of the main areas.  It was also possible to have a one month miniprogram, scheduled on shorter notice, if something new came up. 

At the start of 1993, shortly after I arrived, the two areas were high energy physics and relativity.  The respective subjects were Nonperturbative String Theory and Small Scale Structure of Spacetime.  Effectively this was to be one double-size program, bringing together string theorist and general relativists.  What it developed into was a giant program on the black hole information paradox.

When Hawking's paper first appeared, my reaction was the same as most quantum field theorists.  When we burn a lump of coal, the disorder increases monotonically, so the entropy is maximized at the end.  But this is the coarse-grained thermodynamic entropy.  If we look at the microscopic 
quantum state, for coal that starts out in a pure state, the final state must also be pure and the microscopic van Neumann entropy must end up zero.   
Hawking was saying that for black holes, even the microscopic disorder increased monotonically, so that the final state was no longer pure but mixed, meaning that information had been lost and the Schrodinger evolution of quantum mechanics had to be modified.

So the naive reaction was that Hawking had mixed up the coarse-grained and microscopic descriptions, and a more careful treatment of the quantum state would find Hawking's mistake.  I first learned why Hawking's paradox is so difficult from a talk in Austin by Susskind, who had started thinking about the problem around the time I left SLAC.  As he explained, the difference between the coal and the black hole is that the black hole had an event horizon.  So with the coal, a quantum degree of freedom could rattle around for a while inside and then escape, but for a black hole, once it fell past the horizon there was no escape.

I am not sure why this problem rose to prominence just when it did.  I think Lenny's talks had introduced many string theorists to the difficulty and importance of the problem.  Also, Callan, Giddings, Harvey, and Strominger (CGHS) had recently presented a seemingly solvable two dimensional model of black holes, which allowed explicit studies of the problem.  This was probably the most active subject of the program.  But it seemed that there were ambiguities in the definition of the model, so Jeff Harvey said that it was like a Rorschach test: you could get whatever you expected.

That ITP workshop did not solve the problem.  The first big step forward was seven years later, with AdS/CFT, and even today there are key issues to resolve.  But it did succeed in communicating what the problem was and what the possible resolutions might be.  Essentially, almost all ideas fell into one of three categories: a) information is lost in the way that Hawking argued; b) information escapes from the black hole interior, seemingly requiring traveling faster than light; c) black holes do not evaporate all the way, but end up as Planckian remnants.  Each of these seemed to have unacceptable consequences.  

At the conference at the end of the program, I ran a discussion in which I took a vote as to which alternative people expected.  There were a few `remnants' and a few `none of the above,' but the bulk broke down 60-40 for information escape versus information loss.  This just reflected the fact that the audience was 60\% field theory/string theory and 40\% general relativity.  The former were more ready to give up relativity, and the latter to give up quantum mechanics.  As for myself I was a natural agnostic, going back and forth among the possibilities, looking for a resolution.

Another highlight of the conference was Susskind, who gave two talks.  At the start of the week he introduced the idea of black hole complementarity, but by the end of the week he had refined it so much that he insisted on speaking again, and the organizers (me, mostly) extended the session. 

Looking at my own work from this period (I do rely heavily on INSPIRE to make up for my memory), the program led me to several papers about the information problem:
1) Constructions of string theory black holes, with Giddings, Harvey, Shenker, and Strominger; 2) An argument (with Strominger, who later expanded it) that models where degrees of freedom from the black hole interior escape into baby universes do not actually destroy information, but are examples of remnants.  This was similar to the Coleman-Giddings-Strominger analysis of baby universes; 3) With Lowe, Susskind, Thorlacius, and Uglum, a project initiated by David Lowe to determine whether string theory is local.  If it were nonlocal, there might be no information problem.  Lenny interpreted our result as saying that it was indeed nonlocal.  I thought that it was inconclusive: it was not clear whether we were looking at the right observables.  This is still an open question.  We also analyzed the Nice Slice, a coordinate system first introduced by Robert Wald, where slices pass into the black hole but never get near the singularity.

\subsection{Working with Matthew}

Things were about to change in a big way, but first a little condensed matter interlude.  Matthew Fisher, newly arrived at the ITP, and Charley Kane, an assistant professor at Penn, were interested in the edge currents in the fractional quantum Hall system.  Even for the simplest case of $\frac{2}{3}$, there was a discrepancy between the observations, which showed charges moving only in one direction, and the theory, which had charges moving in both directions.  Fisher and Kane had the idea that disorder in the system would produce an interaction between the right- and left-movers, which would then flow to a new phase.

They did not see how to solve the resulting Hamiltonian, but thought that I might have some ideas.  It resembled a conformal field theory such as one encounters with strings, but there were two complications: the disorder, and the absence of Lorentz invariance.  But surprisingly, a bit of fiddling revealed an unexpected symmetry, which allowed the 
model to be solved.  It had the desired feature that currents moved only in one direction, but also an unexpected feature: a neutral mode moving in the opposite direction.

This was one of my few measurable predictions.  Unfortunately the neutral mode did not seem to be there.  Apparently it was observed recently, more than 20 years after the prediction, but the story was more complicated.
I did learn that in condensed matter, authors are not determined alphabetically but often younger to older.  And I did not meet Kane at the time, Fisher was the intermediary, but he has become quite distinguished for his work on topological insulators.
	
\subsection{Strings '95}

This was a slow time in string theory.  The excitement from the first superstring revolution had passed, and many directions had been explored without clear result.  But beneath the surface, something was brewing.  Besides D-branes and supermembranes, there were the black branes of Horowitz and Strominger, the weak/strong duality conjecture of Font, Ibanez, L{\"u}st, and Quevedo, the 5-brane conjectures of Duff and Strominger, the study of duality effective actions and spectra by Schwarz and Sen, and in gauge theory the tests by Vafa and Witten and by Seiberg, and more.  
All of these were weak/strong dualities, dubbed $S$-duality by Font et al., as opposed to the easy $T$-dualities.
But in the `fog of war,' the connections between these different strands were not obvious.  As Andy Strominger has reminded me, he, Gary Horowitz and I had lunch together nearly every day for three years, without realizing that their black $p$-branes and my D$p$-branes were the same.

One thing that did strike me was Seiberg's paper on strongly coupled ${\cal N}=1$ gauge theories.  His ${\cal N}=2$ papers with Witten a few months later have been much more extensively followed up, because the higher symmetry allows more calculations.  But I was impressed that even for the presumably more physical  ${\cal N}=1$ theories one could do exact calculations.  I could not have done what Seiberg did.  I would have needed to see something proven, probably this would require first figuring out what cutoff to use, and this would get nowhere.  But what Seiberg said was that if a strong coupling duality passed several nontrivial tests, it had to be for a reason, and duality should be the default.  He wanted to know what was true, not what could be proved.  And as more dualities were found, they formed consistent webs.

For myself, I revisited D-branes a bit.  With fellow aficionado Michael Green we looked at a possible new interpretation of D-branes, but it went nowhere.  I also thought about Shenker's argument that nonperturbative string effects should be of order $e^{-C/g}$, and realized that D$-1$ branes (D-instantons) were exactly of this order.  This was a bit nontrivial: one had to realize that D-objects were independent, so one had to sum not only over string worldsheets, but also over D-brane degrees of freedom.  I also gave a set of Le Houches lectures entitled ``What is String Theory?''  These consisted of several introductory chapters from my book (remember that?), and some of the recent attempts to go further. I included a short section on the various duality ideas noted above.

When Strings '95 began at the University of Southern California that March, there was some feeling of gloom.  This was both professional and scientific.  The first problem was that there were large numbers of excellent postdocs who were not getting jobs.  The organizers ran a session at which the postdocs could express their unhappiness.  Somehow I was asked to moderate, but I had no ideas to offer.  The scientific problem was the post-revolution slowdown noted above, as well as the lack of prospects for experiment.  Susskind addressed these in his after dinner speech.  His theme was that we did not need experiment, that we could figure string theory out without it.  He supported this by looking at various past theoretical discoveries, and showing how they could have been reached by thought alone.  But I was not convinced that we could have figured out quantum mechanics without experiment.

However, one of Susskind's points has stayed with me.  He said that he wanted to know the mathematics of the equations, not the mathematics of the solutions.  Techniques for solving a problem, as in geometry, can be much more elaborate than those originally needed to define the problem.
Like Susskind, I always wanted to find the simplest example that would make a point.

The irony was that both problems had been transformed just two days earlier, when the second superstring revolution began.  Everyone had heard Witten's talk, but the full magnitude of it took some time to absorb.  My recollection is that  the talk began with Witten saying that the organizers had asked the speakers to think about big questions, so he was going to give the strong coupling behavior for every string theory in every dimension.  The strategy was remarkably simple (he noted that he was building on recent work of Hull and Townsend).  Essentially, assume that weak/strong duality ($S$-duality) is true and see if this is consistent, as Seiberg and others had recently done to great effect in SUSY gauge theories.

Taking any of the five known string theories in ten dimensions, there was a dilaton field $\phi$ and an effective action $S(g_{\mu\nu}, \phi, \ldots)$.  Duality would mean that under $\phi \to -\phi$, perhaps with some other changes of variable, the action would again be that of some string theory.  Indeed, this worked for the Type IIB string --- it was dual to itself.  For the heterotic $SO(32)$ string, the dual theory was the Type I theory (and vice versa), which had the same gauge groups and supersymmetries.  Type IIA theory threw a bit of a curve: the strongly coupled limit is eleven-dimensional supergravity.  Finally, the dual of heterotic $E_8 \times E_8$ was a puzzle in the initial talk, but Witten and Petr Horava soon identified it as eleven-dimensional supergravity on an orbifold.  For each of the five ten-dimensional string theories, and eleven-dimensional supergravity, there was a unique candidate strong coupling dual.  Moreover, these dualities extended to BPS excited states, and to lower dimensional states, in a highly connected and consistent way.  And combining these $S$-dualities with the $T$-dualities, it seemed that all string theories were connected.\footnote{In his talk, Witten introduced the term `M theory' to denote the eleven-dimensional theory dual to Type IIA and heterotic $E_8 \times E_8$.
The `M' might turn out to be mystery, magic, or membrane, when the theory was better known.  BFSS (9.2) would add `matrix.'  The term M theory is   refer also the full quantum theory, with all the dual limits.}
It was rather overwhelming, for me and the rest of the audience.  Even Witten described his conjectures as shaky, as compared to $T$-duality.  But just in the previous year the work of Seiberg and Witten had finally made this kind of reasoning convincing in supersymmetric gauge theories, and so it did not take long to believe this for the more mysterious regime of strings.

At the end of Witten's talk, Mike Green and I looked at each other and said `that looks like D-branes.'  For me what was most striking was that the $e^{-C/g}$ effects that I discussed above occurred extensively in Witten's dualities.  But there were a lot of things to think about.  Witten's talk ran to roughly 60 slides, and there was so much in it that was new.  I ended up with a list of 14 homework problems just to understand the talk.  There was a disproportionate number of questions about K3's, as there had been in the talk.  And the open string questions were near the end, in the talk and in the list.  

\subsection{D-branes}

The start of the second superstring revolution did not go well for me.  Counting Witten's talk as day zero, day two was my ineffective discussion with the postdocs.  (My one conference talk, on day -1, was a review of the black hole information problem).  On day 14 I arrived at the ICTP in Trieste to give a set of spring school lectures on string theory.  My plan had been largely to repeat my lectures `What is String Theory?' from Le Houches.  These had gone well when I presented them eight months earlier, but now they felt years out of date.  I did not have the time to absorb the new understanding.  A set of lectures on D-branes would have been great, but their significance still had not emerged.  So I did repeat the earlier lectures, feeling quite depressed as other speakers like Seiberg and Sen were zooming forward into the new era.  Between this and the jet lag I slept very badly, so one time I fell asleep during my own lecture.  (If you find this hard to believe, you can ask Seiberg.)

But I gradually caught up.  With Shyamoli Chaudhuri I studied dualities in some models she had developed, heterotic string compactifications with maximal supersymmetries.  With my first UCSB student, Eric Gimon, I studied type I string compactifications.  Both projects had some K3's in them --- I was doing my homework.  And I had gotten past my natural skepticism, after seeing weak/strong predictions work.

By August I had learned enough that I could give reviews of string duality to nonexperts, and I was at the Yukawa Institute for Theoretical Physics to present one.  My former student Natsuume had just returned to Japan, and gave me an extended tour of Kyoto, including a traditional bath house and many scenic points.\footnote{Though I always had the feeling that I was embarrassing Makoto, as when I wandered off the usual paths, or lost my train ticket and had to try to explain to the conductor where I was supposed to be.}  But the most interesting part of the trip was going to a tiny laundromat in Kyoto by myself.  After loading the washer, I spent a little time looking at the Japanese magazines, and then sat down to do some physics.  Next up was the homework about open strings.  And immediately I ran into a problem.

The weak/strong duality between heterotic $SO(32)$ strings and Type I strings from Witten's talk passed basic tests like the action, but overall it had been little studied.  These strings were less clearly connected to phenomenology, and had the additional complications of orientation reversal and open strings.  One of the first checks would be putting the system on a small circle, do a T-duality on both sides, and see if it made sense.  And it didn't: there was a range of parameter space where both sides were weakly coupled.  But this would be a contradiction: two different results for the same theory.

When I got back to the ITP, I emailed Witten and said that I thought there was an inconsistency with his conjecture.  So we first did some calculations of non-BPS states (the BPS states are automatic and do not provide a test), and there were indeed states in the heterotic theory which did not seem to have images in Type I.  We figured out what the problem had to be.  Because of the Type I orientifolding, the T-dual Type I$'$ theory was not translation invariant.  But that would mean that the dilaton is a function of position, and with a nonzero mean value it could still blow up in places, producing additional states.  So we calculated this and it worked exactly.  

It was an unconventional calculation, with discontinuities along eight-dimensional planes.  I referred to these as D8-branes, and Witten asked, `How can these be D8-branes, they are supersymmetric?'   And I explained that D-branes are BPS states, carrying Ramond-Ramond (R-R) charges.  Witten seemed astonished, and said that I should write this up (I don't thing we met in person, and don't recall whether his astonishment was conveyed by phone or email).  So I dropped everything and wrote.

The paper took just a little over a week to write.  Most of it was a careful presentation of what was in the papers with Cai, and Dai and Leigh.  But there was one new calculation that I felt was needed.  D-branes whose dimension sum to six form an electric-magnetic pair, and so their charges had to satisfy the Dirac-Nepomechie-Teitelboim quantization, $q q' = 2\pi n$ for integer $n$.  Otherwise, the theory could not be consistently quantized.  The calculation was of my favorite kind, a vacuum amplitude, but now with a cylinder bounded by two D-branes.
On my first pass through, $n$ came out to be $\pi \sqrt 2$, and on my second it was $1/\sqrt 2$.  Either of these would be inconsistent, but these are the standard kind of error that one gets with a new calculation.  On the third it was exactly $n=1$.  The D-branes exactly saturated the quantized charges, strongly suggesting that these were the sources of R-R fields, which previously had no weakly coupled limit and were just identified as charged singularities.

Even as I was doing this, I started getting messages from people who had heard from Witten that I was writing an important new paper.  And so I began to realize that I had finally, at the ripe old age of 41, done something that had changed the direction of science.  More than that, it was a shock wave, for me and the rest of the field.  I had been living with D-branes for eight years, but never taking it too seriously because of the lack of heterotic D-branes (still true at weak coupling, but now they were needed to understand strong coupling).  But for almost everyone else, it was a new thing: string theory was no longer just string theory, it had D-branes as well.  These made many new calculations possible, and rather suddenly string theory became D-brane theory.  Of course, two years later it became AdS/CFT theory, which is still our most complete picture.\footnote{My history has not had much about my personal life, but hear is an odd note. My son Steven was playing roller hockey, and a coach was needed.  I had no aptitude for this, but neither did anyone else, so I volunteered.  This was very stressful for me, an it came at the same time as the D-branes.  What I remember is that the coaching consumed much more of my attention and energy than the D-branes.}

For eight years, D-branes had belonged to me and a few other fans, but now it was out in the world for everyone.  Within weeks, people found implications that I had never expected.  
Witten used them to calculate the bound states of strings and branes.  Douglas, and Witten, connected D-branes to instantons.  Witten and I finished our paper, which also demonstrated the duality between type I D1-branes and heterotic strings.   Strominger determined the rules for branes to end on other branes.  Townsend, and Schwarz, worked out the connection between D-branes and the eleven-dimensional M2-branes (the M now added by Witten to denote the eleven-dimensional theory).  Vafa, with Bershadsky and Sadov and with Ooguri, did something topological with them that I did not understand and did not fully approve of.\footnote{I have told Vafa that one of my life goals is to understand one of his papers, but no success yet.}  Bachas determined the scattering amplitudes for D-branes.  Strominger and I wrote a paper on D-branes and Calabi-Yau manifolds which was mostly his (he is always generous); my main contribution was an early study of brane creation at brane crossings.  And, a few days into the new year, Strominger and Vafa (SV) used D-branes to give a statistical interpretation to the black hole entropy for the first time.

You might wonder, how long had I known the answer to Witten's question?  I had known that D9-branes were BPS states that coupled to R-R fields since my work with Cai, and that all D$p$-branes were equivalent under $T$-duality since my work with Dai and Leigh.  But I think that I only fully connected them when Witten asked the question.

\subsection{Family time}

This may seem like an odd point for this change of subject.  But in fact it is fitting, because just as my physics career was taking this spectacular jump, most of my mental energy was actually being spent on coaching Steven and Daniel's roller hockey team.

Both Dorothy and I were mostly unathletic before college, her from going to Catholic schools and me from general nerdiness.  But in college we both enjoyed sports, and we met playing volleyball in our first graduate year.  When our first son, Steven, came along this accelerated.  From the age of one or so he wanted me to be throwing or kicking a ball to him all the time.  Daniel seemed more easygoing, but he also joined in, and so life for us largely centered on sports.

Steven started playing roller hockey when he was six, and after a few years I was asked to coach.  This did not come naturally to me.  Even teaching physics had always made me anxious, and here I had no expertise.  I took it on, and so spent the quarter  mostly figuring that out.  But this was the exact same time D-branes came along: somehow it all worked out.

Just to finish this subject, even Dorothy started playing roller hockey.  For a while, all four of us were on the same team.  Eventually, I switched to biking in the heights around Santa Barbara, Steven switched to ice hockey, and Daniel to wrestling and martial arts.\footnote{Though Dorothy and I have tunneled to a new fixed point, pickleball.}

The whole family graduated from Berkeley, in spite of Paul Martin's advice.  Steven graduated in Economics and Statistics.   After several years working in the financial industry, he is pursuing work in Psychology.  Daniel graduated in Molecular and Cell Biology, became a Doctor of Pharmacy at UC San Fransisco, and is now a Pharmacist at the Stanford Medical Center.

\section{The CC and the discretuum, 1996-2000}

\subsection{Following up}

So next came a lot of lectures and colloquia.  D-branes were a fun story to tell: basically just systematic application of $T$-duality, and from it one gets so much, but it was all new to most people.  Right after I wrote my D-brane paper, I gave a lecture series at the ITP, transcribed by Clifford Johnson and Shyamoli Chadhuri; a few months later I gave an expanded version of this at TASI.  For the colloquia, the most notable point was the duality diagram (shown).  This emphasized that the five string theories and M theory are connected, and each is the limit of the moduli space of a single quantum theory.  The theories are supposed to depict dual theories close together, separated alternately by $S$ and $T$ dualities.  But I somehow switched the two heterotic theories, both in the colloquia and in my book.\footnote{At Fermilab, one of the audience, presumably a hunter, said that the diagram looked like a deerskin, so I always think of it as the deerskin diagram.}
\begin{figure}[!ht]
\begin{center}
\vspace {-5pt}
\includegraphics[width=5in]{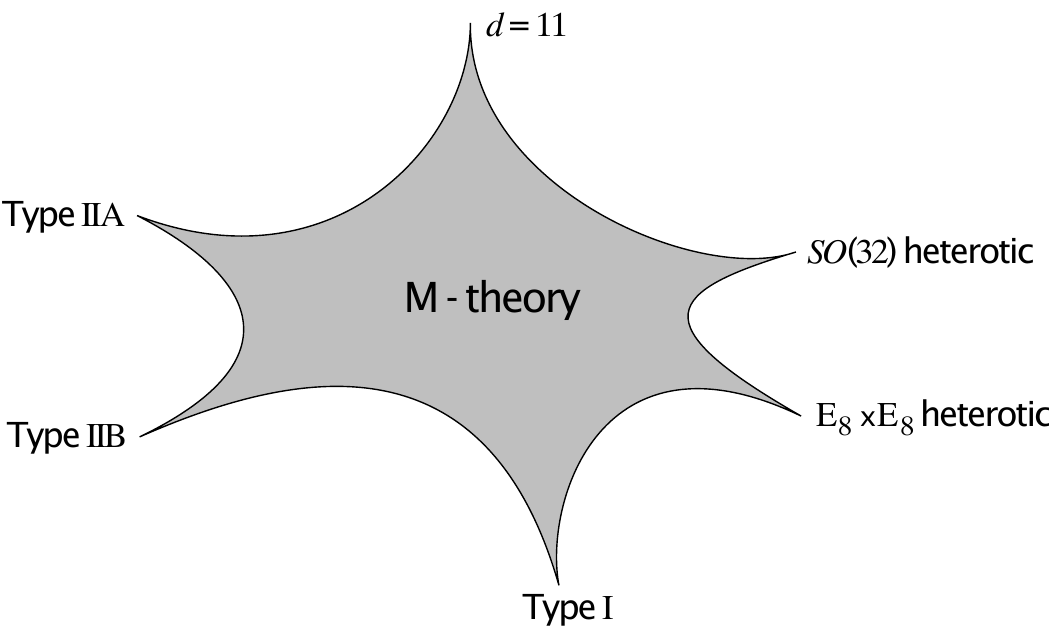}
\end{center}
\vspace {-10pt}
\caption{The duality diagram, with $SO(32)$ and E$_8 \times $E$_8$ accidentally transposed.  Correctly drawn, the edges alternate between S-dualities and T-dualities.}
\label{fig:radii}
\end{figure}

So what to work on next?  Looking at INSPIRE, I see that my next three papers, besides the reviews, were about orientifolds and K3's.  This is more geometric than I would normally like, but orientifolds were a bit of fun that was richer than D-branes alone.  And K3's, the simplest Calabi-Yau manifolds, had an orbifold limit that satisfied my preference for no curvature.  So Gimon and I finished his paper, which was rather more extensive than when it began before D-branes.\footnote{Gimon also had two nice papers on this subject with Clifford Johnson.  He went on to postdocs at Caltech, Princeton, and Berkeley, writing quite a few nice papers.  He now works on energy policy, sustainability, and philanthropy.}
Berkooz, Leigh, Schwarz, Seiberg, Witten, and I made a study of six-dimensional K3 solutions that I think began as a collaboration with Edward and then grew when other groups were working on the same problems.  My last paper was single-author again about K3 puzzles, which I liked because I got to use an idea of Michael Douglas of using D-branes as probes.  

Strings 96 was held at Santa Barbara.  Strominger had said to me a year earlier, ``We should run Strings next year," by which he meant ``Joe, you should run Strings next year.''  My light teaching load made it hard to object, though I have never been a good organizer.  But with the ITP postdocs (Shyamoli Chaudhuri, Clifford Johnson, and Katrin Becker) and especially the excellent ITP staff, it went well.  But I was not happy with my own talk.  I felt that I should have some ringing program for the future, after my world-changing paper of eight months earlier, but all I had was some subtle inconsistencies of certain orientifolds.

I recall two notable events from the meeting. The first was an email signed Steven Hawking, with title ÒI Have Changed My Mind. Information is Not Lost.Ó But this was a spoof, and soon we heard from the real Hawking, with title ÒWhy I Have Not Changed My Mind.Ó The second event was the announcement of Matrix theory, by Banks, Fischler, Shenker, and Susskind
98
(BFSS). I believe that Susskind claimed that they had discovered it at the meeting (and I think he once again demanded a second talk).

\subsection{Revolutions three and four}

Almost immediately after my D-brane paper, Strominger came to me excited that he would be able to calculate the microscopic density of states of black holes.  Having learned GR from Weinberg, I had not given this question much thought, but Strominger, a more gravitational physicist, told me that this was just as important as the information problem.  His calculation was just off by a constant, and he was looking for help.  This was all too new to me, and I had nothing to contribute.  But he found Vafa, who had the right tools, and they  
got the first precise counting of black hole states.  They had connected string theory to a new aspect of quantum gravity.

Gary Horowitz also had a long-standing interest in the black hole entropy.  He kept coming back to the question, how do we count the states for ordinary Schwarzschild black holes, not just the highly supersymmetric Strominger-Vafa black holes.  We could not get as sharp and answer as SV, but we did get a crude but useful result, extending an idea of Susskind.  Imagine turning down the string coupling for a black hole.  The black hole gets smaller, and eventually reaches the string length.  At that point, one should match the  black hole density of states to that of the weakly coupled D-branes and strings.  This gave a correspondence principle, matching the approximate counting for various black holes.  In a follow-up we studied the transitions of long strings to black holes.

After the successes of black hole state counting, it was natural to think about comparing the dynamics of black holes and D-branes, with an eye on understanding Hawking radiation and eventually information.  Douglas, Strominger, and I, working at Aspen, studied branes as dynamical probes, interacting with clumps of D-branes and also with the dual black hole.  These agreed up to one loop, a dynamical correspondence between D-branes and black holes.  At two loops they disagreed.  We looked for many solutions to this, but I think in the end we must have been calculating a quantity that was BPS only up to one loop, an issue that was confusing in the early days of duality.  In any event, the comparison of the dynamics of D-branes and black holes became an active issue around then, and the unexpected agreements between the two sides was one of the clues that led Maldacena to AdS/CFT.

The BFSS matrix model, presented at Strings 16, was fascinating.  It combined D-branes, eleven-dimensional supergravity, and matrix models into what was argued to be a complete description of M theory, the mysterious theory that lived in eleven-dimensions and was dual to string theory.  There was, at the time, a bit of confusion about what this meant.  Now we would understand it as a duality, two distinct descriptions of the same theory that are weakly coupled in distinct regimes.  But the way the theory was presented, many of us interpreted it more as a weak-weak duality, where one obtained the dual gravitational interaction from an explicit calculation in the matrix theory.

Thus it was interesting to see how far the calculations of BFSS could be taken.  The BFSS construction treated the longitudinal and 
transverse directions differently, but they had to combine in a Lorentz invariant way.  The initial calculations of graviton scattering were all for transverse momenta.  Longitudinal processes were harder, involving instantons rather than loops in the matrix theory.  But we had a new postdoc, Philippe Pouliot, who had some experience with instantons, and together we found that the instanton matched the longitudinal process.  This was fun for me, my first detailed instanton calculation, and a foothold into M-theory.

Two other postdocs, the sisters Katrin and Melanie Becker, were also extending the BFSS calculations, from one loop to two.  It was the kind of hard calculation that they enjoyed.  Arkady Tseytlin and I joined in to understand how the result should match on to the gravitational theory.  We found that again they matched.  

The BFSS theory was unusual in several ways, one of which was the need for infinite dimensional matrices.  Susskind argued that even for matrices of finite dimensional $N$, there was a physical interpretation of the theory and its dual.  The longitudinal direction becomes periodic, with quantized charge $N$.  This was an interesting quantum system, a periodic null direction, and with my second UCSB student, Simeon Hellerman, we unraveled some of its subtleties.
I would have liked to take these subjects further, but other obligations intervened.

I see the second superstring revolution as five waves in succession: the first four were Witten's Strings talk (and the preceding Hull-Townsend paper), D-branes, the SV black hole counting, and the BFSS matrix model.  AdS/CFT would be the fifth and crowning glory.  Each built on the ones before it, and each greatly expanded our understanding of string theory.

\subsection{Ch-ch-changes}

I had taken a year off from the book, but it had reached the point where I had to finish, no matter what.  And so I resolved to do nothing else until I was done.  People tell me that I was a zombie during this period, that they knew that there was no point in trying to talk to me.  

To give myself discipline, I wrote down the number of pages I wrote each day.  The list has been tacked to my office bulletin board 
for twenty years, but I had not looked at it in that time, until writing this chapter.  When I did I was stunned.  In nine years, I had written what is now volume 1, less than half the present book.  This is consistent with Candelas's maxim, it is never too late to give up on a book.  And I think I would have, if I had not told everyone that I was writing it.  At least the superstring revolution did not make things a lot worse.  There was a chapter on Superstring Duality, and one on D-branes, but after giving colloquia and lectures these were easy to write.  Then there was just a section each for black hole entropy and on Matrix theory, just to give an idea, and a paragraph on AdS/CFT, which just made the deadline.

One reason for the time spent on volume 1 was that I rewrote the first few chapters several times.  I reduced the amount about BRST symmetry, which had seemed like it might be the key principle at the beginning, and correspondingly I added to the conformal field theory.  I had the idea that I would write a book so clear that a student would pick it up one night, be unable to put it down, and in the morning they would know string theory.  I never got it to a point that satisfied me, but people seem to find it useful.

According to my tally, it took 83 days of writing, spread over 6 months, to produce the 500 page Volume 2: a little over 3 days per week for 6 months.  This does not include the time spent researching the many subjects in the book, most of which I had not worked on myself.
There were also three short breaks, to finish the papers with Becker, Becker, and Tseytlen, with Horowitz, and with Hellerman.  I had some life: the record shows a four day family trip to the Sequoia forest and the Kern river, and I continued to coach my sons' hockey teams.  But I am sure that I was usually a zombie.

The decision to split the book in two came during this period, when the length became clear.  The overall title, simply `String Theory,' had been in place for a long time.  Initially I had used `A Modern Introduction to String Theory,' signifying the use of the Polyakov description, but I realized how quickly such a title could look dated.  Though if I were to write it today, it is not obvious how else to start.  I also started using `Joe's Big Book of String' as an informal title very early; I should have fought harder to make this the official title.

And after the writing, there was another six months of drudgery: proof reading, checking equations, designing exercises, making 
copyeditor's corrections, writing a glossary, references, and an index.  And at each step I had to go through all 800 pages.  The index at least was fun.  There is a right way to do an index, which is to go over each page and see if there is anything on the page that a reader would need to find.  But finally I was done.

I had fallen short on my easy-reading goal, and I also fell short on my goal of no typos.  I had gone through every equation, but I have to face the fact that being detail oriented is not one of my strengths.  And it is especially hard to keep consistent notation on an 800 page book with many interlocking subjects.  So there are now more than 400 errata, at least 200 coming from Bank's then-student Lubos Motl.

Finally, I had planned to include a proof of the finiteness of superstring theory.  I think I had done a good job explaining why the bosonic string is finite, modulo the tachyon.  But there was no proof in the literature, and after a few attempts I realized that it was beyond me in any reasonable time.  Indeed, this was done only recently, by Witten, in several long papers.

Earlier I mentioned Heavy Quark Theory as a missed opportunity from working on my book.  A second one was the chance to work more with Simeon Hellermen.  He was an outstanding student, with a unique approach to life (for example, his current seminars consist of 3000 slides, shown in stop motion).  We wrote two papers getting into Matrix theory, but then I had to go into zombie mode for a year at a key time for him.  He wrote two nice papers with Sean Carroll (then an ITP postdoc) and Mark Trodden on domain walls.
He then went on to postdocs at SLAC/Stanford and the IAS and then a faculty position at IPMU in Tokyo, writing novel papers all the way.\footnote{
Hellermen was the kind of wiseacre who would tell me that going to Stanford was a step up for him.  And I was the kind of wiseacre who would respond ``Yes, and I'm the one who had to lie to get you in there.''  It was a comeback worthy of Sidney Coleman.}

I don't recall any particular celebration, just a chance to get back to work and catch up with all the latest excitement.  The royalties started coming in, which was a nice bonus but of course not the reason I wrote the book.  Years before, David Jackson had a party, and showed us the house in the Berkeley hills that his E\&M book had paid for.  Some time later, in Florida, Pierre Ramond showed me the nice telescope that he had bought with the royalties of is quantum field theory book.  Well, over time, my book paid for a BMW, including taxes: the root mean square of the house and the telescope.

So freedom from the book was the first big change.  At about the same time, David Gross arrived to become the new ITP director, the second big change.  After Kohn's five years, and Schrieffer's five years, Jim Langer had stayed for seven years to try to find his successor, and then Jim Hartle became interim director, and continued the search.  Their patience was rewarded when David Gross accepted.

I had not known Gross very well before, but his reputation as a force of nature was quickly justified.  The ITP had been running in its original mode for 19 years, and was still regarded as a model for the world.  But there was a need for renewal,  and Gross came in with a bang.  Right from the start there were changes.  An espresso maker was the first symbolic step, but then there were new programs for graduate students, physicists doing research at universities with heavy teaching loads, high school teachers, artists, and journalists, and new or expanded areas of science such as biophysics, mathematical physics, and geophysics.  

Most significantly, he reorganized the programs.  These had been running on the same  $2 \times 2 \times 5$ month annual schedule since the beginning.  But the new building was not being fully utilized.  Also, it did not make sense that every program should have the same length: some fields and subfields are bigger than others.  And changes in families and universities made the idealized five month stay impractical.  So programs became 50\% larger, but with variable length.  It took a while to convince the staff, some had spent years with the old system, but Gross gets his way.  And we still had Boris Kayser at the NSF to help fund the expansion (but never enough with the NSF).  So there was a new feeling of excitement.

The third big change was Strominger moving to Harvard (where his father taught Chemistry).\footnote{I have to confess that at all my stops I have had the good fortune to be associated with someone of vision --- Weinberg, Strominger, Gross --- because that is a quality that I do not believe I could learn.  So trading Gross for Strominger satisfied my personal conservation law.} 
The second superstring revolution had set off a wave of reshuffling.  Departments could tell that there was something exciting going on, even if they were not sure what it was.  And a large backlog of excellent grad students got jobs.

\subsection{AdS/CFT}

While I was finishing the book, the fifth wave of the revolution struck, AdS/CFT.  But I was in the perfect place.  A program, `Dualities in String Theory,' was scheduled for January to June, 1998, and Maldacena's paper appeared just a month before.  
Moreover, he was attending the program, and spoke about his work in the second week.  Neither the paper nor the talk produced an immediate sensation.  It was too new.  We had internalized field-field dualities, and string-string dualities, but string-field dualities? How could the degrees of freedom match?

So, like many, I went through the Kubler-Ross stages for dualities: disbelief, contradiction, testing, and acceptance.  The immediate contradiction was that string theory had many more degrees of freedom than field theory.  But the large $N$ of the field theory made many things possible.  More specifically, it seemed that one could find non-supersymmetric string states that had no analog in the field theory.  But a closer look identified them as bound states.  And after a few such checks, duality became the simplest explanation for what was happening.  Having come to this point of view, it bothered me that for a long time people would say that AdS/CFT is just a conjecture, rather than a duality.  Of course, dualities are almost all conjectures, but `duality' indicates the further  
tests above.  And AdS/CFT was rapidly subjected to an enormous number of tests, without contradiction.

I got a slow start on AdS/CFT while I finished my book.  But I was used to that.  I usually went into a new area slowly, while I tried to understand what was really going on.  For this reason, I have rarely had to worry about being scooped; if someone else can solve the problem, I am not needed.

With AdS/CFT there was a UV/IR connection, where the AdS radial coordinate scaled as the energy of the CFT.  What puzzled me and postdoc Amanda Peet was that different arguments gave different powers of the string coupling $g$ in the AdS-distance/CFT-energy relation.  What we realized was that the different relations came from using different probes, with different masses.  If the probes were labeled by their size rather than their energy, then the AdS/CFT relation became uniform.  So it was a size-inverse size relation, rather that size-energy.

The next exercise was to obtain the flat spacetime S-matrix as a limit of the AdS S-matrix.  This was a straightforward limiting process, though it required a non-'tHooftian large-$N$ limit.  Coincidentally, Susskind published the same result on the same day, though I think that his had a broader scope.  

This got us talking about AdS/CFT, and he told me about a paradox he was puzzling over.  If one has a quantum scattering at the 
center of AdS, the energy density at the boundary cannot change before a light-travel time.  But at that time it has to change instantaneously to a new distribution.  This seems acausal, but it is not, it is perfectly consistent with quantum field theory.  So we analyzed the bulk and boundary causality, introducing `precursor' for boundary operators that represent bulk states in the interior, a term that is in wide use now.  But the idea went back to the year before, to Banks, Douglas, Horowitz, and Martinec (BDHM) and to Balasubramanian, Kraus, Lawrence, and Trivedi (BKLT).\footnote{The initial paper with Susskind was withdrawn, and a longer paper that included Susskind's student Nicolaos Toumbas among the authors was submitted.  There was nothing wrong with the original draft, Susskind just wanted to submit an expanded explanation.}

\subsection{Strassler}

One of the great experiences I had at the ITP was having a visitor come into my office, explain to me the solution to an important problem, and ask me to help them work it out.

But first some backstory.  Kallosh and Linde had found new solutions to supergravity.  These had negative energy singularities and repelled massive objects, so KL named them `repulsons.'  Postdocs Peet and Johnson 
and I, with the aid of AdS/CFT, deduced that the singularity should expand into a nonsingular shell of branes.  This was a satisfying result, and allowed me to tell relativists that the reason they could not resolve the repulson problem was because they didn't have enough branes.  We named this shell an enha\'con, for its enhanced gauge symmetry.

So Matt Strassler came into my office with a singularity that he needed resolved.  He was interested in four-dimensional AdS duals to confining theories.  These could readily be obtained by giving masses to some or all of the CFT scalars, leading to ${\cal N} = 0, 1$, or 2 SUSY; he called these $0^*, 1^*$, and $2^*$ SUSY QCD.  The problem was that these seemed to lead to singular solutions, whose properties could not be calculated.  Strassler thought that the same ideas that had worked for the repulson might work here as well.

In fact, he recognized the key idea even before he spoke with me, the D3-branes blew up into D5-branes by a beautiful mechanism that had been discovered by Rob Myers.  So Strassler had a rather complete picture of both sides of the duality even when he first came into my office.  For example, he knew that there would have to be NS-5 branes, and bound states, as well.  My main contribution was to identify a small parameter, $gN/n^2$, where $n$ was the number of probe branes, that allowed calculations.  In the end it was a nice picture, with a lot of physics in it. It was also a very long paper, which has never been published.  Strassler is a perfectionist, and we got stuck on one thing, getting the $U(1)$ gauge factor straight.
Impressively, in the same year he found a completely different solution to a very similar problem.  With Klebanov, he found a solution that was purely geometric, without brane sources. 

I was pleased to have several more opportunities to work with Strassler: the combination of our different points of view was productive.  One project began from my memory from SLAC, of Stan Brodsky's work on hard scattering, where all the constituents of a hadron scatter together: it is suppressed, but still power law in field theory.  Could AdS/CFT reproduce this?  
Normally one would expect soft scattering in the string description, but the warping of space converted this to the power law of the field theory.  Because our first paper had run so long, I insisted that we publish in Phys.\ Lett., with its 4-page limit, but it still took a while.

We then extended this to deep inelastic scattering, scattering a hard probe against a hadron.  This was the basic process by which the internal properties of hadrons were seen.  Of course, strings had a very different internal structure, and correspondingly the scattering was very different.  Between weak and strong gauge coupling there was a transition from the operator product dominated by parton operators to one dominated by hadrons.   I had heard about these things in the early days of QCD.  Now we could have our own toy with AdS/CFT, and understand what was the same and what was different.

Our last project, a few years later, was understanding the Regge behavior,  $s^{\alpha(t)}$ at large $s$ and fixed $t$.  In flat-space string scattering, the Regge trajectory $\alpha(t)$ is linear in $t$.  Many years earlier, Charles Thorn had told me that in QCD, the trajectory is linear in the timelike region of $t$ negative, but then bends over toward a constant at spacelike $t$.  These two regions were referred to as the soft and hard (or BFKL) pomerons.  So my motivation was to understand this in AdS/CFT.  The other collaborators, now including Rich Brower and Chung-I Tan, may have had other motivations.  And it worked nicely, thanks to the warping.  The soft pomeron came from the IR region of AdS, and the hard pomeron came from the UV region.  So once again AdS/CFT gave a nice way to think about QCD physics.

I also had a nice but little-known follow-up with Susskind.  He wanted to understand how the string dual of a gauge theory could have local currents, which are impossible in normal string theories.  This discussion came up when we were both at the 60th birthday celebration for John Schwarz, and we solved it there.  As with many aspects of nonconformal field theory duals, it was the warping of the bulk that was responsible.  But I most remember a bit of grandstanding by Susskind, who asked Mimi Schwarz (John's mother, and not a scientist), to adjudicate an issue in the discussion.  
He was making the point (convincingly) that the issue was so clear that he could explain it to Mimi.  So we added her name to the acknowledgements.

\subsection{Bousso}

In 1998, strong evidence was found for a cosmological constant, surprising almost every theorist.  One might have expected string theorists to drop everything and think about this, but there was little reaction.  Certainly a large part of this was that AdS/CFT had just been found, transforming fundamental theory.  We needed to understand the theory better before applying it.

My own reaction was different, from my interactions with Weinberg.  I had half-expected the CC, and had feared it.  Indeed, when the evidence started to come in, I told our postdoc, Sean Carroll, that if the CC turned out to be there, he could have my office.  It would mean that the anthropic principle was here, and I would have to give up physics.  I make a lot of comments like this that I do not remember --- unfortunate, otherwise this memoir would be funnier.  But Sean remembered, and as he introduced me at a meeting two years later, he asked when he was going to get the office.  

Others were also unsurprised, including Linde, Kallosh, Susskind, Banks, Bousso, Silverstein, and Kachru.  Notably, these were all on the West Coast.  It was a new version of the East Coast/West Coast divide.  Those on the East expected an elegant theory, with vanishing cosmological constant and perhaps even a unique ground state; perhaps some small effect would explain away the CC. Those on the West Coast were not so caught up in these myths, though I would prefer if they were true.\footnote{David Gross, having moved from East to West just at this time, was in an odd state: he knew the truth, but could only speak it when pretending to be someone else.}

I thought it might be a good idea to see whether string theory had the right microphysics to allow Weinberg's solution, but I put it off.
Then Raphael Bousso, a former student of Hawking, now a postdoc at Stanford, came to town.  He was interested in the same question, and goaded me to think about it with him.  First, it was clear that the old idea of Hawking, Duff-van Nieuwenhuizen, and Aurilia-Nicolai-Townsend of a continuously variable four-form potential could not work in string theory.  In string theory, the forms are the charges of space-filling D-branes and so are quantized.  The old ideas of Abbott and Brown-Teitelboim used discrete charges, but they needed implausibly small quanta and large charges, $10^{60}$ with SUSY and $10^{120}$ without, to get a small enough CC.  And they did not have a mechanism to get matter.

Bousso and I realized that in string theory there were typically multiple fluxes, which could be incommensurate depending on the topologies.  In this way, much smaller individual charges could combine in many ways.  With 100 fluxes, a typical number for a Calabi-Yau compactification,
charges of order 10 would give us $10^{100}$ states.  This produced a spectrum much more disordered than the single-flux case, which we named a `discretuum' to contrast with `continuum.'  With large compact dimensions, as few as four fluxes might work in large-dimension models.    

Most of this came from one or two conversations.  But when Bousso came back a few months later, he had a complete draft.  He had added an important part of the story, the cosmology that allowed the theory to explore all these states.  It was just Linde's eternal chaotic inflation: given any de Sitter state, all the rest would eventually be produced by expansion and tunneling.  I had always assumed that such a thing would not be part of string theory, but in fact it arose quite naturally.

Of course, at the time we wrote our paper, no de Sitter solutions were known, we were just working with a simplified model.  But this was a natural consequence of string theorists starting with the simplest SUSY solutions, which have negative CC, and working toward the more generic ones.  Now that the second superstring revolution had given us a more complete picture of the theory, young West Coast theorists would soon fill in this gap.  For a while, there was lore that string theory only allowed negative CC, but not on the West Coast.

Bousso's draft had one more important point.  As discussed earlier, many ideas for a vanishing CC led to a spacetime without matter.  But for his (and Linde's) picture this was no problem.  Tunneling could readily produce excited states of the inflationary potential, which would then decay to ordinary matter in the usual way.  So, with a few details soon to be filled in, string theory produced the small nonzero cosmological constant seen in nature.

It was great being at the ITP.  In quick succession, two outstanding young people brought me important ideas and asked me to work with them.  And each was perfectly complementary: Strassler's particle physics and field theory, Bousso's relativity and cosmology, and my string theory.
Indeed, it was an embarrassment of riches.  Bousso came in with his draft a few weeks after Strassler came in with his idea.  I knew that I could not work on two such intense projects at the same time, so Bousso had to wait for what turned out to be a couple of months.

Even worse for Bousso was my aversion to any mention of the anthropic principle.  By the end it was down to one mention in the introduction, and one in the final paragraph.  Even to get me to sign for this much was difficult, but he had a trump card.  We had just offered him a senior postdoctoral position at ITP, and he said that he would accept only if I agreed to be on the paper.
If not for my obstruction, the paper would have looked much more like the later and more open treatment by Susskind.  

It was not that Bousso and I disagreed in any way about the physics.  Just the opposite: I thought it was so compelling that even experimentalists would realize that they were measuring random numbers, and be discouraged.  I did not want to be the cause of that.  But of course I overestimated both the credence that experimentalists gave to theorists, and the ability to make progress even with such obstacles.  As Bousso and Susskind both knew, it is wrong to suppress what you know.  Georgi again:
``Do not hide your light under a bushel basket."  As far as I know, this is the first paper written about string theory and the anthropic principle, a real illustration of the power of anthropic denial.

\section{After the end of physics, 2001-2007}

\subsection{Bena, Grana, Frey}

Having told Carroll that I would give up physics if a cosmological constant were found, how could I go on?  Well, I had just taken on three new grad students after finishing my book, and I had to take care of them.  And, we still needed to see if all those de Sitter vacua were there.  And there were all these cool things about AdS/CFT to look at.  So life went on, and Carroll did not get the office.

Iosif Bena, originally from Romania, was perhaps my most independent student.  I think he is the only one I did not write a paper with while they were a student, although we did work together later.  I gave him only one project, to work out the precursor (\S 9.4) for spaces that were less symmetric than AdS.  We might call this ``non-AdS/non-CFT dualities,'' or today simply ``gauge-gravity duals.''  I wanted to verify that conformal symmetry was not essential, so he did the general D$p$ case, without benefit of conformal symmetry.  I think that my only advice, beyond the idea, was that he speak more slowly and check his work.  He then took an interest in my model with Strassler, extending our D3 $\to$ D5 polarization to many other brane systems, each with their own peculiarities.  He went on to postdocs at USC and Princeton, and then a faculty position at Saclay, where he is an expert in supergravity solutions and their many applications.

When Mariana Grana, from Argentina, was first reading string theory with me, she seemed to have a particular interest in how the Calabi-Yau solutions of Candelas, Horowitz, Strominger, and Witten were fixed by the ${\cal N} = 1$ SUSY conditions.  I had always wanted to do such a calculation, ever since Strominger had told me how much money he had made from it.  So rederiving my solution with Strassler in this way was a great exercise, which we then extended to many other examples.\footnote{Just a few years ago, at my 60th birthday, Grana told me that when I first suggested this, she thought it sounded boring, but it developed into her life's work.}  She then, entirely on her own, went on to find the effective low energy supersymmetry breaking for branes in fluxes, an important and technically difficult exercise.  So she became an expert in flux compactifications in string theory, writing one of the classic reviews.\footnote{When I was writing a report for the Ecole Polytechnic, I noted that her review was the most highly cited paper written there, over a period of years.}  After Santa Barbara,  Ecole Polytechnic, and Ecole Normale Superieure, she ended up at Saclay, just as Bena had, where their strengths are nicely complementary.  

Andrew Frey, from Wake Forest, also started out with a project from the work with Strassler.  The $N=1^*$ theory had an infinite number of supersymmetric vacua labeled by D5 and NS5 quantum numbers.  Different vacua could be connected by domain walls.  Strassler and I had looked at some examples, but Frey found the general case.  He got the surprising result that not every pair of solutions was connected directly; in some cases, one had to go through multiple steps.  Another, more senior, group at the same time missed this. Beyond this, he was active and broad at UCSB.  He and I wrote a paper on ${\cal N}=3$ warped compactifications,\footnote{Warped compactifications, string realizations of the Randall-Sundrum idea, were developed by Strominger and by K. Becker and M. Becker.} which I was interested in only as an oddity, SUSY's usually coming in powers of 2.  He also had work on ${\cal N} = 1$ SUSY, on BPS states, dilaton stabilization,  a careful study of Lorentz breaking in warped space (my suggestion), instabilities of the KKLT model, and new warped solutions.  Several of these involved other students, including Grana,  Matthew Lippert, Brook Williams,  Anupam Mazumdar, and a postdoc Alex Buchel.  After postdocs at Caltech and McGill he is now on the faculty at Winnipeg, and  has moved more toward particle astrophysics.

\subsection{Silverstein and Kachru}

One of the great things about the ITP is getting to work with remarkable young people at key stages in their careers.  The period after the second superstring revolution was particularly fruitful.  I have already written about Strassler and Bousso, and now Eva Silverstein and Shamit Kachru arrived in early 2001 to run an ITP program, along with local Steve Giddings.  The program was nominally about M-theory.  For many of us, the focus was to move on from the highly supersymmetric situations used to understand the theory to less symmetric and more physical ones.

In particle physics, Randall and Sundrum had shown that in five-dimensional theories, warped compactifications gave a new mechanism to produce a large gauge hierarchy in four dimensions.  This had led to widespread interest in the phenomenology of such higher dimensional theories.  Thus, it was very natural to ask whether these models might be more closely connected to string theory.  Herman Verlinde had already pointed out that T-duals of ${\cal N} = 4$ string theories naturally led to warped compactifications.  Giddings, Kachru and I showed that this could be extended to ${\cal N} =1$.  In particular, it automatically accounted for the stabilization of the hierarchy: it arose from the quantization of the fluxes.  Many of the fields could be stabilized, but not all.  To stabilize the K\"ahler moduli would require nonperturbative effects.

At this point, there was still no example of a de Sitter vacuum of string theory, much less the enormous number that would be needed to explain the cosmological constant via the discretuum.  There was even a widely quoted no-go theorem by Maldacena and Nunez, and earlier by de Wit, Smit, and Hari Dass.  If this were true generally, it would mean that either the cosmological constant would have to be wrong, or string theory would.  This result was widely quoted.  But the West Coast group working on the problem knew that it was nonsense.  The no-go theorem held only for classical backgrounds.  One might as well claim that atoms don't exist, because they are classically unstable.  There were other exceptions as well, such as noncritical dimensions as in work of Myers.$^{\ref{Myers}}$

So the group of us discussed this problem intensely.  Already at Strings 2001 in Mumbai, which had been held in January, Silverstein had delineated some of the key features needed.  In particular, there was a universal instability to be dealt with, from the radius of the compact space.  Taking account of scaling from ten dimensions to four, all interactions went to zero as the radius went to infinity.  In order to obtain a de Sitter solution, one needed at least three terms to give a potential with a stable positive minimum.  At the ITP Silverstein completed a model with a stablized radius, taking as her three ingredients orientifolds, fluxes, and a noncritical dimensionality.   

It took a couple of years more to stabilize all the moduli, but Kachru, with a series of collaborators, found completely stabilized de Sitter solutions, using branes, fluxes, and D-brane instantons.
This started with his work with Giddings and me, then with Pearson and H. Verlinde, and ended up with Kallosh, Linde, and Trivedi (KKLT).  And so with this, people began to take seriously the possibility of a large set of de Sitter solutions, aptly named the string landscape by Susskind.\footnote{For many years, the size of the landscape has been crudely estimated as $10^{500}$, based on large Calabi-Yau's.  But Taylor and Wang have recently shown that there is an F-theory geometry of dimension $10^{272,000}$.}

At the ITP program, I also worked with Silverstein and her student Allan Adams on a different project.  Ashoke Sen had shown that open string tachyons represented the decays of unstable D-branes, and he was able to describe the decay using open string field theory.  Closed string tachyons would be much more complicated: rather than decays of branes in a fixed spacetime, they would represent decays of spacetime itself.  Also, the closed string field theory was much less tractable.  But we were able to sort many things out.  

Tachyons like that of the bosonic string, which fill space, presumably have no stable final state.  But we realized that there were closed string tachyons that were localized just like the open string ones, in spacetimes that are flat except for conical singularities.  And so for these we could make sense through a combination of linear sigma models at short distance and the spacetime field equations at large distance.  The result was rather simple: the singularities just spread out in an expanding shell.  

\subsection{Pauli, Heisenberg, Dirac}

Around this time, the centennials for four of the founders of quantum mechanics were celebrated --- Fermi, Pauli, Heisenberg, and Dirac.  I missed Fermi's, but spoke at the other three.  It was fun to review some of the history of each of them, and to notice how modern some of their ideas were.
I think my talk for Pauli was not so interesting.  My theme in Zurich was just to look at all the parts of the standard model that Fermi was responsible for, with a rather silly graphic of miniature Pauli's labeling each one.  But it was interesting hearing some of the other speakers, notably the historian Norbert Straumann.  It was from him that I learned of Pauli's unpublished interest in the cosmological constant --- had I known earlier, the story leading from Pauli to the multiverse would have made a more interesting talk.

The other two talks had a bit more heft. With Heisenberg, the theme was unification, from his Worldformula to our M-theory.  Heisenberg actually had presented his ideas at Caltech when I was a student there.  It seemed very crude at the time, just fermions with a nonlinear interaction.  He also wanted some modification of the uncertainty principle so as to produce a minimum length.  But in retrospect, he was only a few steps away from Matrix theory: just introduce a matrix structure to get nonlinear commutators, and supersymmetrize.  Another foresighted idea was the S-matrix.  This leads to another connection, Heisenberg $\to$ Chew $\to$ Veneziano $\to$ strings.\footnote{At my talk in Munich, Helmut Rechenberg, curator of the Werner Heisenberg archive, informed me that this chain was even more direct than I had guessed.  As early as 1954, Heisenberg wrote in a letter that in Urbana he had met Ôa particularly nice younger physicist with the name Chew,Õ and they continued to correspond.}

The lecture for Dirac was in Tallahassee, Florida, where Dirac retired, and where I got to meet his daughter, Monica.  I started with a review of Dirac's remarkable career even after quantum mechanics, and quoted his very modern point of view,
\begin{quote}
One must be prepared to follow up the consequences of theory, and feel that one just has to accept the consequences no matter where they lead.
\end{quote}
For him it led to antimatter, and for me it led to the string landscape and the multiverse.  It also led Dirac to magnetic monopoles.  The main part of my talk was about this, proposing two principles: (1) In any theoretical framework that requires charge to be quantized, there will exist magnetic monopoles, and (2) In any fully unified theory, for every gauge field there will exist electric and magnetic sources with the minimum relative Dirac quantum.  I illustrated this with five examples: grand unification, Kaluza-Klein theory, lattice gauge theory, the Kalb-Ramond theory, and D-branes.
So I argued that magnetic monopoles are our most certain prediction about physics beyond the standard model, though unfortunately the scale is not predicted.  Notably, the nonobservation of magnetic monopoles led Dirac to recant on his quote, but he should have been more patient: we have only explored a tiny range of scales.\footnote{One of the smaller detectors at the LHC is MoEDAL, searching for magnetic monopoles.  When it was proposed I was asked to write in support, based on this lecture, and I was happy to oblige.}

\subsection{More odds and ends}

Here are a few papers from this period that I do not want to forget, but are not worth a full section.

\subsubsection{Crunching with Horowitz}

The ekpyrotic idea, a universe bouncing at a brane, arose at around this time.  One of the arguments for a consistent bounce was the example of  the resolution of singularities in string theory.  Liu, Moore, and Seiberg had studied the toy model of a null orbifold, and found that the singularity in amplitudes was resolved. But this ignored the back-reaction.  Of course, the well-understood string singularities were timelike, and the ekpyrotic singularity was spacelike, so the issues with backreaction would be more severe.  Horowitz and I studied the backreaction both using general relativity and string theory, and concluded that  a single particle caused the spacetime to collapse to a strong curvature singularity, even in regions arbitrarily far from the particle.  This was not surprising: intuitively, a bounce would require infinite fine-tuning.  This did not rule out all possibilities for a bounce, but emphasized the contrived nature of the idea.

\subsubsection{Emergent gravity?}

I tend to be skeptical, both with my own work and that of others.  I guess, having thought hard about various problems, I am always  wary about new claims about them.  I am not always right --- I have mentioned Rubakov as one who surprised me twice --- but I should have an item on my CV for `Papers prevented.'  Certainly the Weinberg-Witten theorem, restricting emergent gravity, provided one of those potential alarm bells.  Of course, AdS/CFT had shown that the theorem had an exception, but that required a whole extra dimension.

A model by Zhang and Hu, which appeared to get a 3+1 dimensional graviton from the boundary of a 4+1 dimensional quantum Hall system, had to be understood.  So with a grad student, Henriette Elvang, we studied this model to see how it might have evaded the theorem.  Their idea was to take parallel copies of the quantum hall system, e.g. in the 12 and 34 directions, and sum to give an $SU(2)$ symmetry.  Then turning on a magnetic field, there would be a potential that would lead to massless degrees of freedom on the 3+1 dimensional boundary.  Moreover, one could get any massless spin, in particular two.  Note that this is not holographic but anti-holographic: the field theory lived on the higher dimensional space, and the would-be graviton would live on the boundary.

To analyze this, we first took a large-$N$ limit so that the massless degrees of freedom lived in flat space.  We then found that the spectrum was not that of a 3+1 dimensional space, but a cone of 1+1 dimensional theories.  For example, the low energy density of states was larger.  So the system might be interesting, but it was not Lorentz invariant, and not gravity.

Henriette went on to work with Gary Horowitz and has been very successful in gravity and field theory.  She is now a professor at Michigan.  At the time I had too many students to take on, and I felt that Gary was a better fit for Henriette.  I tend to give my students rather ill-defined problems, trusting that there is some gem underneath; this often works out, but a student who likes to calculate would be better off with Gary.\footnote{Two other students whom I had to pass up earlier with Don Marolf, who worked with Bryce Dewitt and is now my own colleague and collaborator, and Scott Thomas, who worked with Willy Fischler and is now a leader in particle phenomenology.  But I think each of these did better than they would have with me as their advisor.}
  
\subsubsection{Integrability}

Our postdoc Radu Roiban, my former student Bena, and I were discussing whether we might be able to get an analytic understanding of confinement, at least at large $N$ and perhaps with SUSY.  Our idea was to use AdS/CFT to rewrite the boundary 4-d field theory in terms of the 2-d string world-sheet (the string action would be complicated, because the conformal invariance was broken).  We would then hope to use two-dimensional methods to solve the theory.  

As a warmup, we studied the conformal $AdS_5 \times S^5$ theory.  We knew that there was a method due to Luscher and Pohlmeyer to find infinite symmetry algebras for a wide class of nonlinear sigma models.  Indeed, as we were working on this we learned that Mandal, Suryanarayana and Wadia had recently found this for the bosonic $AdS_5 \times S^5$ string.  So we generalized this to the superstring, and this also worked.

Having an infinite dimensional symmetry seemed like it would enable us to do many new calculations.  But we quickly learned that this kind of algebra, the Yangian, was different from the usual algebras of physics, and we still had a lot of work ahead.  We also learned that this same method, integrability, had been applied a few months earlier by Minahan and Zarembo on the CFT side of the theory, where we were on the AdS side.  So integrability became an active area, and Roiban ended up as one of the 26 joint authors of a massive review.

My own further attempts were not so successful.  It was not my way to learn the necessary technical machinery for this subject.  I figured, `this is a symmetry, I know all about symmetries,' and set out to figure this out, with the help of my latest student Nelia Mann.  We both worked quite hard, and were able to do a few interesting calculations, but in the end it was not a good approach.  One of my limitations is that I am best working on problems where the physics and the math are close together.  When one has to start dealing with objects whose physical content is not evident, I lose my way.
But Nelia went on to write several nice papers with other students and postdocs, worked with Jeff Harvey on pomeron phenomenology at Chicago, and is now a junior faculty member at Union College.

\subsection{Gross, Kavli, Nobel, and the Future}

This section is a bit outside the main flow here.  But I had to explain how ITP programs suddenly became KITP programs, and one thing led to another.  The original ITP had grown more than 50\% under Gross, taking full advantage of the new building.  Over time he brought in three new permanent members (Lars Bildsten, replacing Doug Eardley,  Leon Balents replacing Matthew Fisher, and Boris Shraiman representing a new effort in Biophysics.  

This led to an increased presence in astrophysics and biophysics, and before long even the new building was too small.  And so Gross went looking for a donor, and came back with Fred Kavli.  I do not know how they connected, or the details of their negotiations: Gross keeps his cards close to his chest.  But in the end, we had the funds to build a nice extension of the building, ending up with perhaps 2.5 times the size of the original ITP before 1993.  Moreover, it made the building more connected and added many new public and working areas.  So in 2002, the ITP became the KITP.  At first it seemed strange, but after a few years ITP seemed naked without the K.

We had hoped to raise enough to enlarge the building and begin an endowment, but construction prices were rising fast.  We did expect though that this was a beginning of a relationship with Kavli, that would lead to an endowment in the future.  So it was a bit of a shock when, a year later, Kavli gave an equal donation to establish a Kavli institute in astrophysics at Stanford: we were being franchised!  Over time there were 20 Kavli institutes.  We did get some additional Kavli support, but still need an endowment to protect our activities from the indefiniteness of NSF support. 

But we did have the new building, which was a beautiful and stimulating place to work.  To celebrate, in October 2004 we had a three day conference in Gross's grand style, bringing together the leaders of all areas of physics to discuss the future.  And delightfully, two days before the conference Gross, Wilczek, and Politzer received the Nobel Prize.  So the meeting became a celebration, for 25 years of the (K)ITP, for the new building, and for David's long-awaited Prize.  Wilczek also attended, expressing his satisfaction at winning his first Nobel Prize.

\subsection{Cosmic strings}

One of the discoverers of the Higgs mechanism, Tom Kibble also pioneered the idea of topological defects in cosmology.  For example, solitonic strings could form at a phase transition and then expand with the expansion of the universe, growing to a cosmic length.  Early in the first superstring revolution, Witten noted that fundamental strings might do the same.  If so, they would be a spectacular observational signature.

But Witten noted that there were several obstacles to this.  Fundamental strings generally had tensions close to the Planck scale, which would lead to excessive fluctuations in the CMB.\footnote{For heterotic strings there was a simple relation between the string tension, the gauge coupling, and the Planck scale, $\mu = g^2 / 16 \pi^2 G$.  This was too large by several orders of magnitude.}  Also, there were potential instabilities.  Heterotic strings carry axion charge, so the strings will actually bound axion domain walls, producing a confining potential that prevents their growth.  Type I strings were unstable against rapidly breaking up into small open strings.  And type II strings were confined by NS5-brane instantons.

But at a KITP program on string cosmology in 2003, Silverstein was giving a talk about superstring vacua, and made reference to F strings and D strings (fundamental and Dirichlet).  For cosmologist Ed Copeland, the idea of having two kinds of cosmic string, and their bound states, was a novel one, and he became excited about the possibilities.  I was aware of the problems with this idea, but it was a good time to revisit it, with the first fully compactified theory in KKLT, and its cosmology in KKLMMT.  And indeed, the problems potentially went away.

First, KKLT was a warped compactification, reducing the scale of the string tension: it could easily fit with the constraints.  Further, any of the instabilities might be suppressed due to separation in the compact dimensions, or warping.  And Tye and collaborators had shown that cosmic strings arose naturally in KKLMMT-like brane inflation models.   So Copeland and I, together with brane maven Rob Myers, worked out the phenomenology of these potential strings.

Like Copeland, I became quite excited by this: a potential direct signature of physics near the Planck scale.  And so this became a large part of my research for the next few years.  The first major question was, if cosmic strings were found, could we distinguish fundamental strings from field theory solitons?  Indeed we could.  As we noted in our discussion of Matzner in $\S$6.5, when two strings cross each other, if they are solitons then their ends always reconnect, a classical process.  But for two fundamental strings, the crossing is a quantum process: the strings may reconnect, or pass through.  I had worked this out as an exercise many years before for the bosonic string, in part with Jin Dai.  Now, with visiting KITP grad fellows Nick Jones and Mark Jackson, I extended this to supersymmetric F- and D-strings, and their bound states.  So if cosmic strings were ever found, we could indeed make direct measurements of properties at the string scale.

But, following this up, the job would not be quite so simple.  What we would measure would likely be some correlators on string networks.  These had been studied numerically, and different groups had gotten radically different results.  One key quantity was the typical radius of the string loops that broke off from the main network; these were the main source of the gravitational waves emitted from the network.  
Remarkably, estimates ranged from the horizon scale (the only obvious scale in the problem) down to the Planck scale, for what was a well-posed classical problem.   So with my latest student Jorge Rocha, and in part with a visiting postdoc Florian Dubath, we made a scaling model which explained why there were actually two scales: there was an infrared divergence.  We tried to get more detail out of this, but improved numerical methods from the Tufts group eventually gave the sharpest picture, and if cosmic strings are found these simulations  will be essential.

I think that Jorge was more interested in working on black holes than on cosmic strings.  Fortunately I found him a good black hole problem before he left.  This was to study the dual of an ${\cal N}=4$ theory coupled to another field that could carry away energy.  In this way, one could study decaying black holes even in AdS.  I think this was the first study of such a hybrid system.
(Jorge is back in Spain, and still happily working on black holes).

At this point, it seemed that we had to wait for the experimentalists.  Unfortunately, so far improved measurements from WMAP and then from Planck have only lowered the upper bound on cosmic strings.  But it was good to have an opportunity to think about observation.\footnote{A couple of times, the observations of apparently paired galaxies, produced by the gravitational field of a cosmic string, have led to some excitement.  But they have so far always turned out to be coincidences.}

Thinking about cosmic strings led to a surprising observation: open heterotic strings could actually exist, in the $SO(32)$ theory but not in the $E_8 \times E_8$.  This came from thinking about a general classification of strings.  Originally, cosmic strings were referred to as `global' or `local.'  Global strings had fluxes running in their cores, and local strings did not.  But from thinking about all the examples that arose in string theory, I realized that there were two more, `Aharonov-Bohm' (AB) and quasi-AB.  Further, the right way to distinguish these was not by what was in their cores (which could, after all, depend on duality), but by their properties at long distance, which controlled their stability.  Thus, global strings had a long-range axion field, and so were confined.  Local strings could break into short open strings.  AB strings were stable due to discrete symmetries, unless the same charges were carried by massless fields (quasi-AB), in which case they could tunnel and then expand in the perpendicular direction.

Applying this classification to the heterotic string, one finds that for the $SO(32)$ string, it can be any of local, AB, or quasi-AB, depending on the compactification.  The $E_8 \times E_8$ string had to be global.  The puzzle of an open heterotic string is that the degrees of freedom moving on the two sides are very different, and had no consistent boundary condition.  Indeed, what happens is that when a worldsheet field on the $SO(32)$ string reaches the boundary, it does not reflect but rather leaves the string and becomes a spacetime degree of freedom.  This odd picture could be described explicitly in open string field theory.  Unfortunately, I have never found anything useful to do with it.  David Morrison did notice that I had posted it, coincidentally, on the tenth anniversary of the D-brane paper.\footnote{Morrison came to UCSB from Duke about ten years ago, with a joint position in math and physics.  He plays a unique role in tying these subjects together.  He and I have an ongoing friendly dispute about whether I know much math (I claim not).  I think that the difference goes back to Susskind's distinction between the mathematics of the equations and the mathematics of the solutions, where I care only about the former.}

\subsection{Down time}

The last few years in this period were a bit of a down time for me.  Certainly I had plenty of fun and joy, and some good physics.  But also I had extended periods of anxiety.  Sometimes these just plagued the early hours of sleep, but other times they took over the day, and my work.  One that I remember clearly was my induction into the National Academy of Sciences in 2005. This should have been a time of great celebration, and there was some, but throughout I was filled with an ill-defined anxiety.

Part of this was still a hangover from finding the anthropic principle in string theory.  I feared that most of the routes to the discovery of the fundamental theory were blocked by it.  Also, this was the time of the well-publicized anti-string books.  I got caught up in this because Rosalind Reid, a KITP visiting journalist who was editor at American Scientist, asked me to review the book by Smolin.  I felt that it was a good thing to do: many people, including some of our colleagues from other fields, were taking it uncritically.  I tried to use it as an opportunity to present the positive argument for string theory, but I do not think that I was very effective.  As Mark Twain said, ``A lie can travel halfway around the world while the truth is putting on its shoes.''  Smolin tried to avoid outright lies, but my reaction to reading the book, which I wish I had stated more directly, was ``This is not the way a scientist writes."  Facts were twisted to create the impression desired, rather than being a way to reach truth.

Beyond this, I think that my anxiety was a long-standing aspect of my biology, going back even to my childhood shyness.  I think that many of the decisions I have made over time have been driven more by anxiety than by positive emotions.  Matthew Fisher, who had some experience here, was a strong believer in the value and effectiveness of psychiatric drugs.  And indeed, after some experimentation, a small regular dose of Lexapro has kept me balanced.  So one can say that the anthropic principle drove me to drugs.  But one must follow science where it leads. 

Over the years, I have noted one visiting speaker prepare with a dose of Adderall, and one with a dose of Valium (and offer one to me also).  A doctor has told me that half her colleagues are on Lexapro, it is an occupational hazard for high-functioning people.  And, following the lead of Paul Erdos, who is said to have written all of his 1500(!) papers under the influence of stimulants, I tried working a few times with Adderall.  I had a couple of very effective days, but overall did not like its effect.

\section{Before the firewall, 2007-2011}

\subsection{Quantum gravity: wormholes, black hole models, bubbles of nothing, loops}

Having spent most of the last few years on cosmic strings, AdS/QCD, integrability, and other odds and ends, I wanted to focus more on the fundamental question, `what is quantum gravity.'  Even with the anthropic principle looming, the problem of finding the theory of quantum gravity remained one that needed to be solved.  Solving this might lead to any number of wonders. Moreover, it was the kind of problem that might be solved by theoretical reasoning alone.  And, we had this remarkable tool, AdS/CFT, or more generally gauge/gravity duality, which we had certainly not applied to its fullest.

The first order of business was sorting out the old confusion about the Euclidean wormholes of Coleman: do they appear in the path integral for quantum gravity?  This came up when Nima Arkani-Hamed was visiting, and he was interested in the possibility that such wormholes might allow us to study other parts of the string landscape.  With grad student Jacopo Orgera, we first found Euclidean wormholes in spacetimes with known field theory duals; this was really the hard part.  Then, extending an argument by Rey, wormholes would violate cluster decomposition in a way that was inconsistent with the dual field theory.  (Several people suggested that this might be corrected by adding nonlocal operators to the dual field theory, but that would only produce nonlocality at the boundary).  So our conclusion was that these solutions did not appear in the path integral for the evolution in quantum gravity.  This was supported by the observation that all the solutions we could find had actions that were {\it less} than the BPS action.\footnote{At an event where we were both a bit inebriated, Andy Strominger told me that he thought this paper was a negative contribution to physics.  I found this hilarious, and remind him of it at every opportunity, but it reflected Andy's very positive point of view, that every wormhole must be good for something.}

According to INSPIRE, it was at this point more than ten years since I had written a paper on black holes and the information problem.  Like many of those who had worked on this, I regarded it as essentially solved by gauge/gravity duality, in the BFSS matrix form and in the AdS/CFT form.  Of the three options --- information loss, information emission, and remnants --- only emission was consistent with duality to gauge theory.  There remained the question, how does the information escape?  But this seemed to fit nicely with the principle of black hole complementarity, enunciated by Susskind, Preskill, and 't Hooft: the information could be both inside the black hole and outside, as long as no single observer could see both copies.  And various thought experiments supported this.

Still, our understanding seemed to be incomplete.  For example, we only had a nonperturbative construction of the CFT side of the duality.  We could in principle calculate the black hole S-matrix by making a duality to the CFT and solving numerically.  But in the bulk regime, where the black hole radius was large compared to the Planck length, it seemed that there should be a nonperturbative construction in the bulk.  So even though I was not actively working on the problem, I was often thinking about it.  The 2001 paper by Maldacena, recasting the problem in terms of the long-range two point function, struck me as a particularly clear way to formulate it.

Thus, a nice paper by Festuccia and Liu (FL) caught my eye.  They wanted to make a toy model of the CFT which captured what seemed to be the main feature of the black hole as discussed by Maldacena.  That is, at infinite $N$ the long-term behavior of the black hole two-point function falls off exponentially forever, but at finite $N$ there is a minimum, past which the two-point function is disordered.  FL argued that this behavior could be seen even in the weakly coupled limit on the CFT side.

FL's argument was based on truncation to a simple subset of graphs.  With postdoc Norihiro Iizuka, we wanted to find a solvable model that exhibited this behavior.  After some experimentation, we found a matrix model that worked. In particular, we showed that at $N \to \infty$, there was a range of parameters for which the asymptotic two-point function fell exponentially, as with a black hole, while the finite-$N$ correlators had to exhibit exponential decay and then disorder.   

Perhaps the most notable thing about this paper is that it is the only time that I have used Mathematica, in this case to solve a nonlinear recursion equation.\footnote{Prior to that, I used Fortran to get a mass spectrum on one paper with Wise.  He was impressed.}
 This is aside from a few simple integrals, and even there I preferred Gradshteyn and Ryzhik.  I guess I'm a Luddite (Dorothy, who is my IT manager as well as my wife, would agree), but until recently I have always been more accurate than the students and postdocs with which I was working.  Up until now, I was able to find problems that did not need more.
 
In a follow-up, with postdoc Takuya Okuda, we found a larger set of models, including a simple one that could be solved analytically at large $N$.  We were also able to obtain the $1/N^2$ correction, each of us doing it a different way: Nori by direct Feynman sum, Takuya by sum over Young tableaux, and me using loop equations.  Unfortunately, this was complicated enough that we could not see getting the general term, or summing for an exact expression.

One other paper from this period dealt with the stability of nonsupersymmetric orbifolds in AdS.  With Horowitz and Orgera, we addressed the question of whether any nonsupersymmetric vacua could be stable.\footnote{Orgera had started out as Gross's student, but I took over when Gross became busy with the Nobel, and we wrote two nice papers together.  After his Ph.D. he returned to the private sector in Italy.}  We focused on nonsupersymmetric orbifolds (the same as Adams, Silverstein, and I had looked at in unwarped spaces).  These had tachyons at weak coupling but not at strong, but we suspected that there would still be some instability.  Indeed, it was the Witten bubble of nothing, now wrapped around the twisted direction of the orbifold.

One final note in this section is a comment on loop quantum gravity.  It was widely believed by those working in that field that it predicted violations of Lorentz invariance at high energy. 
But for those familiar with quantum field theory, this did not make sense: renormalization would spread the symmetry breaking to all operators allowed by dimensional analysis, and these included relevant operators visible at low energy.  I had thought about writing this argument out, as had several others, but did not care to get caught up in it.  Fortunately Collins, Perez, Sudarsky, Urrutia and Vucetich (CPSUV) did write out the argument, which I think had a large impact.  So I was happy.  

Gambini and Pullin, two of the original authors of the loop Lorentz-breaking idea, wrote a paper proposing two ways to evade the CPSUV argument.  So I studied the paper, and found that it failed to do what it intended.  One of the models depended on being on a Euclidean lattice, and the other depended on the Lorentz symmetry being weak at all scales.  I did realized that there was a way to make it work, though: supersymmetry!  (I then looked around and found that Nibbelink, Pospelov, Jain, and Ralston had already noted this).  So if Lorentz violation is found, we can say that SUSY is predicted, though not the reverse.

\subsection{Understanding AdS/CFT}
 
Any AdS string theory will have at least three scales: the Planck scale $l_{\rm p}$, the string scale $l_{\rm s}$, and the AdS scale $l_{\rm}$.  In order to have a spacetime that is smooth on the string and Planck scales, $l_{\rm}$ must be much larger than the others.  In terms of the dual CFT, these correspond to a large number of fields $N$ and large dimension for all nontrivial operators of spin three or more (since these spins cannot appear in the low energy field theory).  It seemed plausible that the reverse was true as well: any CFT with a large gap above two in its operator spectrum, and a large number of fields, would have a spacetime dual.

I had a chance to clarify this when Joao Penedones came to the KITP as a postdoc.  He first worked with Giddings and student Michael (Mirah) Gary to expand on my work on the flat space limit of AdS scattering.  This was very nicely done, and it seemed to me that it could be applied to a proof of the sufficiency of large $N$ and a large gap of dimensions in generating spacetime.  So, with my students Idse Heemskerk and Jamie Sully, we investigated the simplest CFT model and the simplest nontrivial observable, the four-point function.  

We solved for the most general CFT with the given spectrum of states by solving the bootstrap equation.  On the AdS side, we then found the most general Hamiltonian, with given spins and dimensions.  There was a one-to-one match between the possible bulk actions and the possible CFT's: there were no CFT's with large $N$ and a gap 
that did not have a candidate bulk dual.  Here, Penedones's calculational abilities were essential, the first (but not the last) exception to my record of beating the computers.  We would have liked to take this further, but our approach of counting was clumsy and difficult to generalize. But we could say we had proven one nontrivial aspect of AdS/CFT.

I had thought that the idea that many fields and a large gap were sufficient to give a large spacetime was general lore, and I thought that it came from Tom Banks, who generally contributes such deep insights.  But I checked with him and he denied it.  So is seems that this idea was immaculately conceived.  

With Heemskerk we looked at other ways to understand AdS/CFT.  We figured that the scale-radius relation should be interpreted in a Wilson renormalization group form, so the bulk fields should be integrated out a radius at a time.  There were other papers with an RG interpretation for the radius, but I think we were different than most (Faulkner, Liu, and Rangamani was very similar) in being more faithfully Wilsonian.  Our formalism reflected the fact that double-trace operators arise necessarily even in the planar limit - a surprise to us.\footnote{Some time later, with another student Eric Mintun, we applied this to higher spin theories.}  

In the end, our paper struck me as a new formalism, but not a new insight into the nature of AdS/CFT.  However, it turned out to be useful in AdS/Condensed Matter (AdS/CM) studies.  Indeed, almost all of my work at this time was based on AdS/CFT, and this connected all of physics, from black holes to condensed matter to conformal field theory.  So in writing this, I have to separate the subjects for clarity, but sometimes the right separation is not clear.

\subsection{AdS/CM}
 
I had some small history in condensed matter physics, with my interpretation of Fermi liquids, my dabbling in non-Fermi liquids, and my brief collaboration with Charlie Kane and Matthew Fisher.\footnote{To date I have not collaborated with any Nobel laureate, even though I have been a colleague for 28 years combined with Weinberg and Gross.  I guess our styles are a bit different, though our goals are much the same.  But this drought is likely to be broken soon, when Kane receives the Prize for topological insulators.}  With AdS/CFT, I thought about ways that it might produce a non-Fermi liquid.  Around 2003, I recall that Matthew Fisher announced that high $T_c$ was about to be solved.  He had his own new idea in mind, but I responded, ``Yes, and AdS/CFT will solve it.''  But so far neither of our approaches has succeeded.

It was Subir Sachdev, together with Christopher Herzog, Pavel Kovtun, Dam Son, Sean Hartnoll, and Markus Muller, who first found a useful role for 
AdS/CM.  Sachdev was the world expert on quantum critical phenomena, critical points that sit at zero temperature, but with an interesting approach to that zero.  AdS gave one of the few tools for studying such a strongly coupled fixed point, and moreover high-$T_c$ seemed to lie close to such a point.  But, having tried AdS/CM before, I was happy to sit back and let the large group of excited young people go with it.

But, as we have seen, being at KITP means that lively visitors were always pulling me in new directions.  In this case, it was a month-long `miniprogram' on AdS/CM in summer 2009.  This program was designed and run by Sean Hartnoll.  When he took it to the KITP advisory board in early 2008, there were only around six papers on the subject, mostly by him, but it was clearly a good thing to run, and by the time it ran a year later it was one of our most oversubscribed programs.  Sachdev and I signed on as co-organizers to give the program some heft, since Hartnoll was just a postdoc.  But because I was running a five month string theory program just before then, I got Hartnoll to agree to do all the work.  I think that he was peeved when I stuck to my word, especially as Sachdev seemed to have the same deal.

But it was an outstanding program, and it pulled me back into the subject.  First, with Hartnoll, Silverstein, and David Tong, we studied backgrounds with Lifshitz symmetry, in a probe limit for the charged fields in a thermal background.  I was skeptical that a probe approximation could capture high-$T_c$, but it was notable that for Lifshitz dimension $z=2$ one obtained the correct anomalous dimension for the conductivity.  I think my main focus in the project was the interesting RG flows.

A different approach to high-$T_c$/CFT, due to Thomas Faulkner, Hong Liu, John McGreevy, and David Vegh, was based on an $AdS_2 \times R_{d-1}$ black hole.  In trying to understand their construction, I realized that it separated into a short distance part that was universal, and a long distance part that was not.  Faulkner, who had just joined KITP as a postdoc after getting his Ph.D. from MIT, was thinking along similar lines.  We realized that there was a simple way to extract the universal behavior, which one could think of as an IR $AdS_4$ space (the dimension relevant to high $T_c$) coupled to a d=3 UV field theory that had no bulk dual.  So we called it `semiholographic,' since only part of the CFT had a bulk dual.  Working this out led to a lot of interesting issues both for renormalization and for condensed matter.\footnote{I had never thought that I would write a paper about spin-orbit couplings.  I always thought that they were an annoying breaking of symmetry, not knowing that they had become the key to topological insulators.}

Yet another approach, with Kristan Jensen, Kachru, Andreas Karch, and Silverstein, looked at models with branes moving in two directions, corresponding to a lattice of fixed charges coupled to itinerant charges.  This simulated the marginal Fermi liquid phenomenology of high-$T_c$.  As with many other AdS/CM attempts, the effect of backreaction was not fully controlled.  Much work on AdS/CM was `phenomenological,' meaning that one postulated a bulk theory without a known CFT dual.  We preferred `top-down' constructions, with a known dual theory.  But there was a downside: such theories had extra fields, in particular scalars, that were prone to instabilities.  For example, there was what Silverstein called `Fermi sea-sickness,' where scalars that were supposed to remain at the origin became tachyonic, and developed expectation values.

With my latest grad student Ahmed Almheiri,\footnote{His first two papers he signed, `Almuhairi'.} I looked at an alternative approach to stable top-down Fermi and non-Fermi liquids.   Working on AdS/CM seemed to dismay Almheiri, he wanted to work on quantum gravity, but I told him it was a good project.  After all, everything was dual.
The idea was simply to take a familiar duality like $AdS_5 \times S^5$ or $AdS_4 \times S^7$ and turn on magnetic fields carrying $S^5$ or $S^7$ charges.  Using a magnetic field for the symmetry breaking tended to be more stable than breaking by scalars or electric fields.  This idea originated from d'Hoker and Kraus, who studied one example; we looked at the general case in the search for stability.  It was a fun system to work out.  We found that in a neighborhood near the space of supersymmetric values of the charge, there were stable solutions (modulo a possible fix for the dilaton).  In our first draft we missed one instability, so this was completed by Donos, Gauntlett, and Pantelidou; this reduced but did not eliminate the region of stability.

One last condensed matter motivated idea, with Silverstein, was to reinterpret a vacuum state as a finite density theory in higher dimensions.  For example, the F1+NS5 system, is normally interpreted as a vacuum of a field theory in two dimensions.  Instead, the F1 strings could be interpreted as excitations in the NS5 vacuum, so the state would be six-dimensional.  Of course by T-dualities and compactifications one could vary the dimension.  Our main goal was to find holographic systems with the kind of `2$k_F$' singularities that arose in Fermi and non-Fermi liquids.  These had not been seen in holographic models previously, but they were here.

\subsection{More odds and ends}  

\subsubsection{AdS hierarchies}

In $AdS_5 \times S^5$, the $AdS_5$ and $S^5$ lengths are equal.  More generally, in all simple examples, the $AdS$ and compactification radii are of the same order.  But in the landscape, there should be a dense spectrum of compactifications with positive and negative cosmological constants.  The latter would include some with AdS radius much larger than their compactification radii.  Silverstein and I set out to find such solutions.  After discussing general constraints, the strategy we hit upon was to add 7-branes to the compactification, because they add a negative term to the energy density.  We found some ansatze of the desired form, but I do not know if we really succeeded: there were some singularities that went beyond my limited understanding of F-theory.  Silverstein was optimistic, but I was by nature more skeptical.  It may be that the solutions we sought were more sporadic, disordered sums of positive and negative energies.

\subsubsection{Wilson loops}

Ray, Yee, and Maldacena had shown that a Wilson loop in the CFT was dual to a string worldsheet ending on the boundary loop.  More precisely, this was true for a BPS Wilson loop, which has a scalar piece as well as the vector potential.  I had mused over the fact that the ordinary Wilson loop, with vector potential only, was a perfectly good operator, and so should be calculable on the AdS side as well.  Perhaps in connection with AdS/CM, I pursued this with my student Sully.  It was fairly easy to figure out what was going on.  The string for the BPS loop satisfied Dirichlet conditions, the position being fixed by the direction of the scalar on the loop.  It was then easy to guess that the loop without the scalar was dual to a string with Neumann boundary conditions.  Indeed, that fit with all the symmetries.  As a further check, we considered loops that interpolated between the two limits, and showed that there was a nice flow between the simple loop operator in the UV and the BPS loop in the IR.

Unfortunately, when we put our short paper on the arXiv, we learned that Alday and Maldacena had noted this some time before.  The renormalization part was new, however.  So we rewrote the paper, expanding the RG part (adding in strings that were Dirichlet in some directions and Neumann in others) and resubmitted.  But I did not have any application in mind for this, and indeed the paper has only reviewed four citations. But I liked it, as a new application of the RG, and a new corner of AdS/CFT. 

\subsubsection{Scale and conformal}

Almost 25 years earlier, I had built on Zamolodchikov's 1+1 RG irreversibility theorem to prove that under broad conditions, scale symmetry would imply conformal symmetry in 1+1 dimensions as well.  In the time since, people had occasionally tried, without success, to generalize Zamolodchikov's result to four dimensions.  But around this time, there was a renewed focus on quantum field theory, and Komargodski and Schwimmer (KS) succeeded in proving the  3+1 dimensional irreversibility theorem.  So it was natural to ask whether in 3+1 dimensions scale invariance again implied conformal invariance.  Fortunately, there were two outstanding quantum field theorists at the KITP for an LHC workshop, Markus Luty and Riccardo Rattazzi.  We first spent some time understanding the KS derivation, which was much more intricate than the 1+1 Zamolodchikov theorem.  We then examined how the KS theorem might be used to generalize my 1+1 argument. 

It was an enjoyable project, with various twists and turns and with all three of us contributing key insights.  In the end, we did obtain a theorem, but it was not quite as general as in 1+1: it held for perturbative theories, but for nonperturbative theories a technical assumption was needed, though it seemed plausible.  A bit of excitement arose when another group at the same time announced a counterexample, a perturbative theory that was scale invariant but not conformal invariant.  After some time, it was recognized that their theory was actually conformal.  It was an impressive calculation, though, and in sorting this out we much improved our own analysis.  Also, the two groups jointly managed to understand a classic paper by Osborn, which turned out to have gotten to many of the key results long before.

\section{Firewall days, 2012-2015}

\subsection{Schrodinger's cat in a black hole}

If Schrodinger's cat were behind the horizon of an AdS black hole, not yet fallen to the singularity, could we determine its state by a measurement in the dual CFT?\footnote{I am talking about a one-sided black hole that formed from collapse of ordinary matter, so we know its initial state.}  A gauge/gravity dualist would naturally answer `yes.'  The CFT is a complete description of the dual black hole, so this information should be available.  Indeed, with my students Heemskerk and Sully, and my colleague Don Marolf, we showed how to get it.\footnote{I first met Marolf when he was a 16 year old student looking for a grad school in physics.  I was happy he chose Austin, though I missed the chance to take him as a student.  But it worked out quite well, as he became my colleague at UCSB,  in time to do some great work together.  But I still think of him as 16.}  

The basic idea, integrating the field equation in the AdS bulk, was worked out by Bena and others noted earlier.   It had been refined in a series of papers by Hamilton, Kabat, Lifschytz, and Lowe (HKLL).  Our group was the first to apply it to a normal (one-sided) black hole, but it seemed to work just fine.  You take the operators inside the black hole, and integrate first backward through the horizon and then spatially to the boundary to get the CFT operators.

There were various subtleties, the most notable perhaps being that some of our construction required boundary operators out of time order, which mapped to `time folds' in the bulk.  We were not thinking about chaos at the time, but now these play a wide role.  A second version of the paper had two additions.  It was at this point that Sully joined, and added the appendix working out the Green's function in detail.  And, we added a note that the construction worked only before the Page time, because of the firewall, which had just been found.\footnote{Heemskerk had a nice followup on this, extending it from scalar fields to gauge fields.  When he first took QFT with me, he had a clear interest in the fundamentals, and he got to do some nice work in this area.  But he also wanted to make contact with experiments, and I could not promise him that.  So after his Ph.D. he moved to biophysics, studying the development of cells with Shraiman.}

\subsection{Bits, branes, black holes}

In spring 2012 the KITP ran a ten week workshop on `Bits, Branes, and Black Holes.'  This was directed at the basic questions of quantum gravity: the emergence of spacetime, the connection of area with entropy, the black hole information problem, and so on.  For me, I would have said that the central problem was to find the nonperturbative construction of the bulk theory, with the black hole information problem as a key clue.

At the beginning of the program, Ted Jacobson and I were asked to present our perspectives on the information problem.  I presented it much as I have noted in the last chapter: gauge/gravity duality showed us that information was not lost, and black hole complementarity (BHC) showed us there was no paradox.  But there was still the problem of finding Hawking's mistake: how exactly does the information get out?\footnote{Just to be clear, I like to refer to `Hawking's mistake,' but it is meant to be ironic.  He may have been wrong about the answer, but he was right about the importance, and the subtlety, of the question.  And his `mistake' has challenged countless theorists for 40 years.}

I thought that what I had said was common lore, but I was surprised.  The conservation of information was indeed nearly universally held.  But the second part, black hole complementarity, was not.  Perhaps the widest response on this was simply not knowing precisely what BHC meant.  So I set as my immediate goal, to make a simple model of BHC that would answer this.

There were some nice models of the quantum mechanics of black holes that seemed useful here.  These `bit models,' due to Samir Mathur and Giddings, were just bits in a line, with rules for what happens when a bit evaporates from the black hole.  These simple models showed the original paradox: either information could not escape from the black hole, or it had to travel much faster than light.  So I sent my now-seasoned students, Sully\footnote{Sully did not always seem enthusiastic about some of our earlier projects, but jumped into emergent spacetime and quantum information.  He has done some excellent work since going as a postdoc to Stanford and now to McGill.} and Almheiri, to find a bit model that accounted for BHC.  The idea was to break up the bit system into smaller systems, each of which was as much as a single observer could see, and with some kind of junction condition between them.  But this failed almost at once. 

The problem was that there was a single observer who could see both copies of the information, the one inside and the one outside the black hole, thus violating QM.  The original thought experiments that went into BHC had seemed convincing, but a striking paper by Hayden and Preskill, bringing in ideas from quantum information theory, led people to think more clearly about the possible measurements that can be made.  So my students and I became more puzzled each week.  I was certain that such a basic violation of black hole complementarity must be ruled out, and surely someone at the KITP program could straighten us out.  But no one could.  

In fact, our own colleague, Don Marolf, had come to the same conclusion by a somewhat different route, thinking about `mining' the black hole rather than just throwing things into it.   
So AMPS joined forces.  The fact that we had come to the same result by different arguments, and that no one could easily rebut it, increased our faith in it.  Eventually we tested it on two of the originators of BHC, Preskill at the program conference and Susskind by email.  I put off contacting Susskind because I expected the response ``yes, I thought about this ten years ago, and here is what you're missing'' --- I had gotten that from him on other points before.  But Preskill and Susskind had the same reaction that we had had: first `this can't be true,' and then realizing a week or two later that there was no easy rebuttal. 

So we wrote up our results.  Three of the principles of BHC cannot all hold: (i) Hawking radiation ends up pure, (ii) there was no drama (violation of effective field theory) outside the horizon, and (iii) there was no drama for an observer falling through the horizon.  So what gives?  Not (i): None of us thought that this gave any evidence for information loss, given the problems.  My conservative inclination was that some subtle breakdown of effective field theory at distances of order the black hole scale would fix things, violating (ii).  Marolf was sure this did not work, and there had to be drama at the horizon, which he named the `firewall,' violating (iii).  I tried to make models of how (ii) might break down, but I failed.  So I had to go along with Marolf conclusion, that perhaps the most conservative resolution was that the infalling observer `burns up' at the horizon.  Another intuition he had was that the interior stopped when the black hole's `quantum memory' became full.  So perhaps a `bit wall' would have been more accurate.

\subsection{Personal notes}

Before continuing with the physics, a few personal notes.

The three times that I shook up the field --- D-branes, the string multiverse, and firewalls --- might give you the impression that I am a radical, but it is not by design.  Rather, I think I am more like Dirac, with a knack for how theory fits together and the philosophy `one must be prepared to follow up the consequences.'  But of course we know that even Dirac did take some time to accept what he had predicted, and so did I{}.  It took me nearly ten years of playing with D-branes before recognizing their importance.  And with both the multiverse and the firewall, my inclination was to soft-peddle the results, and it took brilliant young collaborators, Bousso and Marolf, to push things forward.

The second note is  a mention of the many others who have proposed modifications of the black hole interior.  Chapline, Hohlfeld, Laughlin, \& Santiago and Mazur \& Mottola made such arguments, but I don't think their physics made sense.  Braunstein had an argument and conclusions resembling ours, but his black hole Hilbert space was not correct.  But the one who certainly did something correct, and important, was Samir Mathur.  

Mathur had devoted a large part of his career to the information problem, even after most string theorists accepted `AdS/CFT + BHC' and moved on.  He was most known for the idea of fuzzballs, modifications of the black hole horizon from higher dimensional brane configurations.  It was proposed that this was the resolution of the information problem.  The issue for me was that almost all his arguments were based on nearly supersymmetric black holes.  It did not seem that there was any extension to Schwarzschild black holes.  But along the way, Mathur sharpened the information paradox.  

In particular, he originated the argument that black hole complementarity violates strong subadditivity of the entropy, which was one of the arguments that AMPS gave.  I am sorry that our first and second versions did not acknowledge him on this; I think that because we found his central story about fuzzballs unpersuasive, we did not pay careful enough attention to the rest.  Indeed, you might wonder what is the difference between a `fuzzball' and a `firewall'.  What we had in mind was the horizon ending as a sea of bits, rather than some geometric structure that extends further.

\subsection{Following up}

It was fun to have once again kicked over the hive and watched the bees swarm.  Though I was a bit peeved that, after we had spent three months looking for flaws, within two weeks people were writing papers explaining why we were wrong without having fully thought through our arguments.  But happily the best of them, Raphael Bousso and Daniel Harlow, each recognized their error and withdrew their paper.  Susskind did the same, then changed back again, and by now is out on some perpendicular axis.

Of course, we only had an argument by contradiction, not a proof or even a calculation of what happens in the interior.  Even on the question of what time the firewall forms, we had only an upper bound, the Page time.  I had no good ideas for this, so I spent the next year or more reading what everyone else wrote in response to us.  Various alternatives to the firewall were developed and ideas were exchanged, with a workshop every few months: Stanford, CERN, KITP.  

AMPSS, the original AMPS group plus Douglas Stanford, a KITP grad fellow from Stanford, wrote a followup in which the arguments were clarified and sharpened.  We also pointed out problems with various alternatives to the firewall.  One advance was to put the black hole in AdS, where the boundary conditions gave greater control.  Black holes normally do not decay in AdS space, but by coupling to an additional heat bath (as in the earlier work of my student Rocha) one could do controlled thought experiments.  We also found a simplified argument for the firewall.  In its original form a very fast quantum computer was needed.  We showed that even without this there was a paradox, using the butterfly effect.

In a second followup, Marolf and I added some new arguments and observations.  There was a common question, does the firewall invalidate the calculation of Hawking radiation?  The usual calculation does depend on the geometry behind the horizon, but causality would seem to say that events behind the horizon could not affect the radiation.  By a statistical argument, we showed that the radiation was unaffected.  We also returned to the Schrodinger's cat question.  We showed that the firewall argument also made it impossible to see what is behind the horizon of an AdS black hole in the dual CFT.  This seems to contradict the general assumption about the CFT seeing the whole interior, but makes sense if spacetime ends at the firewall.

What surprised me was how many were willing to modify quantum mechanics in order to avoid the firewall. QM was not one of the explicit assumptions of black hole complementarity, it was implicit.  So I thought of this as a new alternative, `quantum drama.'  They differed from Hawking's modification of QM, which was visible in measurements outside the black hole.  These, including final state conditions (Lloyd and Preskill), limits on quantum computation (Harlow and Hayden), ER=EPR (Maldacena and Susskind), and state-dependent observables (Papadodimas and Raju), could be restricted to observations behind the horizon.  It is possible that one of these is true, but there are issues with each.  I particularly had an issue with state-dependence.   It sounds benign: aren't observables always state dependent?  Well, not in this way, which required that the Born rule of quantum mechanics be modified.  So Don and I wrote another paper, making this clear.  

Any of these modifications of quantum mechanics might turn out to be correct, but if you are going to modify QM you have a lot of explaining to do.  Of course the other alternative, a modification of the geometry, also needs a lot of explaining.  There were various ways this might be implemented: fire (AMPS), fuzz (Mather), strings (Silverstein), `nonviolent nonlocality' (Giddings).  As I have noted before, I am a natural agnostic, willing to examine any possibility.  

\subsection{Branes}

The black hole information problem still seemed like the best insight into the nature of quantum gravity, but after a while the issues seemed to solidify.  Perhaps we had to wait for a new insight, as with AdS/CFT.  So I was ready for a break.  Happily, D-branes still had their puzzles.  

My collaborator Karch was still working on AdS/CM, and his student Sichun Sun came to the KITP as a grad fellow with a puzzle.  Consider intersecting 0123 and 0145 D3-branes.  The 01 intersection degrees of freedom carry a $U(1)$ charged scalar.  Duality then requires also a $U(1)$ magnetically charged degree of freedom, but where could it come from: was it a independent field on the intersection, or a solitonic monopole?  Neither seemed to make sense: there was no independent magnetic degree of freedom on the branes, but a 1+1 dimensional intersection did not seem to leave room for a 3+1 dimensional magnetic soliton.  

So, together with my latest student Eric Mintun, we figured this out in four steps: (1) the ${\cal N} = 2$ implied that on a 1+1 dimensional intersection the scalar couples to the magnetic dipole in addition to the electric potential; (2) the 1+1 dimensions allowed higher dimensional interactions, and SUSY required them, thus leading to a 1+1 dimensional soliton; (3) there was a log divergence in the classical action, which could be treated by the usual process of renormalization (always nice to learn new wrinkles in renormalization, this from Goldberger and Wise); (4) because of the log, the effective field theory did not make sense up to infinite energy, one needed the branes.  So lots of cool field theory in a simple system.

Thinking about this new application of renormalization led to clarifying an old puzzle of mine.  The motion of a brane depends on its interactions with other branes.  But how does one treat the self-interaction, which is often divergent?  What is often done is ignore it, introducing the notion of a probe brane.  This was not a controlled approximation.  But having understood the classical renormalization of branes, it became clear that they should be understood in the language of effective field theory, with no probe approximation needed.
And as I was working this out with Mintun and an excellent undergrad Philip Saad, the perfect application came along.

Having followed the development of the KKLT model, and participating in part of it, I was puzzled by claims that it was unstable.  I tried to understand the arguments (which, incidentally, were largely due to my own former students Bena and Grana) but I could not.  So when their student (my grand-student) Andrea Puhm came to Santa Barbara as a postdoc, I tried once again to understand the issues.  And, they were nicely resolved by the new interpretation of branes: there was no way for a dangerous singularity to arise.  So Mintun, Puhm, Saad, a new student Ben Michel, and I wrote up both the correct interpretation of branes and the stability of KKLT.

Getting involved in KKLT led to much more correspondence, and eventually an invitation to speak at SUSY15.  Many people had arguments, or intuitions, that these de Sitter vacua could not exist.  The stakes were high.  If string theory had no such vacua, perhaps string theory was wrong.  If it had too few vacua, perhaps the anthropic argument would be ruled out.  But looking at the objections, most of them were clearly wrong; some appealed to no-go arguments that were known to be irrelevant even when they were first written down.  The most interesting objection had to do with the fact that the KKLT construction required both 10-dimensional and 4-dimensional analysis.  By careful treatment of scales, we showed that this could be justified in effective field theories.\footnote{It is worth noting, though, that without a nonperturbative construction of string theory (a.k.a. quantum gravity), the KKLT construction is still a conjecture.  It is another argument, beyond the information problem, that we are missing a nonperturbative construction of gravity.}

One more brane puzzle began with Michel studying different duality frames in string moduli spaces.  When Puhm joined us, she brought the Saclay point of view on fuzzballs as well as KKLT.  In hearing about the simplest (2-charge) case, we realized that the duality frames had not been fully taken into account.  So this became a nice exercise for Michel, Puhm, another postdoc Fang Chen, and me: the Journey to the Center of the Fuzzball.  We followed the different duality frames as we went down the throat, ending up with a geometry different from that previously assumed.\footnote{We later learned that Martinec and Sahakian had done this first for the zero spin case, not in the context of the fuzzball.}  Sticking to the highly supersymmetric two-charge geometry meant that we were not close to addressing the fundamental questions, but perhaps we learned something that will be useful down the road.

As an aside, some speakers will refer to their work as a game.  I have never liked this: physics is never a game for me. Everything is directed at the big questions, even if circuitously.  This is why I am in this field, not to play games.  And why is the public paying us?

\subsection{Precursors and Chaos}

Though the firewall puzzle was largely on hold for me, there were many related questions to follow now.  Some were motivated by the firewall, but many were motivated by the AdS/CM connection, and by ideas from quantum information.  The discovery of the Ryu-Takanagi (RT) formula generated an enormous wave of interest in the relation between entropy and geometry.  Ryu and Takanagi did their work at the KITP, and even came into my office at an early stage to ask what it might mean.  But having little intuition for entropy, and perhaps some skepticism about the result, I was of little help, and I missed my chance to be an early adopter.  Actually, I have not worked on RT yet.  Many people jumped into it, and I avoid doing things that other people could do.  Perhaps I will wait until they move on, and then look around for what might have been missed.

So my last paper with Almuhairi was motivated by AdS/CM, but ended up having some relevance to quantum gravity - it's all connected.  I had been puzzled by 0+1 conformal theories, which often came up in finite density systems.  When the transverse directions were compact, the symmetry implied that the low energy density of states had to be of the form $A \delta(E) + B/E$.  But the $A$ term comes only from zero energy, and the $B$ term has a divergence and can't continue down to zero energy.  So how could there be dynamical states in such systems, as there seemed to be?  So we looked at a simple model, based on the CGHS model.  I did not work on the first wave of CGHS, twenty years earlier, so I was happy to get a chance to study it.  What we found was that the interactions broke the conformal symmetry.  This has some current relevance because it happens in the SYK model (below).\footnote{Almuhairi went to Stanford as a postdoc, and did some remarkable work just below.}

The bulk to boundary operator map was an ongoing interest for me, most recently in the Schrodinger cat question.  I had played with it many times, as had others, but I had a sense that all we were doing was to rewrite the AdS/CFT dictionary of Gubser, Klebanov, Polyakov, and Witten.  So I was excited by a paper of Almheiri, Harlow, and Dong, which presented something new, perhaps for the first time in twenty years, casting it in terms of quantum information rather than differential equations.  In studying the paper, postdoc Vladimir Rosenhaus, Mintun, and I realized that in their nice toy model the quantum information argument could be rewritten in terms of gauge symmetry.  I think now that our result was just a special case, but sometimes one just has to throw one's hat into the ring.\footnote{Mintun is now a postdoc at British Columbia.}

Following up on the firewall paradox, Shenker and Stanford began studying the growth of small perturbations to black holes and the butterfly effect (chaos).  This was interesting, and I followed their work for a while until I had an idea of my own.  Their papers, like many others, focused on equilibrium black holes, shown by Israel and Maldacena to correspond to two-sided black holes.  I was used to the original information problem, with its one-sided state, and so I wanted to see how chaos would manifest there.  I quickly realized that it explained something that I had wondered about for a long time.

For more than twenty years, 't Hooft had been presenting what he said was the black hole S-matrix.  This did not make sense to me: it was in the framework of the quantum field theory of GR, with no information from string theory or other completion.  But Susskind had told me that one should always pay attention to 't Hooft, so I kept this in mind.  
In fact what he had calculated was not the S-matrix but the butterfly effect, the change in observables under a small change of state.  This could be  seen from the time-ordering of the operators.  As a side effect, it gave a new and more physical derivation of the firewall.\footnote{I was pleased to learn recently from Stanford that his work with Shenker had been spurred by our discussion of the butterfly effect and the firewall during our work on AMPSS.}

The subject became more interesting when Alexei Kitaev showed that the chaos in black holes had a characteristic Lyapunov exponent, and that there was a 0+1 matrix model, the SYK model, that exhibited this.  This had some similarity with my old 0+1 models with Iizuka and Okuda, so with postdoc Rosenhaus, student Michel, and visiting KITP grad fellow Josephine Suh, we looked at whether these models, designed to capture some of the behavior of black holes, might exhibit chaos with the right Lyapunov exponent.  Not surprisingly they did not, being too simple.

Kitaev is famous for not publishing his work, or delaying for years.  There was a lot of interest in it, which had appeared only in talks.  But Rosenhaus was a dogged calculator, and began to reproduce Kitaev's results, pulling me in.  So we obtained the spectrum and four-point functions of the SYK model, reproducing Kitaev's work and getting some new results.  Rosenhaus went on with Gross to develop a variety of extensions and variations.  They are well-matched, liking to talk and calculate for hours on end.

With a new student, Alex Streicher, I was trying to understand the latest from Papadodimas and Raju.  This led to a study of the analytically continued partition function.  On a trip to Stanford, we found that Shenker and his students were working on the same thing.  Eventually, with the addition of numerical types, the group grew to nine, Cotler, Gur-Ari, Hanada, Polchinski, Saad, Shenker, Stanford, Streicher, and Tezuka.

\subsection{Well, that sucks}

On Nov.~30, 2015 I gave a talk ``General Relativity and Strings'' at the meeting to celebrate the 100th anniversary of GR.  It was held at Harnack House in Berlin, where Einstein often worked and spoke.
I was scheduled to speak also the following week in Munich, at a rather different meeting.  This was to address whether such theories as strings and inflation were in fact theories.  I was looking forward to it, I felt that there were important points that were long overdue to be put forward.  My paper, `String Theory to the Rescue,' presented the case that string theory, though often criticized, was in fact a great success.  

Unfortunately I never gave the second talk, because three days after my talk at Harnack House I suffered a seizure that sent me to the hospital.\footnote{David Gross graciously presented my talk, but it was not the same.  You can read the original at https://arxiv.org/pdf/1512.02477.pdf, with follow-up https://arxiv.org/pdf/1601.06145.}  I was found to have brain cancer.  After many months of surgery, treatment, and recovery, I can write, as you see, but I still do not know whether I will be able to do physics again.

So Rosenhaus finished our last two papers, doing an outstanding job.  The other members of the group of nine finished their work and graciously kept my name on the paper, though I was only involved early on.  My student Michel developed collaborations with co-advisor Srednicki as well as several other faculty, students, and postdocs, and will be moving to UCLA as a postdoc.  My youngest students, 
Streicher and Milind Shyani both found new advisors, at Stanford, who graciously stepped in.  I have always thought highly of Stanford, forming sort of a West Coast axis with us in our interest in the important questions.

\section{Epilogue}

It is interesting to go through one's life like this.  It has taken a rather linear path, from the How and Why Wonder Books to today, with few deviations.  I have not achieved my early science fiction goals, nor explained why there is something rather than nothing, but I have had an impact on the most fundamental questions of science.  But it was a close thing: at the age of 40 you could say that I had not lived up to my potential.  And if someone else had stepped in during the six or more years between my finding D-branes and figuring out what they were good for, that might still be true.

How far are we from finding the fundamental theory of physics, and what will we learn from it?  Again, I am an agnostic, and not good at predicting things.  I only follow my nose.  Happily my nose is very busy, with the firewall, chaos, entanglement, and quantum information.  So we may be close, or we may still have big steps ahead.  I hope to help figure this out.


\end{document}